\documentclass[12pt,onecolumn]{IEEEtran}
\usepackage{amssymb}

\usepackage{amsmath,epsfig}

\def\dspace{\multiply\normalbaselineskip 160
            \divide\normalbaselineskip 100 \normalbaselines
            \csname @@normalbaselineskip\endcsname\normalbaselineskip}

\def\sspace{\multiply\normalbaselineskip 200
            \divide\normalbaselineskip 300 \normalbaselines
            \csname @@normalbaselineskip\endcsname\normalbaselineskip}

\newcommand{\bh}{\mbox{\boldmath $h$}}
\newcommand{\bb}{\mbox{\boldmath $b$}}
\newcommand{\br}{\mbox{\boldmath $r$}}
\newcommand{\btheta}{\mbox{\boldmath $\theta$}}

\newcommand{\by}{\mbox{\boldmath $y$}}

\def\bee{\mbox{\boldmath $e$}}

\def\sc{{\scriptstyle {\rm c}  }}
\def\be{\begin{equation}}
\def\ee{\end{equation}}
\def\half{{\textstyle{1\over 2}}}

\def\Exp{{\rm E}\,}

\def\R{\mbox{I{\kern-0.2em}R}}

\def\spark{\mathop{\rm spark}\nolimits}

\def\Exp{{\rm E}\,}

\def\newsection#1{\section{#1}}

 \def\be{\setcounter{subequation}{0}
        \begin{equation}}
\def\ee{\end{equation}}

\def\dspace{\multiply\normalbaselineskip 150
            \divide\normalbaselineskip 100 \normalbaselines
            \csname @@normalbaselineskip\endcsname\normalbaselineskip}

\def\sspace{\multiply\normalbaselineskip 200
            \divide\normalbaselineskip 300 \normalbaselines
            \csname @@normalbaselineskip\endcsname\normalbaselineskip}

\def\Exp{{\rm E}\,}

\def\Psiit{{\mathit \Psi}}
\def\Phiit{{\mathit \Phi}}

\def\half{{\textstyle{1\over 2}}}

\def\1over4{{\textstyle{1\over 4}}}
\def\oneover8{{\textstyle{1\over 8}}}

\def\sN{{\scriptstyle {N}}}

\def\schi2{{\scriptscriptstyle {\chi^2}}}

\def\sA{{\scriptstyle {A}}}

\def\cE{{ \cal E }}
\def\cQ{{ \cal Q }}
\def\cH{{ \cal H }}

\def\sML{{\scriptscriptstyle {\rm ML}}}
\def\sMN{{\scriptscriptstyle {\rm MN}}}

\def\sMAX{{\scriptscriptstyle {\rm MAX}}}

\def\sMIN{{\scriptscriptstyle {\rm MIN}}}

\def\bh0{\mbox{\boldmath $h_0$}}

\def\bb{\mbox{\boldmath $b$}}

\def\bh{\mbox{\boldmath $h$}}

\def\bb{\mbox{\boldmath $b$}}

\def\bz{\mbox{\boldmath $z$}}

\def\bee{\mbox{\boldmath $e$}}

\def\bn{\mbox{\boldmath $n$}}
\def\ba{\mbox{\boldmath $a$}}

\def\bmu{\mbox{\boldmath $\mu$}}

\def\bx{\mbox{\boldmath $x$}}
\def\br{\mbox{\boldmath $r$}}
\def\bs{\mbox{\boldmath $s$}}
\def\btheta{\mbox{\boldmath $\theta$}}

\def\by{\mbox{\boldmath $y$}}

\def\bx{\mbox{\boldmath $x$}}

\def\b0{\textbf{0} }

\def\Phiit{{\it \Phi}}

\def\Psiit{{\it \Psi}}

\def\Sigmait{{\it \Sigma}}

\def\sbs{{ \scriptstyle \boldsymbol{s}  }}
\def\sbz{{ \scriptstyle \boldsymbol{z}  }}

\def\sby{{ \scriptstyle \boldsymbol{y}  }}
\def\sbtheta{{ \scriptstyle \boldsymbol{\theta}  }}
\def\sTheta{{ \scriptstyle \Theta  }}

\newtheorem{definition}{Definition}
\newtheorem{lemma}{Lemma}

\newtheorem{theorem}{Theorem}

\renewcommand{\IEEEQED}{\IEEEQEDopen}

\def\appendix{\par
    \setcounter{section}{0} \setcounter{subsection}{0} \renewcommand
    {\theequation}
    {\Alph{section}.\arabic{quation}\alph{subequation}}
    \renewcommand{\thesection}{\Alph{section}}}

\def\appsection#1{
    \renewcommand{\thesection}{Appendix \Alph{section}.}
    \section{#1}
    \addtocounter{section}{-1}
    \renewcommand{\thesection}{\Alph{section}}
    \refstepcounter{section}
    \setcounter{equation}{0}
    \setcounter{subsection}{0}}

\def\newsection#1{\section{#1} \setcounter{equation}{0}}

\def\be{\setcounter{subequation}{0}
        \begin{equation}}
\def\ee{\end{equation}}

\newcounter{subequation}
\newcounter{oldsub}
\renewcommand
   {\theequation}
   {\arabic{section}.\arabic{equation}\alph{subequation}}

\def\baselabel#1
   {\setcounter{oldsub}{\value{subequation}}
    \setcounter{subequation}{0}
    \label{#1} \setcounter{subequation}{\value{oldsub}}}

\def\bses{\setcounter{subequation}{1}
          \begin{equation}}

\def\eses{\end{equation}
          \setcounter{subequation}{0}}

\def\bse{\stepcounter{subequation}
         \addtocounter{equation}{-1}
         \begin{equation}}

\def\ese{\end{equation}}

\def\refsetcounter#1#2{
   \setcounter{#1}{#2}}
\def\bsea{\begin{eqnarray}
          \refsetcounter{subequation}{1}}

\def\nosenumber{\nonumber
                \stepcounter{equation}
                \addtocounter{subequation}{-1}}

\def\Nextline{\\ \stepcounter{subequation}\addtocounter{equation}{-1}}

\def\esea{\end{eqnarray}\refsetcounter{subequation}{0}}

\def\df{\; \stackrel \triangle = \;}

\def\adots{\mathinner{\mkern1mu\raise1pt\vbox{\kern7pt\hbox{.}}
                      \mkern2mu\raise4pt\hbox{.}
                      \mkern2mu\raise7pt\hbox{.}\mkern1mu}}

\def\R{\mbox{I{\kern-0.2em}R}}

\def\boxit#1{\vbox{\hrule\hbox{\vrule\kern3pt
             \vbox{\kern3pt#1\kern3pt}\kern3pt\vrule}\hrule}}

\newcommand{\bzbar}{\mbox{\boldmath $\bar{z}$}}

\newcommand{\bstilde}{\mbox{\boldmath $\widetilde{s}$}}

\def\sADORE{{\scriptstyle {\rm ADORE}  }}

\def\PSNR{\mbox{PSNR}}
\def\Exp{{\rm E}\,}

\def\sinit{{\scriptstyle {\rm init}}}

\newcommand{\bthetahat}{\mbox{\boldmath $\widehat{\theta}$}}
\newcommand{\bthetatilde}{\mbox{\boldmath $\widetilde{\theta}$}}

\newcommand{\bshat}{\mbox{\boldmath $\widehat{s}$}}

\newcommand{\bztilde}{\mbox{\boldmath $\widetilde{z}$}}

\newcommand{\sigmahatsq}{\mbox{$\widehat{\sigma}^2$}}

\def\scalS{{ \scriptstyle {\cal S}  }}

\usepackage{url}

\makeatletter
\def\url@leostyle{%
  \@ifundefined{selectfont}{\def\UrlFont{\sf}}{\def\UrlFont{\small\ttfamily}}}
\makeatother
\begin{document}

\title{ECME Thresholding Methods for Sparse Signal Reconstruction
}
%
% Single address.
% ---------------
\author{ \normalsize   {\em Kun Qiu}\/ and  {\em Aleksandar Dogand\v{z}i\'c} \\
ECpE Department, Iowa State University\\ 3119 Coover Hall, Ames,
IA 50011\\ email:  \texttt{\{kqiu,ald\}@iastate.edu}
}

\date{}

\maketitle

\begin{abstract} We propose a probabilistic framework for interpreting
and developing hard thresholding sparse signal reconstruction methods
and present several new algorithms based on this framework. The
measurements follow an underdetermined linear model, where the
regression-coefficient vector is the sum of an unknown deterministic
sparse signal component and a zero-mean white Gaussian component with
an unknown variance. We first derive an {\em expectation-conditional
maximization either (ECME) iteration}\/ that guarantees convergence to
a local maximum of the likelihood function of the unknown parameters
for a given signal sparsity level.  To analyze the reconstruction
accuracy, we introduce the {\em minimum sparse subspace quotient
(SSQ)}, a more flexible measure of the sampling operator than the
well-established restricted isometry property (RIP).  We prove that,
if the minimum SSQ is sufficiently large, ECME achieves perfect or
near-optimal recovery of sparse or approximately sparse signals,
respectively.  We also propose a {\em double overrelaxation (DORE)}\/
thresholding scheme for accelerating the ECME iteration. If the signal
sparsity level is {\em unknown},\/ we introduce an {\em unconstrained
sparsity selection (USS)}\/ criterion for its selection and show that,
under certain conditions, applying this criterion is equivalent to
finding the sparsest solution of the underlying underdetermined linear
system. Finally, we present our {\em automatic double overrelaxation
(ADORE)} thresholding method that utilizes the USS criterion to select
 the signal sparsity level.  We apply the proposed schemes to
 reconstruct sparse and approximately sparse signals from tomographic
 projections and compressive samples.
%and
%compare them with existing approaches that are feasible for
%large-scale data.]
%Our numerical examples show that DORE is
%significantly faster than the ECME and related iterative hard
%thresholding (IHT) algorithms and that ADORE's reconstruction
%performance is comparable to that of the
%methods that require the prior knowledge about the sparsity
%level.
\end{abstract}

\begin{keywords}
%Compressive sampling,
Expectation-conditional maximization either (ECME) algorithm,
iterative hard thresholding, sparse signal reconstruction,
sparse subspace quotient, unconstrained sparsity selection, overrelaxation.
\end{keywords}

\dspace 12pt

\newsection{Introduction}
\label{Introduction}

\noindent
Sparsity is an important concept in modern signal processing.  Sparse
signal processing methods have been developed and applied to biomagnetic
and magnetic resonance imaging, spectral estimation, wireless sensing,
 and compressive sampling, see
\cite{GorodnitskyRao}--\cite{BrucksteinDonohoElad}
and references therein.
For noiseless measurements, the major sparse signal reconstruction task is
finding the sparsest solution of an underdetermined linear system $\by =
H \,
\bs$ (see e.g.\ \cite[eq.\ (2)]{BrucksteinDonohoElad}):
\be
\label{eq:P0}
({\rm P}_0): \quad \quad    \min_{\sbs} \| \bs \|_{\ell_0} \quad
\mbox{subject to} \,\, \by = H \, \bs
\ee
where $\by$ is an $N \times 1$ measurement vector,
$H$ is a known $N \times m$ full-rank {\em sensing matrix}\/ with $N \le m$,
 $\bs$ is an $m \times 1$ unknown {\em signal vector}\/,
and $\| \bs \|_{\ell_0}$ counts the number of nonzero elements in the signal vector $\bs$.
The $({\rm P}_0)$ problem requires combinatorial search and is known to be
NP-hard \cite{Natarajan}.

A number of tractable approaches have been proposed to find sparse
solutions to underdetermined systems. They can be roughly divided into
three groups: convex relaxation, greedy pursuit, and probabilistic
methods.  Convex methods replace the $\ell_0$-norm penalty with the
$\ell_1$-norm penalty and solve the resulting convex optimization
problem. Basis pursuit (BP)
 directly substitutes $\ell_0$ with $\ell_1$
in the $({\rm P}_0)$ problem, see \cite{ChenDonohoSaunders}. To combat
measurement noise and accommodate for approximately sparse signals,
several methods with various optimization objectives have been
suggested, e.g.\ basis pursuit denoising (BPDN)
\cite{CandesRombergTao}, \cite{ChenDonohoSaunders} and Dantzig selector
\cite{CandesTaoDantzig}. The gradient projection for sparse reconstruction
(GPSR) algorithm in \cite{FigueiredoNowakWright} solves the
unconstrained version of the BPDN problem in a
computationally efficient manner.
%it can be viewed as a
%gradient-projection algorithm for the convex quadratic program
%obtained by splitting the signal $\bs$ into positive and negative
%parts.
Greedy pursuit methods approximate the $({\rm P}_0)$ solution
in an iterative manner by making locally optimal choices. Orthogonal
matching pursuit (OMP)
\cite{MallatZhang}, \cite{Tropp}, \cite{TroppGilbert}, compressive
sampling matching pursuit (\textsc{CoSaMP}) \cite{NeedelTroppCoSaMP},
and iterative thresholding schemes
\cite{HerrityGilbertTropp}--\cite{BlumensathDavies2} belong to this category.
Probabilistic methods utilize full probabilistic models and
statistical inference tools to solve the sparse signal reconstruction
problem. Examples of the methods in this group are: sparse Bayesian
learning (SBL)
\cite{WipfRao}, Bayesian compressive sensing
(BCS) \cite{JiXueCarin} and expansion-compression variance-component
based method (\textsc{ExCoV}) \cite{QiuDogandzic10}.  Most existing
sparse signal reconstruction schemes require {\em tuning}\/
\cite{MalekiDonoho}, where the reconstruction performance depends
crucially on the choice of the tuning parameters.

%Amongst the above state-of-the-art sparse reconstruction approaches,
Iterative hard thresholding (IHT)
and normalized iterative hard thresholding  (NIHT) algorithms in
\cite{BlumensathDavies0}--\cite{BlumensathDavies2}
(see also \cite{HerrityGilbertTropp})
have attracted significant attention due to their low computation and
memory requirements and theoretical and empirical evidence of good
reconstruction performance. The IHT and NIHT methods require only
matrix-vector multiplications and {\em do not}\/ involve matrix-matrix
products, matrix inversions, or solving linear systems of equations.
The memory needed to implement IHT and NIHT is just
$\mathcal{O}(N m)$, and can be further reduced to $\mathcal{O}(m)$ if
the sensing operator $H$ is realized in a function-handle form.
%Both the computation and memory requirements
%are minimal for the sparse reconstruction problem.
%It is shown in
%\cite{BlumensathDavies0} that, for a specified sparsity level,
%IHT {\em converges}\/ to a local minimum of the squared residual error
%in \cite[eq.\ (1.6)]{BlumensathDavies0}, see
%\cite[Theorem 4]{BlumensathDavies0}. However, the
%convergence analysis in
%\cite{BlumensathDavies0} requires the spectral norm of the sensing
%matrix $H$ to be strictly less than one,  implying that IHT is
%{\em sensitive}\/ to the scaling of $H$.  In fact, if this
%spectral-norm condition fails, IHT may become unstable and diverge.
%The NIHT approach in \cite{BlumensathDavies2} overcomes this problem
%by introducing a normalizing coefficient in each IHT iteration
%\cite[eqs.\ (12)--(14)]{BlumensathDavies2}.  To guarantee that the
%squared residual error is monotonically nonincreasing, NIHT monitors
%the normalizing coefficient and adjusts it in each iteration, see
%\cite[eq.\ (14)]{BlumensathDavies2}. However, such monitoring and
%adjustment require additional computation, making NIHT typically
 % slower than IHT, see the
%simulation results in Section \ref{NumEx}.
However, the IHT and NIHT methods
\begin{itemize}
\item converge slowly, demanding a fairly large number of iterations,
\item require the knowledge of the signal sparsity level, which is a
      tuning parameter, and
\item  are sensitive to
scaling of the sensing matrix (IHT) or require elaborate adjustments in each
iteration to compensate for the scaling problem (NIHT).
%do not offer a universal (monitoring- and adjustment-free)
%      solution that is invariant to the scaling of $H$.
\end{itemize}
IHT and NIHT guarantee good recovery of the
underlying sparse signal if the sensing matrix satisfies the {\em
restricted isometry property (RIP)}\/ and a modified non-symmetric
RIP; see \cite{BlumensathDavies} and
\cite{BlumensathDavies2}, respectively. The restricted isometry
property was introduced in \cite{CandesTao} to measure how well
 sparse vectors preserve their magnitudes
 after being transformed by
the sensing matrix $H$. To preserve this magnitude for a sparsity
level $r$, any $r$ columns of $H$ must be
approximately orthonormal, which corresponds to $H$ having a small
{\em restricted isometry constant (RIC)}\/. We refer to
this requirement as the {\em RIP condition}\/. (See Section
\ref{ExactRecovery} for the definition of the RIC for
 sparsity level $r$, statement of the corresponding RIP condition, and
 further discussion.)  Besides being used to analyze the IHT and NIHT
 schemes, the RIP condition is a common ingredient of reconstruction
 performance analyses of many
sparse reconstruction methods, e.g.\ convex
methods \cite{CandesTao},
\cite{CandesRombergTao}, \cite{CandesTaoDantzig} and
\textsc{CoSaMP} \cite{NeedelTroppCoSaMP}. However,
\begin{itemize}
\item
the RIP condition is quite restrictive: a simple linear transform or even a scaling of $H$ by a constant can
easily break the equilibrium required by RIP.
\end{itemize}

The contribution of this paper is four-fold.

\textbf{\textit{1.\ Probabilistic model.}}
We propose a probabilistic framework for generalizing iterative hard
 thresholding (IHT) algorithms and interpreting them as
{\em expectation-conditional maximization either} (ECME) iterations,
see also \cite{QNDE09}. If the rows of the sensing matrix $H$ are
orthonormal, the signal update of the ECME iteration is equivalent to
one IHT step. Note that IHT is a greedy pursuit scheme whereas ECME is
a probabilistic scheme; hence, our framework blurs the boundary
between the two categories.

\textbf{\textit{2.\ Analysis.}}  We prove that our ECME iteration
monotonically converges to a fixed point corresponding to a local
maximum of the marginal likelihood function under our probabilistic
model. The conditions that we use in this convergence analysis are
{\em invariant}\/ to invertible linear transforms of either the rows
or the columns of $H$, which indicates that the convergence of our
ECME iteration is robust to linear transforms of $H$.  Such a
convergence robustness to linear transforms and scaling of $H$
is in contrast to the IHT convergence analysis
in \cite[Theorem 4]{BlumensathDavies0} that requires the spectral norm
of $H$ to be strictly less than one.

We also provide perfect and near-optimal guarantees for the
recovery of sparse and approximately sparse signals, respectively.
Our signal recovery analysis {\em does not}\/ rely on the common assumption
that $H$ has a sufficiently small RIC; rather, we
introduce new measures of $H$ useful for reconstruction analysis: the
$r$-sparse subspace quotient ($r$-SSQ) and minimum $r$-SSQ.  The
minimum $r$-SSQ measures how well sparse vectors with sparsity level
$r$ preserve their magnitudes after being projected onto the row space
of $H$, see Section \ref{ExactRecovery}.
 Unlike the RIC, the minimum $r$-SSQ is
{\em invariant}\/ to invertible linear transforms of the rows of $H$.
  We prove that, if the minimum $2r$-SSQ of the
sensing matrix is larger than $0.5$, our ECME algorithm for sparsity
level $r$
\begin{itemize}
\item
{\em perfectly}\/ recovers the true $r$-sparse signal
 from noiseless measurements and
\item estimates the best $r$-term approximation of an arbitrary {\em
non-sparse}\/ signal from {\em noisy}\/ measurements within a bounded
error.
\end{itemize}
Due to the row transform invariance of the minimum $r$-SSQ, our reconstruction analysis allows for
sensing matrices that violate the RIP condition: the columns of the sensing matrices can have arbitrary
magnitudes and be highly correlated. Therefore, our results {\em
widen}\/ the scope of sensing matrices that allow perfect or
satisfactory sparse reconstruction performance via tractable
algorithms.

\textbf{\textit{3.\ Convergence acceleration.}}  We develop a {\em
double overrelaxation (DORE)} thresholding method that interleaves two
overrelaxation steps with ECME steps, see also \cite{CISSDORE}. DORE
significantly {\em accelerates}\/ the convergence of the ECME
algorithm (and therefore of the IHT method as well, which is its
special case).  The line searches in the overrelaxation steps have {\em
closed-form}\/ solutions, making these steps computationally efficient.
The theoretical convergence and reconstruction
properties of ECME in {\bf 2} (above) apply to the DORE method as well.

\textbf{\textit{4.\ Signal sparsity level selection.}}  Finally, we propose
an {\em automatic double overrelaxation (ADORE)} thresholding method
that {\em does not} require the knowledge of the signal sparsity level.  To {\em
automatically}\/ select the sparsity level (i.e.\ estimate it from the
data), we introduce an {\em unconstrained sparsity selection (USS)}
model selection criterion.  We prove that, under certain mild
conditions, the unconstrained criterion USS is {\em equivalent}\/ to
the constrained $({\rm P}_0)$ problem (\ref{eq:P0}).  ADORE combines
the USS criterion and DORE iteration and applies a golden-section
search to maximize the USS objective function.

In Section \ref{MM}, we introduce our two-stage hierarchical
probabilistic model and the ECME thresholding algorithm (Section
\ref{ECME}). Our convergence and
near-optimal reconstruction analyses of the ECME iteration are
presented in Sections
\ref{Analysis} and \ref{ExactRecovery}, respectively.  In
Section \ref{DORE}, we describe the DORE thresholding method for
accelerating the convergence of the ECME iteration.
 In Section \ref{USSADORE}, we introduce the USS criterion
 and our ADORE thresholding scheme (Section \ref{ADORE}).
 In Section \ref{NumEx}, we compare the performances of the proposed and existing
large-scale sparse reconstruction methods via numerical
experiments. Concluding remarks are given in Section
\ref{Conclusion}.

\subsection{Notation and Terminology}
\label{Notation}

\noindent
We introduce the notation used in this paper:
\begin{itemize}
\item
${\cal N}( \by \,;\, \bmu, \Sigmait )$ denotes the multivariate
probability density function (pdf) of a real-valued Gaussian random
vector $\by$ with mean vector $\bmu$ and covariance matrix $\Sigmait$;
\item $| \cdot |$, $\| \cdot \|_{\ell_p}$, $\det( \cdot )$,
       ``$^T$'' denote the absolute value, $\ell_p$ norm,
       determinant, and
       transpose, respectively;
\item
  the smallest integer
larger than or equal to a real number $x$ is
  $\lceil x \rceil$;
\item  $I_n$, ${\bf 0}_{n \times 1}$, and  $0_{n \times m}$
are the identity matrix of size $n$, the $n \times 1$ vector of zeros,
and the $n \times m$ matrix of zeros, respectively;
\item $\lambda_{\min}(X)$ and  $\lambda_{\max}(X)$ are the minimum and
      maximum eigenvalues of a real-valued symmetric square matrix
      $X$;
\item $\spark(H)$ is the smallest number of  linearly dependent
      columns of a matrix $H$ \cite{BrucksteinDonohoElad};
\item $H_{\sA}$ denotes the
{\em restriction}\/ of the matrix $H$ to the index set $A$,
e.g.\ if $A = \{ 1,2,5 \}$, then $H_{\sA} = [ \bh_1 \, \bh_2 \, \bh_5 ]$,
where $\bh_i$ is the $i$th column of $H$;
\item
$\bs_{\sA}$ is the restriction of a column vector $\bs$
to the index set $A$, e.g.\ if $A = \{ 1,2,5 \}$, then $\bs_{\sA} = [
s_1, s_2, s_5 ]^T$, where $s_i$ is the $i$th element of $\bs$;
%\item $\frac{\partial f}{\partial \bs}(\ba)=\frac{\partial f (\bs)}{\partial \bs} \big|_{\sbs=\sba}$ and
%      $\frac{\partial^2 f}{\partial \bs \partial \bs^T}(\ba)=\frac{\partial^2 f (\bs)}{\partial \bs \partial \bs^T} \big|_{\sbs=\sba}$
%      denote the gradient and Hessian of multivariate function $f(\bs)$ evaluated at point $\ba$,
%    and $\frac{\partial f}{\partial s_i}(\ba)$ represents the $i$-th element of $\frac{\partial f}{\partial \bs}(\ba)$, namely the partial derivative
%    of $f(\bs)$ in the direction of $s_i$ evaluated at $\ba$;
\item ${\rm dim}(A)$ denotes the size of a set $A$;
\item ${\rm supp}(\bx)$ returns the support set of a vector
$\bx$,  i.e.\ the index set
      corresponding to the nonzero elements of $\bx$, e.g.\
${\rm supp}( [0,1,-5,0,3,0]^T ) = \{2,3,5\}$;
\item the thresholding operator ${\cal T}_r(\bx)$ keeps the $r$
largest-magnitude elements of a vector $\bx$ intact and sets the
rest to zero, e.g.\  ${\cal T}_2( [0,1,-5,0,3,0]^T  ) = [0,0,-5,0,3,0]^T$.
\end{itemize}

We refer to an $N \times m$ sensing matrix $H$ as {\em proper}\/ if it
has full rank and
\be
\label{eq:Nlessthanm}
N \leq m
\ee
which implies that the rank of $H$ is equal to $N$. Throughout this
paper, we assume that sensing matrices $H$ are proper, which is
satisfied in almost all practical sparse signal reconstruction scenarios.

\newsection{Probabilistic Model and the ECME Algorithm}
\label{MM}

\noindent
We model a $N \times 1$ real-valued
measurement vector $\by$ as
\bsea
\baselabel{eq:model}
\label{eq:model_1}
    \by = H \, \bz
\esea where $H$ is an $N \times m$ real-valued proper sensing matrix,
$\bz$ is an $m \times 1$ multivariate Gaussian vector with pdf
\addtocounter{equation}{-1}
\bsea
\stepcounter{subequation}
\label{eq:model_2}
p_{ \sbz \,|\, \sbtheta }( \bz \,|\, \btheta ) = {\cal N}( \bz \,;\, \bs, \sigma^2 \, I_m )
\esea
$\bs = [ s_1, s_2, \ldots, s_m ]^T$ is an {\em unknown} $m \times 1$
real-valued sparse signal vector
containing {\em at most} $r$ nonzero elements ($r \le m$), and
 $\sigma^2$ is an unknown {\em variance-component parameter}; we refer
 to $r$ as the {\em sparsity level}\/ of the signal and to the signal
 $\bs$ as being {\em $r$-sparse}\/. Note that $\| \bs \|_{\ell_0}={\rm dim} ({\rm supp} (\bs))$ counts the
number of nonzero elements in $\bs$; we refer to $\| \bs
\|_{\ell_0}$ as the {\em support size}\/ of $\bs$.
Therefore, the
support size  $\| \bs \|_{\ell_0}$ of the $r$-sparse vector $\bs$ is
less than or equal to the sparsity level $r$.
The set of unknown parameters is
\be
\btheta = ( \bs, \sigma^2 ) \in \Theta_{r}
\ee
with the parameter space
\bsea
\label{eq:parameterspace}
\Theta_r = {\cal S}_r \times [0,+\infty)
\esea
where
\addtocounter{equation}{-1}
\bsea
\stepcounter{subequation}
\label{eq:signalparameterspace}
{\cal S}_r =  \{ \bs \in {\cal R}^m: \, \| \bs \|_{\ell_0} \leq r \, \}
\esea
is the sparse signal parameter space.  The marginal likelihood function of
$\btheta$ is obtained by {\em integrating $\bz$ out}\/ [see
(\ref{eq:model})]:
\bsea
\label{eq:likelihoodfunction}
    p_{ \sby \,|\, \sbtheta }( \by \,|\, \btheta  ) = {\cal N}(  \by \,;\,  H \, \bs, \sigma^2 \, H \, H^T )
\esea
where the fact that $H$ is a proper sensing matrix ensures that
$H \, H^T$ is invertible and, consequently, that the pdf
(\ref{eq:likelihoodfunction}) exists.
For a given sparsity level $r$, the maximum likelihood (ML)
estimate of $\btheta$ is
\addtocounter{equation}{-1}
\bsea
\stepcounter{subequation}
\label{eq:thetahatr}
\widehat{\btheta}_{\sML}(r) = \big( \widehat{\bs}_{\sML}(r),
\widehat{\sigma}^2_{\sML}(r) \big) = \arg \max_{ \sbtheta \in \sTheta_{r} } p_{ \sby \,|\, \sbtheta }\big( \by \,|\, \btheta \big).
\esea
For any fixed $\bs$, the
marginal likelihood (\ref{eq:likelihoodfunction}) is maximized by
\be
\label{eq:sigma2_hat}
\widehat{\sigma}^2(\bs) = ( \by - H \, \bs )^T \, (H \, H^T)^{-1} \, ( \by - H \, \bs ) \,/\, N.
\ee
Therefore, maximizing (\ref{eq:likelihoodfunction}) with respect to
$\btheta$ is equivalent to first maximizing the {\em concentrated
likelihood function}\/
\be
\label{eq:profilelikelihood}
    p_{ \sby \,|\, \sbtheta }( \by \,|\, \bs, \widehat{\sigma}^2(\bs)  )
= \frac{1}{ \sqrt{ \det(2 \, \pi \, H \, H^T) } } \,  [ \widehat{\sigma}^2(\bs) ]^{-0.5 \, N} \, \exp(-0.5 \, N)
\ee
with respect to $\bs \in {\cal S}_r$, yielding
$\widehat{\bs}_{\sML}(r)$, and then determining the ML estimate of
$\sigma^2$ by substituting $\widehat{\bs}_{\sML}(r)$ into
(\ref{eq:sigma2_hat}).  Obtaining the exact ML estimate
$\widehat{\btheta}_{\sML}(r)$ in (\ref{eq:thetahatr}) requires a
combinatorial search and is therefore infeasible in practice. We now
present a computationally feasible iterative approach that aims at maximizing
(\ref{eq:likelihoodfunction}) with respect to $\btheta \in \Theta_{r}$
and circumvents the combinatorial search.

\subsection{ECME Algorithm For Known Sparsity Level $r$}
\label{ECME}

\noindent We treat $\bz$ as the {\em missing (unobserved) data}\/ and
present an ECME algorithm for approximately finding the ML estimate in
(\ref{eq:thetahatr}), assuming a fixed sparsity level $r$.  Since the
sparsity level $r$ is assumed known, we simplify the notation and omit
the dependence of the estimates of $\btheta$ on $r$ in this section
and in  Appendix \ref{Appder}.
An ECME algorithm maximizes {\em either}\/ the expected complete-data
log-likelihood function (where the expectation is computed with
respect to the conditional distribution of the unobserved data given
the observed measurements) {\em or}\/ the actual observed-data
log-likelihood, see
\cite[Ch.\ 5.7]{McLachlanKrishnan}.

Assume that the parameter estimate $\btheta^{(p)} = \big( \bs^{(p)},
(\sigma^2)^{(p)} \big)$ is available, where $p$ denotes the iteration
index.  {\em Iteration $p+1$} proceeds as (see Appendix \ref{Appder}
for its derivation):
\begin{itemize}
\item
update the sparse signal estimate using the
expectation-maximization (EM) step, i.e.\ the expectation (E) step:
\bsea
\baselabel{eq:ECMEonestep}
\label{eq:zpplusone}
\bz^{(p+1)} = \Exp_{\sbz \,| \sby,  \,  \sbtheta }[\bz \,|\, \by,
\btheta^{(p)}] = \bs^{(p)}
+ H^T \, \big( H \, H^T )^{-1} \, ( \by - H \, \bs^{(p)} )
\esea
followed by the maximization (M) step, which simplifies to
\addtocounter{equation}{-1}
\bsea
\stepcounter{subequation}
\label{eq:ECMEzdeltapplusone}
\bs^{(p+1)} = \arg \min_{  \sbs \in {\cal S}_r } \|
\bz^{(p+1)} - \bs  \|_{\ell_2}^2  = {\cal T}_r\big( \bz^{(p+1)}  \big)
\esea
and
\item
update the variance component estimate
using the following
conditional maximization (CM) step:
\addtocounter{equation}{-1}
\bsea
\stepcounter{subequation}\stepcounter{subequation}
\label{eq:ECMEdeltapplusone}
(\sigma^2)^{(p+1)} = ( \by - H \, \bs^{(p+1)} )^T \, (H
\, H^T)^{-1} \, ( \by - H \, \bs^{(p+1)} ) \big/ N
\esea
obtained by maximizing the marginal likelihood
(\ref{eq:likelihoodfunction}) with respect to $\sigma^2$ for a
fixed $\bs = \bs^{(p+1)}$, see (\ref{eq:sigma2_hat}).
\end{itemize}
In (\ref{eq:zpplusone}), $\Exp_{\sbz \,|\, \sby,\sbtheta}[ \bz \,|\,
\by, \btheta ]$ denotes the mean of the
pdf $p_{\sbz \,|\, \sby,\sbtheta}(\bz \,|\, \by , \btheta)$, which is
the Bayesian minimum mean-square error (MMSE) estimate of $\bz$ for
{\em known}\/ $\btheta$ \cite[Sec.\ 11.4]{Kay}.  Note that $( H \, H^T
)^{-1}$ can be pre-computed before the iteration starts or well
approximated by a diagonal matrix; hence, our ECME iteration {\em does
not}\/ require matrix inversions. See Section \ref{Complexity} for
detailed discussion on the complexity of the ECME method.  If the rows
of the sensing matrix $H$ are orthonormal:
\be
\label{eq:orthsensingmatrixcond}
H \, H^T = I_{\sN}
\ee
then the EM step
in (\ref{eq:zpplusone})--(\ref{eq:ECMEzdeltapplusone}) is
equivalent to one {\em iterative hard-thresholding (IHT) step}\/ in
\cite[eq.\ (10)]{BlumensathDavies}.

The above ECME algorithm does not satisfy the general regularity
conditions assumed in standard convergence analysis of the EM-type
algorithms in e.g.\ \cite{McLachlanKrishnan} and \cite[Theorem
2]{LiuRubin}. In particular,
\begin{itemize}
\item
the complete-data and conditional unobserved data given the observed
data distributions $p_{\sbz, \sby \,|\, \sbtheta}( \bz, \by \,|\,
\btheta )$ and $p_{\sbz \,|\, \sby ,\sbtheta}( \bz \,|\, \by , \btheta
)$ are both degenerate, see (\ref{eq:model_1}) and Appendix
\ref{Appder};
\item
the parameter space $\Theta_r$ is non-convex and its interior is
empty;
\item
in $\Theta_r$, the partial derivatives of the marginal likelihood
(\ref{eq:likelihoodfunction}) with respect to the components of $\bs$
do not exist for most directions.
\end{itemize}
Therefore, we establish the convergence of our ECME iteration afresh
in the following section.

%However, the
%convergence of the ECME algorithm (\ref{eq:ECMEonestep}) is fairly
%low, requiring a large number of iterations; consequently, the IHT
%algorithm converges slowly as well. In the following section, we
%propose our
%DORE method that accelerates the convergence of the ECME
%iteration.

\newsection{Convergence Analysis of the ECME Algorithm}
\label{Analysis}

\noindent We now answer the following questions.  Does the ECME
iteration in Section \ref{ECME} ensure monotonically non-decreasing
marginal likelihood (\ref{eq:likelihoodfunction}), does it converge to
a fixed point and, if yes, is this fixed point a local or the global
maximum of the marginal likelihood function? How do we define a local
maximum in the parameter space $\Theta_r$ in
(\ref{eq:parameterspace})?
Since the sparsity level $r$ is fixed, we omit
the dependence of the estimates of $\btheta$ on $r$ in this section
and in Appendices \ref{AppProofThm1} and \ref{AppProofThm2} that contain
the proofs of the results of this section.

%As in Section \ref{MM}, we adopt the mild
%assumption that the $N \times m$ sensing matrix $H$ has full rank and
%that (\ref{eq:Nlessthanm}) holds.

Maximizing the concentrated
likelihood function (\ref{eq:profilelikelihood}) with respect to $\bs
\in {\cal S}_r$ is
equivalent to minimizing the weighted squared error
\be
\label{eq:cE}
    \cE(\bs) =  N \, \widehat{\sigma}^2(\bs) = ( \by - H \, \bs )^T \, (H \, H^T)^{-1} \, ( \by - H \, \bs ).
\ee The following identity holds for all $\bs \in {\cal R}^m$ and
$\bs' \in {\cal R}^m$:
\bsea
\label{eq:cE_identity}
\cE(\bs) = \cQ(\bs \,|\, \bs') - \cH(\bs \,|\, \bs')
\esea
where
\addtocounter{equation}{-1}
\bsea
\stepcounter{subequation}
\label{eq:Qfunction}
\cQ(\bs \,|\, \bs') &=& \|   \bs' + H^{T} (H H^T)^{-1} (\by - H \bs'
)  - \bs  \|^2_{\ell_2}
\Nextline
\label{eq:Hfunction}
\cH(\bs \,|\, \bs') &=& (\bs - \bs')^T \, [ \, I_m - H^T \, ( H  H^T)^{-1} \, H \, ] \, (\bs - \bs').
\esea
This identity follows by rewriting (\ref{eq:Qfunction}) as
$\cQ(\bs \,|\, \bs')=\|   (I_m - H^{T} (H H^T)^{-1} H) (\bs'- \bs) +
H^{T} (H H^T)^{-1} (\by - H \bs )  \|^2_{\ell_2}$ and expanding the
squares. Observe that $\cH(\bs \,|\, \bs')$ is
minimized at $\bs=\bs'$.

Denote by $\bs^{(p)}$ the estimate of $\bs$ obtained in {\em
Iteration $p$} of our ECME iteration.
When we set $\bs'=\bs^{(p)}$, $\cQ(\bs \,|\, \bs^{(p)}) = \|
\bz^{(p+1)} - \bs \|_{\ell_2}^2$ becomes exactly the expression that is
minimized in the M step (\ref{eq:ECMEzdeltapplusone}) and, consequently,
\bsea
\label{eq:Qineq}
    \cQ(\bs^{(p+1)} \,|\, \bs^{(p)}) \leq \cQ(\bs^{(p)} \,|\, \bs^{(p)}).
\esea
Since $\cH(\bs \,|\, \bs^{(p)})$ is
minimized at $\bs=\bs^{(p)}$, we have
\addtocounter{equation}{-1}
\bsea
\stepcounter{subequation}
\label{eq:Hineq}
    \cH(\bs^{(p+1)} \,|\, \bs^{(p)}) \geq \cH(\bs^{(p)} \,|\, \bs^{(p)}).
\esea
Subtracting (\ref{eq:Qineq}) from (\ref{eq:Hineq}) and using
(\ref{eq:cE_identity}) yields
\be
   \cE(\bs^{(p+1)})   \leq  \cE(\bs^{(p)})
\ee
and, therefore, our ECME iteration (\ref{eq:ECMEonestep}) ensures
a {\em monotonically non-decreasing} marginal likelihood
(\ref{eq:likelihoodfunction}), see also
(\ref{eq:profilelikelihood}). Monotonic convergence is also a key
general property of the EM-type algorithms
\cite{McLachlanKrishnan}. Furthermore, since (\ref{eq:cE}) is bounded
from below by zero, the sequence $\cE(\bs^{(p)})$ must converge to a
limit as the iteration index $p$ grows to infinity.

However, the fact that $\cE(\bs^{(p)})$ converges does not necessarily imply
that $\bs^{(p)}$ converges to a fixed point. The
following theorem establishes convergence of
 the ECME signal iterates $\bs^{(p)}$.

\begin{theorem} Assume that
 the sparsity level $r$ satisfies
\bsea
\label{eq:rcondfixedpoint}
r \leq \half \, (m-N)
\esea
and that the sensing matrix $H$ satisfies the unique representation
property (URP) \cite{GorodnitskyRao} stating that all $N
\times N$ submatrices of $H$ are invertible or, equivalently, that
\addtocounter{equation}{-1}
\bsea
\stepcounter{subequation}
\label{eq:URP}
    \spark(H) = N+1.
\esea
Then, the ECME signal iterate $\bs^{(p)}$ for sparsity level
$r$ converges monotonically to its fixed point as the iteration index
$p$ grows to infinity.
%$\btheta^{\star}=(\bs^{\star},(\sigma^2)^{\star})$.
\label{theorem1}
\end{theorem}
\begin{IEEEproof}
See Appendix \ref{AppProofThm1}.
\end{IEEEproof}

Note that (\ref{eq:rcondfixedpoint}) is a mild condition.  In
practice, $N \ll m$ and (\ref{eq:rcondfixedpoint}) specifies a large
range of sparsity levels $r$ for which
the ECME iteration converges to its
fixed point.

Theorem \ref{theorem1} guarantees the convergence of our ECME
iteration to a fixed point. However, can we guarantee that
 this fixed point is a local or
the global maximum of the marginal log-likelihood function
(\ref{eq:likelihoodfunction})? To answer this question, we first
define the local maximum of a function over the parameter space ${\cal
S}_r$ in (\ref{eq:signalparameterspace}).

\begin{definition}
\label{def1}
\textbf{\textit{$\br$-local maximum and minimum.}} For a function $f(\bs): {\cal
R}^m \rightarrow {\cal R}$, a vector $\bs^{\star} \in {\cal S}_r$ is
an $r$-local maximum point of $f(\bs)$
if there exists a $\delta > 0$,
such that, for all $ \bs \in {\cal S}_r$ satisfying $ \| \bs -
\bs^{\star} \|_{\ell_2} < \delta $, we have
\be
\nonumber
f(\bs^{\star}) \ge f(\bs).
\ee
Then, $f(\bs^{\star})$ is the corresponding $r$-local maximum of $f(\bs)$.
We define $\bs^{\star} \in {\cal S}_r$ and $f(\bs^{\star})$ as an
$r$-local minimum point and the corresponding $r$-local
minimum of $f(\bs)$ if $\bs^{\star}$ is an $r$-local maximum point for
the function $-f(\bs)$.
\end{definition}

\begin{figure}
\begin{center}
\parbox{5in}{
    \centering \includegraphics[width=5in]{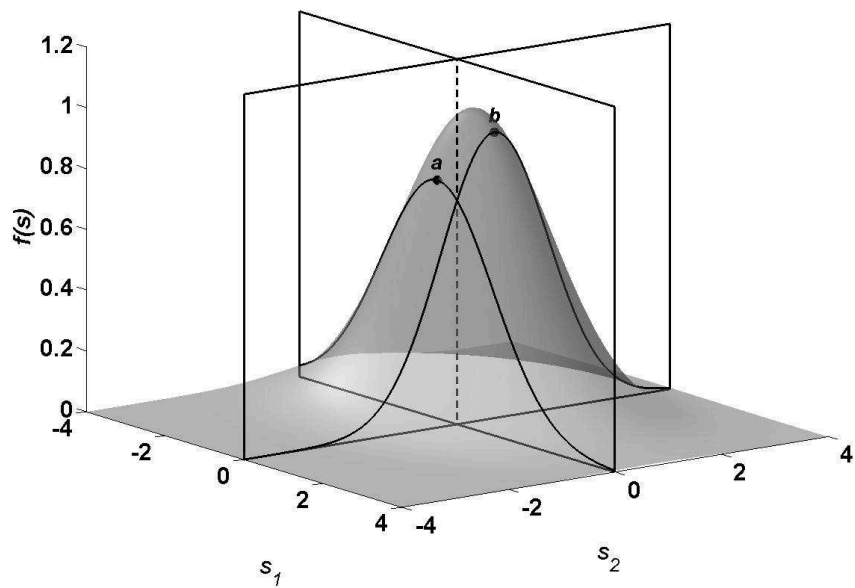}
}
\end{center}

\caption{Function $f(\bs)=\exp\{-0.5 \, [(s_1+0.5)^2+(s_2-0.7)^2]\}$
with $\bs = [s_1,s_2]^T$ and two $1$-local maxima of $f(\bs)$.}
\label{fig:rlocalmax_example}
\end{figure}

Definition \ref{def1} states that an $r$-sparse vector is a $r$-local
maximum (or minimum) point of a function $f(\bs)$ if, in some small
neighborhood, this vector attains the largest (or smallest) function
value among all the sparse vectors within that small neighborhood.
Fig.\ \ref{fig:rlocalmax_example} illustrates this concept using $\bs
= [s_1,s_2]^T$ (i.e.\ $m=2$) and $f(\bs)=\exp\{-0.5 \,
[(s_1+0.5)^2+(s_2-0.7)^2]\}$.  For the sparsity level $r=1$, the points
$\ba=[-0.5,0]^T$ and $\bb=[0,0.7]^T$ are the only two $1$-local
maximum points of $f(\bs)$. Observe that $\ba$ and $\bb$ are {\em
not}\/ local maximum points of $f(\bs)$ when $\bs$ is unconstrained in
${\cal R}^2$.

The following lemma provides a necessary condition for an $r$-local
maximum or minimum point of a differentiable function.

\begin{lemma}
\label{lemma2}
If an $r$-sparse vector $\bs^{\star} \in {\cal S}_r$ is an $r$-local
maximum or minimum point of a differentiable function $f(\bs): {\cal R}^m
\rightarrow {\cal R}$, then, for all $i \in \{1,2,\ldots,m\}$ such
that
\bsea
\label{eq:lemma2cond}
{\rm dim} \big( \{i\} \cup {\rm supp}(\bs^{\star}) \big) \le r
\esea
we have
\addtocounter{equation}{-1}
\bsea
\stepcounter{subequation}
\label{eq:lemma2}
\frac{\partial f(\bs)}{\partial s_i}\Big|_{\bs=\bs^{\star}} = 0.
\esea
\end{lemma}
\begin{IEEEproof}
See Appendix \ref{AppProofThm2}.
\end{IEEEproof}

The condition (\ref{eq:lemma2cond}) of Lemma \ref{lemma2} implies
that, instead of checking that all partial derivatives of our function
are zero (which is required in the standard first-derivative test for
finding local maxima and minima), we only need to check its
derivatives along a few {\em allowed}\/ coordinate axes, where the
allowed coordinate axes are defined by the property that perturbing
along these axes does not violate the sparsity requirement, see
(\ref{eq:lemma2cond}). If $\bs^\star$ has exactly $r$ nonzero
elements, then $i$ in (\ref{eq:lemma2cond}) must belong to ${\rm
supp}(\bs^{\star})$, and we should only check the $r$ partial
derivatives that correspond to the nonzero components of
$\bs^{\star}$. For example, consider Fig.\
\ref{fig:rlocalmax_example}: to determine if $\ba=[-0.5,0]^T$ is a
$1$-local maximum point, we only need to check that the partial
derivative of $f(\bs)$ with respect to $s_1$ is zero at $\bs=\ba$;
the direction along the $s_2$ axis is not allowed because the perturbation
along this direction violates the sparsity requirement. However, when
$\bs^\star$ has less than $r$ nonzero elements, we must check all
partial derivatives, because perturbing along any axis
will not exceed the sparsity requirement.

We now provide a sufficient condition for an $r$-local
maximum or minimum point of a twice differentiable function.

\begin{lemma}
\label{lemma4}
An $r$-sparse vector $\bs^{\star} \in {\cal S}_r$ is an $r$-local
maximum or minimum of a twice differentiable function $f(\bs): {\cal R}^m
\rightarrow {\cal R}$ if
\begin{description}
\item[(1)] for all
$i \in \{1,2,\ldots,m\}$ such that ${\rm dim} \big( \{i\}
\cup {\rm supp}(\bs^{\star}) \big) \le r$, we have
\be
\label{eq:cond1}
\frac{\partial f(\bs)}{\partial s_i}\Big|_{\bs = \bs^{\star}} = 0
\ee
and
\item[(2)] there exists a
$\delta > 0$, such that, for all $\bs \in {\cal S}_r$ satisfying
$\| \bs - \bs^{\star} \|_{\ell_2} < \delta$, the Hessian
matrix
\[
\frac{\partial^2 f(\bs)}{\partial \bs \, \partial \bs^T}
\]
is negative semidefinite (for a maximum) or positive semidefinite (for
a minimum).
\end{description}
\end{lemma}
\begin{IEEEproof}
See Appendix \ref{AppProofThm2}.
\end{IEEEproof}

In the example depicted in
Fig. \ref{fig:rlocalmax_example}, both points $\ba$ and $\bb$ satisfy
the two conditions of Lemma \ref{lemma4} and are therefore the $r$-local
maxima of $f(\bs)$. Lemma \ref{lemma4} is useful in developing the following
theorem stating that our ECME algorithm actually converges to an
$r$-local maximum point of the concentrated marginal likelihood
function (\ref{eq:profilelikelihood}).

\begin{theorem}
\label{theorem2}
If the sensing matrix $H$ is proper and
ECME iteration in Section \ref{ECME} converges to a fixed point
$\btheta^{\star}=(\bs^{\star},(\sigma^2)^{\star})$, then $\bs^{\star}$
is an $r$-local maximum point of the concentrated marginal likelihood
function (\ref{eq:profilelikelihood}).
\end{theorem}
\begin{IEEEproof}
See Appendix \ref{AppProofThm2}.
\end{IEEEproof}

Based on Theorems \ref{theorem1} and \ref{theorem2}, we claim that, if
$H$ satisfies the URP condition (\ref{eq:URP}) and for a sufficiently
small sparsity level $r$, the ECME algorithm in Section
\ref{ECME} converges to a fixed point that is an $r$-local maximum of
the concentrated marginal likelihood function
(\ref{eq:profilelikelihood}).

The conditions of Theorems \ref{theorem1} and \ref{theorem2} hold even
when the sensing matrix $H$ is pre- or post- multiplied by a full rank
square matrix.
In contrast, the IHT algorithm converges to a local
minimum of the squared residual error for a specified sparsity level
%\cite[eq.\
%(1.6)]{BlumensathDavies0}
%\footnote{Note that
%\cite{BlumensathDavies0} adopts
% (\ref{eq:squaredresidual}) as the cost function for its
%convergence analysis of IHT, assuming a fixed sparsity level, see \cite[eq.\
%(1.6)]{BlumensathDavies0}.}
%\be
%\label{eq:squaredresidual}
%\| \by - H \bs \|_{\ell_2}^2
%\ee
{\em only}\/ if $H$ is appropriately scaled.
Indeed, Theorem 4 in \cite{BlumensathDavies0} demands that the
spectral norm of the sensing matrix $H$ is less than unity.
If the spectral norm condition is violated, the IHT
iteration may become unstable and diverge, see
\cite[Sec.\ II-D]{BlumensathDavies2}. To overcome such scaling
requirements and ensure convergence for an arbitrary scaled $H$, a
normalized IHT (NIHT) method has been proposed in
\cite{BlumensathDavies2}, where a scaling term is introduced to the
original hard thresholding step; this term must be monitored and
adjusted in each iteration so that it does not exceed a certain
threshold (see \cite[e.q.\ 14]{BlumensathDavies2}); otherwise, the
squared residual error \cite[eq.\ (1.6)]{BlumensathDavies0}
 is not guaranteed to decrease during the iteration. However, this
 monitoring and adjustment consume CPU time and typically {\em slow
 down}\/ the resulting algorithm, see the numerical examples in
 Section \ref{NumEx}. In contrast, Theorems \ref{theorem1} and
 \ref{theorem2} assert that the monotonic convergence of our ECME
 iteration is not affected by the
 pre- and post-multiplication of $H$ by any full-rank
square matrix of appropriate size,
 thus removing the need for monitoring and adjustment
within the iteration steps.

%For the sparsity level not exceeding the number of measurements (i.e.\ $r
%\leq N$), if we fix the support set of the signal $\bs$, there
%is at most one unique $r$-local minimum point of $\cE(\bs)$.
%Since there are $m \choose r$ choices of the support set, the total number of $r$-local minima of $\cE(\bs)$
%is at most $m \choose r$.

\newsection{Sparse Subspace Quotient and Near-optimal ECME Reconstruction}
\label{ExactRecovery}

\noindent We now study theoretical guarantees for near-optimal ECME
reconstruction.  We first define the $r$-sparse subspace quotient
($r$-SSQ) as a normalized squared magnitude of the projection of an
$r$-sparse signal onto the row space of the sensing matrix. We
introduce minimum $r$-sparse subspace quotient of the sensing matrix
as a separability measure for arbitrary $r$-sparse signals, discuss
its properties, compare it with the existing popular measures such as
restricted isometry and coherence, and use it to establish a condition
for uniqueness of the solution to the $({\rm P}_0)$ problem.
We then show that, in the absence of noise and if the minimum
$2r$-sparse subspace quotient is sufficiently large, our ECME
algorithm estimates the true unknown $r$-sparse signal {\em
perfectly}\/ from the linear measurements. We also give an example of
the existence of low-dimensional matrices that satisfy our perfect
recovery requirement (Section
\ref{lodwimensionalmatrixforperfectrecovery}). We finally show that,  for
non-sparse signals and noisy measurements,
the ECME iteration for the sparsity level $r$ recovers the best
$r$-term approximation of the true signal within a bounded error.
Both the noiseless and noisy reconstruction guarantees hold regardless
of the initial estimate of the signal parameters $\btheta$ employed by
the ECME iteration.

\begin{definition}
\label{def_SSQ}
\textbf{\textit{$\br$-Sparse Subspace Quotient ($\br$-SSQ) and minimum
$\br$-SSQ.}}
We define the $r$-sparse
subspace quotient of  a nonzero
$r$-sparse vector $\bs$ of size $m \times 1$ (i.e.\ $\bs \in {\cal
S}_r \backslash {\bf 0}_{m \times 1}$)
and a proper $N \times m$ sensing matrix $H$ as
the ratio of the squared magnitude of the
projection of $\bs$ onto the row space of $H$ and the squared
magnitude of $\bs$:
\bsea
\label{eq:def_SSQ_1}
\rho_r(\bs,H) \df
\frac{ \| H^T \, (H \, H^T)^{-1} \, H \, \bs
\|_{\ell_2}^{2}}{\|\bs\|_{\ell_2}^2}
= \frac{   \bs^T \, H^T \, (H \, H^T)^{-1} \, H \, \bs  }{ \bs^T \, \bs  }.
\esea
Define the corresponding minimum $r$-sparse subspace
quotient of the sensing matrix $H$ as
\addtocounter{equation}{-1}
\bsea
\stepcounter{subequation}
\label{eq:def_SSQ_2}
\rho_{r,\min}(H) \df \min_{\bs \in {\cal S}_r  \backslash {\bf 0}_{m \times 1}} \rho_r(\bs,H).
\esea
\end{definition}
Note that $H^T \, (H H^T)^{-1} \, H$ is the projection matrix onto the
row space of $H$ and the second equality in  (\ref{eq:def_SSQ_1})
follows from the fact that the projection matrix is idempotent.

The following lemma summarizes a few useful properties of $r$-SSQ and
minimum $r$-SSQ.
\begin{lemma}
\label{lemma5}
For an $N \times m$ proper sensing matrix $H$, a nonzero $r$-sparse vector $\bs$ of size $m
\times 1$, and a sparsity level $r$ satisfying $0 < r \leq m$,
\begin{description}
\item[(a)]  $\rho_r(\bs,H)$ in
(\ref{eq:def_SSQ_1})
can be equivalently defined as
\bsea
\label{eq:OPQvar}
\rho_r(\bs,H) = \frac{ \bs_{\sA}^T \, H_{\sA}^T \, (H \, H^T)^{-1} \,
H_{\sA} \, \bs_{\sA}}{ \bs_{\sA}^T \, \bs_{\sA}  }
\esea
where $A={\rm supp}(\bs)$ is the support set of $\bs$ and
 $\rho_{r,\min}(H)$
can be
 determined by the following equivalent optimization:
\addtocounter{equation}{-1}
\bsea
\stepcounter{subequation}
\label{eq:lemma5_1_1}
\rho_{r,\min}(H) = \min_{A \subseteq \{1,2,\ldots,m\} , \,  {\rm dim}(A)=r} \lambda_{\min} \big( H_{\sA}^T \, (H \, H^T)^{-1} \, H_{\sA} \big);
\esea
\item[(b)]  $\rho_r(\bs,H)$ and $\rho_{r,\min}(H)$ are
invariant to invertible linear
transforms of the rows of $H$, i.e.\
\be
\label{eq:lemma5_2}
\rho_r(\bs,H) = \rho_r(\bs,G \, H), \quad \rho_{r,\min}(H) = \rho_{r,\min}(G \, H)
\ee
for any full-rank $N \times N$ matrix $G$;
\item[(c)] $\rho_r(\bs,H)$ and $\rho_{r,\min}(H)$ are bounded as follows:
\be
\label{eq:lemma5_3}
0 \le \rho_{r,\min}(H) \leq \rho_r(\bs,H) \leq 1
\ee
where $\rho_{r,\min}(H)$ attains
\begin{itemize}
\item the lower bound $0$ when $r > N$
and
\item the upper bound $1$ when $N=m$;
\end{itemize}
\item[(d)]
if and only if $H$ has at least $r$ linearly independent columns, i.e.\
\be
\label{eq:lemma5_4_1}
\spark(H)>r
\ee
the following strict inequality holds:
\be
\label{eq:lemma5_4}
\rho_{r,\min}(H) > 0;
\ee
%\item[(d)]
%assuming $0 < r \leq N$ and that $H$ satisfies the URP condition
%(\ref{eq:URP}), the following strict inequality holds:
%\be
%\label{eq:lemma5_4}
%\rho_{r,\min}(H) > 0,
%\ee
\item[(e)] if $0< r_1 < r_2$, then
\be
\rho_{r_1,\min}(H) \ge \rho_{r_2,\min}(H).
\ee
\end{description}
\end{lemma}
\begin{IEEEproof}
See Appendix \ref{AppProofThm34}.
\end{IEEEproof}

We now compare minimum $r$-SSQ with
the commonly used restricted isometry property (RIP)
\cite{IEEESpMag}, \cite{CandesTao},
\cite{CandesRombergTao}, \cite{CandesTaoDantzig},
\cite{NeedelTroppCoSaMP},
\cite{BlumensathDavies}, \cite{BlumensathDavies2} and coherence
\cite{BrucksteinDonohoElad}, \cite{Tropp}, \cite{TroppGilbert}, \cite{HerrityGilbertTropp}, \cite{DonohoHuo2001},
\cite{DonohoElad2003}. The
idea behind RIP is to upper-bound deviations of the squared magnitude
of $H \, \bs$ from the squared magnitude of $\bs$ for arbitrary
nonzero $r$-sparse vectors $\bs$; therefore, the following quotient
should be close to unity for arbitrary nonzero $r$-sparse $\bs$:
\be
\label{eq:RIP_term}
\frac{ \| H \bs \|_{\ell_2}^{2}}{\|\bs\|_{\ell_2}^2}= \frac{   \bs^T \, H^T \, H \, \bs  }{ \bs^T \, \bs  }.
\ee
%and, therefore, is a
%measure of how well the sensing matrix $H$ preserves the magnitude of
%nonzero $r$-sparse signal vectors after being transformed by $H$.
%Observe
%that (\ref{eq:RIP_term}) has a similar form as $r$-SSQ in
%(\ref{eq:def_SSQ_1}).
The restricted isometry constant (RIC) for sparsity level $r$ can be
written as (see \cite[e.q.\ (1.7)]{CandesTao}):
\be
\label{eq:RIP}
\gamma_r (H) = \max_{\bs \in {\cal S}_r  \backslash {\bf 0}_{m \times 1}} \Big| 1 - \frac{ \| H \bs \|_{\ell_2}^{2}}{\|\bs\|_{\ell_2}^2} \Big|
=\max_{\bs \in {\cal S}_r \backslash {\bf 0}_{m \times 1}} \Big| 1 -
\frac{ \bs^T \, H^T \, H \, \bs }{ \bs^T \, \bs } \Big|
\ee
which quantifies the largest-magnitude deviation of (\ref{eq:RIP_term}) from unity.
Clearly, the smaller the $r$-RIC
is, the closer to orthonormal any $r$ columns of $H$ are. The assumption
that the appropriate RIC is sufficiently small is key for
sparse-signal recovery analyses of the IHT algorithms
\cite{BlumensathDavies},
\cite{BlumensathDavies2},
\textsc{CoSaMP}
\cite{NeedelTroppCoSaMP},
and convex relaxation methods
 \cite{CandesTao},
\cite{CandesRombergTao}, \cite{CandesTaoDantzig}.
%\cite[Theorems 4,5]{BlumensathDavies},
%\cite[Theorem 4]{BlumensathDavies2},
% convex relaxation methods
% \cite[Theorems 1.3, 1.4]{CandesTao},
%\cite[Theorem 1]{CandesRombergTao}, \cite[Theorem 1.1]{CandesTaoDantzig}, and \textsc{CoSaMP}
%\cite[Theorem A]{NeedelTroppCoSaMP}.
The coherence measures the largest-magnitude
inner product of any two distinct columns of $H$.
The assumption that the coherence is small is a basis for
sparse-signal recovery analyses of convex relaxation methods in
\cite{DonohoElad2003} and
of greedy methods (such as OMP) in
\cite{Tropp}.
%\cite[Theorem 2]{DonohoElad2003} and
%of greedy methods (such as OMP) in
%\cite[Theorem 3.5]{Tropp}.
However, both the RIP and coherence requirements are somewhat fragile:
a simple linear transform or even a scaling of $H$ by a constant can
easily break the equilibria required by the RIP or coherence.  In
comparison, the minimum $r$-SSQ in (\ref{eq:def_SSQ_2}) measures the
smallest normalized squared magnitude of the projection of an
$r$-sparse signal onto the row space of the sensing matrix $H$. Here,
it is the {\em row space}\/ of $H$ that matters, rather than $H$
itself. Lemma \ref{lemma5}~(b) states that the minimum $r$-SSQ is {\em
invariant}\/ to invertible linear transforms of the rows of $H$.
Therefore, the sensing matrix $H$ can be pre-multiplied by any $N
\times N$ full-rank matrix,\footnote{Unlike the ECME convergence
analysis in Section \ref{Analysis}, invertible linear transforms of
the columns of $H$ are generally not allowed here.} leading to
arbitrary column magnitudes and highly correlated columns, while still keeping the
same minimum $r$-SSQ value. Hence, minimum $r$-SSQ is a more flexible
property of $H$ than RIP and coherence.

We now utilize the minimum SSQ measure to establish a condition under which
the solution to the $({\rm P}_0)$ problem is unique and leads to exact
recovery under the noiseless scenario. A similar problem is considered
in \cite[Lemma 1.2]{CandesTao} and \cite[Theorem
2]{BrucksteinDonohoElad}, where such uniqueness and exact recovery
conditions have been derived using RIP (\ref{eq:RIP}) and $\spark$.
\begin{lemma}
\label{lemma6}
Suppose that we have collected a measurement vector $\by = H
\bs^{\diamond}$ using a proper sensing matrix $H$,
where $\bs^{\diamond}$ is a sparse signal vector having exactly
 $\| \bs^{\diamond} \|_{\ell_0} = r^\diamond$ nonzero elements.
If the minimum $2 r^\diamond$-SSQ
of the sensing matrix $H$ is strictly positive:
\be
\label{eq:lemma6cond}
\rho_{2 r^\diamond,\min}(H) > 0
\ee
then the solution to the $({\rm P}_0)$ problem (\ref{eq:P0}) is unique
and coincides with $\bs^{\diamond}$.
\end{lemma}
\begin{IEEEproof}
See Appendix \ref{AppProofThm34}.
\end{IEEEproof}

Observe that the condition (\ref{eq:lemma6cond}) implies that
the number of measurements $N$ is larger than or equal to twice the
support size of the true sparse signal $\bs^{\diamond}$, i.e.\
\be
N \geq 2 \, r^\diamond = 2 \, \| \bs^{\diamond} \|_{\ell_0}.
\ee
Indeed, if $N < 2 \, r^\diamond$, $\rho_{2 r^{\diamond},\min}(H)=0$ by
part (c) of Lemma \ref{lemma5}.

Lemma \ref{lemma6} also holds if we replace $r^{\diamond}$
in the condition (\ref{eq:lemma6cond}) with any $r>r^{\diamond}$,
which follows from part (e) of Lemma \ref{lemma5}: if $\rho_{2 r,\min}(H) > 0$
for $r^{\diamond} < r$, then $\rho_{2 r^\diamond,\min}(H) \ge \rho_{2
r,\min}(H) > 0$.  Therefore, (\ref{eq:lemma6cond}) is the weakest
condition on $H$ among all $r \ge r^\diamond$.

In a nutshell, Lemma \ref{lemma6} states that, for a strictly positive
$\rho_{2 r,\min}(H)$, any two distinct $r$-sparse vectors can be
distinguished from their projections onto the row space of $H$, which
furthermore guarantees the uniqueness of the $({\rm P}_0)$ problem.
%In general, Lemma \ref{lemma6} is {\em stronger}\/ than
Note that \cite[Lemma 1.2]{CandesTao} states that
the solution to the $({\rm P}_0)$ problem (\ref{eq:P0}) is unique
and coincides with $\bs^{\diamond}$
if the $2 r^\diamond$-RIC of the sensing matrix $H$ satisfies
\be
\label{eq:RIP_P0cond}
\gamma_{2 r^\diamond}(H) < 1.
\ee
However, for proper sensing matrices, the condition
(\ref{eq:lemma6cond}) of Lemma \ref{lemma6} is {\em weaker}\/ than
(\ref{eq:RIP_P0cond}): (\ref{eq:RIP_P0cond}) implies that $\spark(H) >
2 \, r^\diamond$ [see (\ref{eq:RIP})] and, consequently,
 (\ref{eq:lemma6cond}), but not vice versa.
For example,
the $2 \times 3$ sensing matrix
\be
H=\left(
\begin{array}{ccc}
1 & 0 & 1
\\
0 & 1 & 1
\end{array}
\right)
\ee
satisfies the condition (\ref{eq:lemma6cond}) with $\rho_{2,\min}(H) =
1/3 > 0$, but violates (\ref{eq:RIP_P0cond}), since its 2-RIC is
$\gamma_2(H)=1.618 > 1$. Hence, (\ref{eq:lemma6cond}) {\em does not}\/
imply (\ref{eq:RIP_P0cond}).
%We also found another example where a
%random Gaussian
%$5 \times 6$ sensing matrix with normalized columns (i.e.\ with each
%column having
%unit $\ell_2$%norm) that
% has positive minimum 4-SSQ constant and the 4-RIP constant greater
% than $1$.
%Therefore, Lemma \ref{lemma6} indicates that our minimum $r$-SSQ is a
%less restrictive measure than RIP.

We now develop reconstruction performance guarantees for our ECME algorithms
that employ the minimum $r$-SSQ measure.
%The next theorem shows that, if $\rho_{2r,\min}(H)$ is
%sufficiently large, our ECME algorithm recovers an $r$-sparse signal
%{\em exactly}\/ from noiseless linear measurements.

\begin{theorem}
\label{theorem3}
\textbf{\textit{Exact Sparse Signal Reconstruction From Noiseless Samples.}}
Suppose that we have collected a measurement vector
\bsea
\label{eq:thm3cond1}
\by = H \, \bs^{\diamond}
\esea
where $\bs^{\diamond} \in {\cal S}_r$ is an $r$-sparse signal vector,
i.e.\ $\| \bs^{\diamond}
\|_{\ell_0} \leq r$.  If the minimum $2 r$-SSQ
 of the sensing matrix $H$ satisfies
\addtocounter{equation}{-1}
\bsea
\stepcounter{subequation}
\label{eq:thm3cond3}
\rho_{2r,\min}(H) > 0.5
\esea
then the ECME iteration for the sparsity level $r$ in Section
\ref{ECME} converges to the ML estimate of $\btheta$:
\addtocounter{equation}{-1}
\bsea
\stepcounter{subequation}\stepcounter{subequation}
\label{eq:thm3convergencepoint}
\widehat{\btheta}_{\sML}(r) = (\bs^{\diamond},0)
\esea
and therefore recovers the true sparse signal
$\bs^{\diamond}$ perfectly.
\end{theorem}
\begin{IEEEproof}
See Appendix \ref{AppProofThm34}.
\end{IEEEproof}

Theorem \ref{theorem3} shows that, upon convergence and if the minimum
$2 r$-SSQ of the sensing matrix is sufficiently large, the ECME
algorithm recovers the true sparse signal $\bs^{\diamond}$ {\em
perfectly}\/ from the noiseless measurements.  In this case, the ECME
iteration converges to the {\em global maximum}\/ of the marginal
likelihood (\ref{eq:likelihoodfunction}), which is infinitely large
since the ML estimate of $\sigma^2$ is zero.  This global convergence
is guaranteed regardless of the initial estimate of $\btheta$ used to
start the ECME iteration. In addition, by Lemma
\ref{lemma6}, $\bs^{\diamond}$ is also the unique solution to the
$({\rm P}_0)$ problem. Therefore, under the conditions of Theorem
\ref{theorem3}, the ECME algorithm solves the $({\rm P}_0)$
problem as well.

Next, we consider a more practical scenario where the true signal $\bs^{\diamond}$ is not
strictly sparse and the measurements $\by$ are corrupted by noise.

\begin{theorem}
\label{theorem4}
\textbf{\textit{Near-Optimal Recovery of Non-sparse Signal From Noisy Samples.}}
Suppose that we have collected a measurement vector
\bsea
\label{eq:thm4cond1}
\by = H \bs^{\diamond} + \bn
\esea
where
the signal $\bs^{\diamond}$ is not necessarily sparse
and $\bn \in {\cal R}^N$ is a noise vector. Denote by
$\bs_r^{\diamond}$ the best $r$-term $\ell_2$-norm approximation to
$\bs^{\diamond}$, i.e.\
\addtocounter{equation}{-1}
\bsea
\stepcounter{subequation}
\bs_r^{\diamond} = \arg \min_{ \sbs \in {\cal S}_r }
\| \bs - \bs^{\diamond} \|_{\ell_2} =  {\cal T}_r( \bs^{\diamond} )
\esea
and by $\bs^{\star}$ the $r$-sparse signal estimate obtained upon
convergence of the ECME iteration for the sparsity level $r$ in Section
\ref{ECME}.  If the minimum $2 r$-SSQ of the sensing matrix $H$
satisfies
\addtocounter{equation}{-1}
\bsea
\stepcounter{subequation}\stepcounter{subequation}
\label{eq:thm4cond}
\rho_{2r,\min}(H) > 0.5
\esea
which is the same as the
condition (\ref{eq:thm3cond3}) in Theorem \ref{theorem3}, then
\addtocounter{equation}{-1}
\bsea
\stepcounter{subequation}\stepcounter{subequation}\stepcounter{subequation}
\label{eq:theorem4}
\| \bs^{\star} - \bs_r^{\diamond} \|_{\ell_2} \leq
2 \, \frac{ \| \bs^{\diamond} - \bs_r^{\diamond} \|_{\ell_2} + \|H^T \, (H  H^T)^{-1} \bn \|_{\ell_2} }{\sqrt{\rho_{2r,\min}(H)} - \sqrt{1-\rho_{2r,\min}(H)}}.
\esea
\end{theorem}
\begin{IEEEproof}
See Appendix \ref{AppProofThm34}.
\end{IEEEproof}

Theorem \ref{theorem4} shows that, for a general (not necessarily sparse) signal
$\bs^{\diamond}$ and noisy measurements
satisfying (\ref{eq:thm4cond1}) and sensing matrix satisfying
(\ref{eq:thm4cond}), the ECME estimate is close to the
best $r$-term $\ell_2$-norm approximation of $\bs^{\diamond}$.
 This result holds regardless of the initial estimate of $\btheta$
employed by the ECME iteration.
Observe that, by (\ref{eq:lemma5_3}),
$\rho_{2r,\min}(H) \leq 1$ and therefore the squared roots in
(\ref{eq:theorem4}) are well-defined.  Moreover, since
(\ref{eq:thm4cond}) holds, the denominator on the right-hand
side of (\ref{eq:theorem4})
 is positive and less than or equal to one. When the noise $\bn$
 is zero and
signal $\bs^{\diamond}$ is $r$-sparse, the quantities $\|
\bs^{\diamond} - \bs_r^{\diamond} \|_{\ell_2}$ and $\|H^T \, (H
H^T)^{-1} \bn \|_{\ell_2}$ in (\ref{eq:theorem4}) are zero and, therefore,
$\|\bs^{\star}-\bs^{\diamond}\|_{\ell_2} = 0$,
consistent with Theorem \ref{theorem3}.

Performance guarantees
similar to those in Theorems \ref{theorem3} and  \ref{theorem4} have been developed
for other sparse reconstruction methods. However, these results rely
on either small RIP constants (see e.g.\ \cite[Theorems 1.3,
1.4]{CandesTao}, \cite[Theorem 1]{CandesRombergTao}, \cite[Theorem
1.1]{CandesTaoDantzig} and
\cite[Theorem A]{NeedelTroppCoSaMP}, \cite[Theorems 4, 5]{BlumensathDavies}, \cite[Theorem 4]{BlumensathDavies2})
or small coherence (see e.g.\ \cite[Theorem 2]{DonohoElad2003} and
\cite[Theorem 3.5]{Tropp}). Therefore, all previous results require
that a certain numbers of columns of the
sensing matrix $H$ are approximately orthonormal (RIP) or
orthogonal (coherence). In contrast, our analysis of the ECME method in Theorems
\ref{theorem3} and \ref{theorem4} applies to the cases where the columns of $H$ are not
approximately orthonormal and can be heavily correlated, thus widening
the class of sensing matrices for which it is possible to derive
reconstruction performance guarantees, see also the discussion after
Lemma \ref{lemma5}.

%
%Thanks to the transformation invariance of the $r$-SSQ constant in
%(\ref{eq:lemma5_2}), Theorems \ref{theorem3} and \ref{theorem4} hold
%if the sensing matrix $H$ is scaled or left-multiplied by a full-rank
%$N \times N$ matrix; hence, the columns of $H$ do not need to be
%approximately orthonormal or maintain low coherence.
%In contrast, the similar recovery results for other sparse reconstruction methods
%rely on either small RIP constant or small coherence. See also the discussion
%after Lemma \ref{lemma5}.

%approximate orthogonality of the columns of $H$
%is required by the recovery analysis (of what algorithm?) based on the RIP condition in
%Theorem 4 of \cite{BlumensathDavies}, where the
%sensing matrix must have a small $3 r$-restricted isometry
%constant.

\subsection{An Example of a Low-dimensional Matrix Satisfying The Conditions of
Theorems \ref{theorem3} and  \ref{theorem4}}
\label{lodwimensionalmatrixforperfectrecovery}

\noindent The ongoing search for desirable sensing matrices focuses on
small RIP constants and on asymptotic behavior of {\em large}\/ random
matrices, e.g.\ Gaussian, Bernoulli (with entries equal to $1$ and
$-1$), and Fourier (randomly selected rows of the DFT matrix)
matrices, see e.g.\ \cite{CandesTao} and
\cite{CandesTao_NearOptimal}. We now show that it is possible to find
{\em low-dimensional}\/ sensing matrices that satisfy the condition
$\rho_{2r,\min}(H) > 0.5$ of Theorems
\ref{theorem3} and
\ref{theorem4}.

%We suggest an
%alternative criterion: search for sensing matrices with large
%SSQ constants.
%Here, we illustrate that it is possible to find
%{\em low-dimensional}\/ sensing matrices that satisfy the condition
%$\rho_{2r,\min}(H) > 0.5$ of Theorems \ref{theorem3} and
%\ref{theorem4}.
Consider the $21 \times 32$ sensing matrix $H$ comprised of the 21
rows of the $32 \times 32$ type-II discrete cosine transform (DCT)
matrix (see e.g. \cite[Sec.\ 8.8.2]{OppenheimSchafer}) with
indices
\be
 2, 3, 4, 5, 7, 9, 10, 12, 13, 14, 16, 18, 20, 21, 22, 24, 27, 29, 30, 31, 32.
\ee
It can be verified by combinatorial search that the minimum
$2$-SSQ of $H$ meets the condition (\ref{eq:thm3cond3}):
\be
\rho_{2,\min}(H) = 0.503 > 0.5
\ee
and, therefore, by Theorem \ref{theorem3}, for this $H$,
our ECME iteration {\em perfectly}\/ recovers any $1$-sparse signal $\bs$
from the 21 noiseless linear measurements given by $\by = H \bs$.
(We have checked and confirmed the validity of this statement via
numerical simulations.)
However,  the $2$-RIC
of the same $21 \times 32$ sensing matrix $H$ is
\be
\gamma_2(H) = 0.497.
\ee
which violates the condition required in the theoretical analysis
of the IHT algorithm \cite[Theorems 4 and
 5]{BlumensathDavies}. In particular, Theorems 4 and 5 in
 \cite{BlumensathDavies} require $\gamma_3(H) < 1/\sqrt{32} \approx
 0.177$ for $1$-sparse signals,
but here $\gamma_3(H) \geq \gamma_{2}(H) = 0.497$. We have
checked that the above sensing matrix $H$ also violates the condition
required in the theoretical analysis of the NIHT algorithm
\cite[Theorem 4]{BlumensathDavies2}.  Indeed, for $1$-sparse signals
and the above sensing matrix $H$, the {\em non-symmetric restricted
isometry}\/ constant in \cite{BlumensathDavies2} is at least $0.611$, which is
larger than the upper limit $0.125$, see \cite[Theorems
4]{BlumensathDavies2}.

Due to the invariance property of Lemma \ref{lemma5}~(b), any
invertible linear transformation of the rows of $H$ {\em preserves}\/ the
minimum $r$-SSQ constant. Therefore, upon finding one good sensing
matrix $H$ that satisfies $\rho_{2r,\min}(H) > 0.5$, we can construct
infinitely many matrices that satisfy this condition.

\newsection{The DORE Algorithm for Known $r$}
\label{DORE}

\noindent We now present the DORE thresholding method that accelerates
the convergence of our ECME iteration.  Since the sparsity level $r$
is assumed known, we omit the dependence of the estimates of $\btheta$
on $r$ in this section.

Assume that two consecutive estimates of the unknown parameters
$\btheta^{(p-1)} = ( \bs^{(p-1)}, (\sigma^2)^{(p-1)} )$ and
$\btheta^{(p)} = ( \bs^{(p)}, (\sigma^2)^{(p)} )$ are available from
the $(p-1)$-th and $p$-th iterations, respectively.  {\em Iteration $p+1$}
proceeds as follows:

{\bf 1.\ ECME step.} Compute
\bsea
\baselabel{eq:ECMEDORE}
\label{eq:DOREzhat}
\!\!\!\!\!\!\!\!\!\!\!\!\!\!\!\!\!
\bshat &\!\!\!\!=\!\!\!\!& {\cal T}_r\big( \bs^{(p)} + H^{T} (H H^T)^{-1} (\by - H
\bs^{(p)} ) \big)
\Nextline
\label{eq:DOREdeltahat}
\!\!\!\!\!\!\!\!\!\!\!\!\!\!\!\!\!
\sigmahatsq &\!\!\!\!=\!\!\!\!& ( \by - H \, \bshat
)^T \, (H \, H^{T})^{-1} \, ( \by - H \, \bshat )
\big/ N
\esea
and define
$\bthetahat = ( \bshat, \sigmahatsq )$.

{\bf 2.\ First overrelaxation.}
Compute the linear combination of $\bshat$ and $\bs^{(p)}$:
\bsea
\label{eq:overrelax1}
\bzbar = \bshat + \alpha_1 \, (   \bshat - \bs^{(p)})
\esea
where the weight
\addtocounter{equation}{-1}
\bsea
\stepcounter{subequation}
\label{eq:DOREalpha1}
\alpha_1 = \frac{( H \, \bshat - H \, \bs^{(p)})^T \, (H
 H^T)^{-1} \, (\by - H  \, \bshat)}{ ( H \,
 \bshat - H \, \bs^{(p)})^T \, (H H^T)^{-1} \, (
H \, \bshat -  H \, \bs^{(p)})}
\esea
is the {\em closed-form}\/ solution of the line search:
\addtocounter{equation}{-1}
\bsea
\stepcounter{subequation}\stepcounter{subequation}
\label{eq:DOREalpha1prob}
\!\!\!\!
\alpha_1 \!=\! \arg \max_{\alpha}
p_{\sby \,|\, \sbtheta}\big( \by \,|\, ( \bshat + \alpha \, (   \bshat - \bs^{(p)}), \sigma^2 )  \big)
\esea
with the parameter space of $\btheta$ extended to $\Theta_{r_1}$, where
$r_1={\rm dim}({\rm supp}(\bshat) \cup {\rm supp}(\bs^{(p)}))$
is the sparsity level of $\bshat + \alpha \, (   \bshat - \bs^{(p)})$
and $\sigma^2$ is an arbitrary positive number, see also
(\ref{eq:likelihoodfunction}).
%Here, the optimization problem in
%(\ref{eq:alpha1}) can be rewritten as the following weighted least
%squares (WLS) minimization:
%\begin{eqnarray*}
%\alpha_1 &=& \arg \min_{\alpha}
%\{ \by -  H \, [ \bshat + \alpha \, ( \bshat - \bs^{(p)}) ] \}^T
%\\
%& &
%\cdot  (H \, H^{T})^{-1}
%\, \{ \by -  H \, [ \bshat + \alpha \, ( \bshat - \bs^{(p)}) ] \}.
%\end{eqnarray*}

{\bf 3.\ Second overrelaxation.}
Compute  the linear combination of $\bzbar$ and  $\bs^{(p-1)}$:
\bsea
\label{eq:overrelax2}
\bztilde = \bzbar + \alpha_2 \, ( \bzbar - \bs^{(p-1)})
\esea
where the weight
\addtocounter{equation}{-1}
\bsea
\stepcounter{subequation}
\alpha_2 =
\frac{( H \, \bzbar - H \,  \bs^{(p-1)})^T \, (H \, H^T)^{-1} \,
(\by - H \, \bzbar)}{ ( H \, \bzbar - H \, \bs^{(p-1)})^T \,
 (H \, H^T)^{-1} \, ( H \, \bzbar - H \, \bs^{(p-1)}) }
\label{eq:DOREalpha2}
\hspace{0.05in}
\esea
is the closed-form solution of the line search:
\addtocounter{equation}{-1}
\bsea
\stepcounter{subequation}\stepcounter{subequation}
\label{eq:DOREalpha2prob}
    \alpha_2 \!=\! \arg \max_{\alpha}
p_{\sby \,|\, \sbtheta}\big( \by \,|\, ( \bzbar + \alpha \, ( \bzbar - \bs^{(p-1)})
, \sigma^2 )  \big)
\esea
with the parameter space of $\btheta$ extended to $\Theta_{r_2}$,
where $r_2={\rm dim}({\rm supp}(\bzbar) \cup {\rm supp}(\bs^{(p-1)}))$
is the sparsity level of $\bzbar + \alpha \, ( \bzbar - \bs^{(p-1)})$
and $\sigma^2$ is an arbitrary positive number.
%Here, the optimization problem in (\ref{eq:alpha2}) can be rewritten as
%the following WLS minimization:
%\begin{eqnarray*}
%\alpha_2 &=& \arg \min_{\alpha}
%\{ \by -  H \, [ \bzbar + \alpha \, ( \bzbar - \bs^{(p-1)}) ] \}^T
%\\
%& &
%\cdot  (H \, H^{T})^{-1}
%\, \{ \by -  H \, [ \bzbar + \alpha \, ( \bzbar - \bs^{(p-1)}) ] \}.
%\end{eqnarray*}

{\bf4.\ Thresholding.}
Threshold $\bztilde$ to the sparsity level $r$:
\bsea
\label{eq:DOREztilde}
\bstilde = {\cal T}_r(\bztilde)
\esea
compute the corresponding variance component estimate:
\addtocounter{equation}{-1}
\bsea
\stepcounter{subequation}
\label{eq:DOREdeltatilde}
\widetilde{\sigma}^2 =  ( \by - H \, \bstilde )^T \, (H \, H^{T})^{-1} \, ( \by - H \, \bstilde ) \big/ N
\esea
and define our final overrelaxation parameter estimate $\bthetatilde =
( \bstilde, \widetilde{\sigma}^2 )$.

  {\bf 5.\ Decision (between ECME and thresholded overrelaxation
parameter estimates).}  If $p_{ \sby \,|\, \sbtheta }(\by \,|\, \bthetatilde) \ge p_{ \sby \,|\, \sbtheta }(\by \,|\,
\bthetahat) $ or, equivalently, if
\be
\label{eq:DOREdeltacheck}
    \widetilde{\sigma}^2 < \widehat{\sigma}^2
\ee
assign $\btheta^{(p+1)}=\bthetatilde$; otherwise, assign
$\btheta^{(p+1)}=\bthetahat$ and complete {\em Iteration $p+1$}\/.

Iterate until two consecutive sparse-signal estimates $\bs^{(p)}$ and
$\bs^{(p+1)}$ do not differ significantly.  Since $( H \, H^T )^{-1}$
can be pre-computed, our DORE iteration {\em does not}\/ require
matrix inversion; the line searches in the two overrelaxation steps
have closed-form solutions and are therefore computationally efficient, see
Section \ref{Complexity} for details on computational complexity.

If the rows of the sensing matrix $H$ are orthonormal [i.e.\
(\ref{eq:orthsensingmatrixcond}) holds], Step {\bf 1} of the
DORE scheme reduces to one IHT step.  After Step {\bf 1}, we
apply two overrelaxations (Steps {\bf 2} and {\bf 3}) that utilize the
sparse signal estimates $\bs^{(p)}$ and
 $\bs^{(p-1)}$ from the two most recent completed DORE iterations.  The
 goal of the overrelaxation steps is to boost the
marginal likelihood (\ref{eq:likelihoodfunction}) and accelerate the
convergence of the ECME iteration.  Using a single
overrelaxation step based on the most recent parameter estimate is a
common approach for accelerating fixed-point iterations, see
\cite{HeLiu}. Here, we adopt the idea in \cite[Sect.\ 5.1]{HeLiu} and
apply the {\em second overrelaxation}\/, which mitigates the
`zigzagging' effect caused by the first overrelaxation and thereby
converges more rapidly.
Our algorithm differs from that in \cite[Sect.\ 5.1]{HeLiu}, which
focuses on  continuous parameter
spaces with  marginal likelihood that is differentiable with respect to the
parameters. Unlike \cite[Sect.\ 5.1]{HeLiu}, here we
\begin{itemize}
\item apply overrelaxation steps on parameter spaces with variable
      dimensions (Steps {\bf 2} and {\bf 3}),
\item
threshold the second overrelaxation
estimate  to ensure that the resulting signal estimate is $r$-sparse
(Step {\bf 4}),
and
\item test the
thresholded estimate from Step {\bf 4} versus the
      ECME estimate from Step {\bf 1} and adopt the better of the
      two (Step {\bf 5}).
\end{itemize}
%Note that the numbers of nonzero components
%of $\bzbar$ and $\bztilde$ are generally larger than $r$. Therefore,
%$\bzbar$ and $\bztilde$ are not valid signal estimates under our
%model, which assumes sparsity level $r$. In Step {\bf 4}, we threshold
%$\bztilde$ to the desired sparsity level $r$.
Step {\bf 5} ensures that the resulting new parameter estimate
$\btheta^{(p+1)}$ yields the marginal likelihood function
(\ref{eq:likelihoodfunction}) that is higher than or equal to that of
the standard ECME step (Step {\bf 1}). Therefore, the DORE iteration
(\ref{eq:ECMEDORE})--(\ref{eq:DOREdeltacheck}) ensures monotonically
nondecreasing marginal likelihood between consecutive iteration steps:
\be
p_{ \sby \,|\, \sbtheta }( \by \,|\,  \btheta^{(p+1)} )
\geq    p_{ \sby \,|\, \sbtheta }( \by \,|\,  \btheta^{(p)} ).
\ee
Furthermore, under the conditions of Theorem \ref{theorem1}, the DORE
iteration converges to a fixed point of the ECME iteration.  This
convergence result follows from the facts that each DORE iteration
contains an ECME step and yields the marginal likelihood
(\ref{eq:likelihoodfunction}) that is higher than or equal to that of
the standard ECME step.  To see this, consider two consecutive
DORE signal estimates $\bs^{(p)}$ and $\bs^{(p+1)}$ and the ECME
estimate $\bshat$ in {\em Iteration $p+1$}.  Due to Step {\bf 5} of
the DORE scheme, we have
\bsea
\label{eq:DOREconverge1}
 \cE(\bs^{(p)})-\cE(\bs^{(p+1)}) &\ge&  \cE(\bs^{(p)})-\cE(\bshat)
\Nextline
\label{eq:DOREconverge2}
 &\ge& \big[1-\lambda_{\max} \big( H_{\sA}^T \, (H \, H^T)^{-1} \, H_{\sA} \big) \big] \, \| \bshat - \bs^{(p)} \|_{\ell_2}^2
\esea where $A={\rm supp}(\bs^{(p)}) \cup {\rm
supp}(\bshat)$. (\ref{eq:DOREconverge2}) follows from
(\ref{eq:thm1_proof1}) of the proof of Theorem \ref{theorem1} in
Appendix \ref{AppProofThm1}.  Since the sequence
$\cE(\bs^{(p)})-\cE(\bs^{(p+1)})$ converges to zero and the conditions
of Theorem \ref{theorem1} ensure that the term $1-\lambda_{\max} \big(
 H_{\sA}^T \, (H \, H^T)^{-1} \, H_{\sA}  \big)$ in
(\ref{eq:DOREconverge2}) is strictly positive (see the proof of
Theorem \ref{theorem1} in Appendix \ref{AppProofThm1}), $\| \bshat -
\bs^{(p)} \|_{\ell_2}^2$ converges to zero as well, implying the
convergence of the DORE iteration to an ECME fixed point.  In
addition, by Theorem
\ref{theorem2}, the fixed point that DORE converges to is also a local
maximum of the concentrated marginal likelihood function
(\ref{eq:profilelikelihood}).  The near-optimal recovery results
(Theorems \ref{theorem3} and
\ref{theorem4}) in Section \ref{ExactRecovery} also apply to DORE
and can be easily derived along the lines of the proofs for the ECME
algorithm in Appendix \ref{AppProofThm34} using the facts that each
DORE iteration contains an ECME step and yields the marginal
likelihood (\ref{eq:likelihoodfunction}) that is higher than or equal
to that of the standard ECME step.
%(A: Yes, all I can say is "easily". To see how "easy" it is,
%just replace $\bs^{(p+1)}$ with $\bshat$ in (\ref{eq:thm3_proof2})-(\ref{eq:thm3_proof4})
%and every thing else follows the same as the proof for ECME.
%Remember $\bs^{(p)}$ and $\bs^{(p+1)}$ are DORE estimates and $\bshat$ is the ECME estimate
%in {\em Iteration $p+1$}.
%The reason why the replacement holds is because DORE contains one ECME and results in higher or equal likelihood than ECME. I do not see the point
%explaining all these mumble jumble. They look trivial.)

\textbf{\textit{{DORE Initialization.}}}  The
parameter estimates $\btheta^{(1)}$ and $\btheta^{(2)}$ are obtained
by applying two consecutive ECME steps (\ref{eq:ECMEonestep})
to an initial sparse signal estimate $\bs^{(0)}$.

%
%
%
%\textbf{\textit{{The empirical Bayesian signal estimate.}}} We
%construct the following empirical Bayesian estimate of the missing
%data vector $\bz$:
%\be
%\label{eq:empiricalBayesian}
%\Exp_{\sbz \,| \sby, \,
%\sbtheta }[\bz \,|\, \by, \btheta^{(+\infty)} ] = \bs^{(+\infty)} + H^{T} (H H^T)^{-1} (\by - H  \bs^{(+\infty)} ).
%\ee
%where $\btheta^{(+\infty)} = ( \bs^{(+\infty)}, (\sigma^2)^{(+\infty)}
%)$ denotes the estimate of the unknown parameter set upon convergence
%of the DORE iteration.
%Unlike $\bs^{(+\infty)}$,
%the empirical Bayesian estimate (\ref{eq:empiricalBayesian})
%is not $r$-sparse in general, and is therefore preferable for
%reconstructing approximately sparse signals that have many small
%nonzero signal coefficients.

\subsection{Computational Complexity and Memory Requirements}
\label{Complexity}

\noindent
The major computational complexity of the ECME algorithm lies in the
 matrix-vector multiplications and
sorting of $m \times 1$ vectors. Assuming that the common bubble
sorting is employed, sorting $\bz^{(p+1)}$ in
(\ref{eq:ECMEzdeltapplusone}) requires $\mathcal{O}(m^2)$ operations.
There are three matrix-vector multiplications in one ECME iteration,
namely $H \bs^{(p)}$, $(H H^T)^{-1} [ H \bs^{(p)} ]$ and $H^{T} [(H
H^T)^{-1} H \bs^{(p)} ]$, which requires $\mathcal{O}(Nm)$,
$\mathcal{O}(N^2)$ and $\mathcal{O}(Nm)$ operations, respectively. The
intermediate computation results of $(\sigma^2)^{(p+1)}$ in
(\ref{eq:ECMEdeltapplusone}) can be stored and used to compute
(\ref{eq:ECMEzdeltapplusone}) of the next iteration; therefore, this
step does not cause additional computation.  In summary, the
complexity of one ECME iteration is $\mathcal{O}(m^2+2Nm+N^2)$.  If
$H$ has orthonormal rows satisfying (\ref{eq:orthsensingmatrixcond}),
ECME reduces to the IHT iteration, and in this case $(H H^T)^{-1} [ H
\bs^{(p)} ]$ is simply $H
\bs^{(p)}$. The computation complexity of one IHT step is therefore
$\mathcal{O}(m^2+2Nm)$.

For DORE, there are two sorting operations per iteration, one
in step {\bf 1} and the other in step {\bf 4}, requiring
$\mathcal{O}(2 m^2)$ operations. In one DORE iteration, we
need to compute $H^{T} [(H H^T)^{-1} H \bs^{(p)} ]$, $H \, \bshat$, $(H
\, H^{T})^{-1} \, [ H \, \bshat ]$, $H \, \bstilde$, and $(H \,
H^{T})^{-1} \, [ H \, \bstilde ]$, which require total of
$\mathcal{O}(3Nm+2N^2)$ operations.  Note that $H \bs^{(p-1)}$, $(H
H^T)^{-1} [H \bs^{(p-1)}]$, $H \bs^{(p)}$, and $(H H^T)^{-1} [H
\bs^{(p)}]$ in (\ref{eq:DOREzhat}), (\ref{eq:DOREalpha1}) and
(\ref{eq:DOREalpha2}) can be adopted from the previous two iterations
and do not need to be computed again in the current iteration; in
addition, the quantities $H \bzbar$ and $(H H^T)^{-1} [H \bzbar]$ in
(\ref{eq:DOREalpha2}) are simple linear combinations of computed terms
and do not require additional matrix-vector computations.  To
summarize, one DORE iteration requires
$\mathcal{O}(2m^2+3Nm+2N^2)$, which is slightly less than twice the
complexity of one ECME step. When $H$ has orthonormal rows, we do not need
to compute $(H \, H^{T})^{-1} \, [ H \, \bshat ]$ and $(H \,
H^{T})^{-1} \, [ H \, \bstilde ]$, which brings the complexity down
to $\mathcal{O}(2  m^2 + 3  N  m)$, slightly less than twice the complexity of
one IHT step.

Regarding the memory storage, the largest quantity that ECME (and its special case IHT) and DORE
need to store is the sensing matrix $H$ requiring memory storage of order
$\mathcal{O}(Nm)$. In large-scale applications,
$H$ is typically not explicitly stored but instead
appears in the function-handle form [for example, random DFT sensing
matrix can be implemented via the fast Fourier transform (FFT)].  In this
case, the storage requirement of ECME, IHT and DORE is just $\mathcal{O}(m)$.

Although a single DORE step is about twice more complex than
the ECME and IHT steps, it converges in much fewer iterations than the
ECME and IHT iterations in the numerical examples in Section
\ref{NumEx}, see Fig.\ \ref{fig:PSNRnumiterCPUvsNoverm}~(b) and Fig.\ \ref{fig:numex2}~(c).

\newsection{Unconstrained Sparsity
Selection Criterion for Selecting $r$ and
the ADORE Algorithm}
\label{USSADORE}

\noindent
The ECME and DORE algorithms, as well as most other greedy methods,
require the knowledge of sparsity level $r$ as an input. In this
section, we propose an sparsity selection criterion and an {\em
automatic double overrelaxation}\/ (ADORE) thresholding algorithm that
estimates the signal sparsity from the measurements.

We introduce the following {\em unconstrained sparsity
selection (USS)}\/ objective function for selecting the proper sparsity level
$r$ that strikes a balance between the efficiency and accuracy of signal representation:
\begin{eqnarray}
\label{eq:USS}
{\rm USS}(r)
= - \half \, r \, \ln\big( \frac{N}{m} \big)
 - \half \, ( N - r - 2 ) \, \ln\Big( \frac{ \widehat{\sigma}_{\sML}^2(r) }{
\by^T \, (H \, H^T)^{-1} \, \by / N  } \Big)
\end{eqnarray}
where $\widehat{\sigma}_{\sML}^2(r)$ is the ML
estimate of the variance component $\sigma^2$ in the parameter space $\Theta_r$, see (\ref{eq:thetahatr}).
  ${\rm USS}(r)$ in (\ref{eq:USS}) is
developed from the approximate generalized maximum likelihood (GML)
objective function in \cite[e.q. (13)]{QNDE09}; in particular, when
$\by^T \, (H \, H^T)^{-1} \, \by / N = 1$, the two functions are equal
up to an additive constant.  However, unlike GML, the
USS objective function (\ref{eq:USS}) is scale-invariant: scaling the
measurements $\by$ by a nonzero constant does not change ${\rm
USS}(r)$, which is a desirable property.

Interestingly, the USS objective (\ref{eq:USS}) is closely related to
the (${\rm P}_0$) problem (\ref{eq:P0}), as shown by the following theorem.

\begin{theorem}
\label{theorem5}
Suppose that we have collected a measurement vector $\by = H \,
\bs^{\diamond}$ using a proper sensing matrix $H$, where
$\bs^{\diamond}$ is a sparse signal vector having exactly
 $ r^\diamond = \| \bs^{\diamond} \|_{\ell_0}$ nonzero elements.
If
\begin{description}
\item[(1)]
the sensing matrix $H$ satisfies the unique representation
property (URP) condition (\ref{eq:URP}) and
\item[(2)]  the number of measurements $N$ satisfies
\be
\label{eq:theorem_cond}
N \ge \max\{2 \, r^{\diamond} , r^{\diamond}+3\}
\ee
\end{description}
then
\begin{itemize}
\item
${\rm USS}(r)$ in (\ref{eq:USS}) is {\em globally and uniquely
maximized}\/ at $r = r^{\diamond}$ and
\item
the $({\rm P}_0)$-optimal solution and ML sparse signal estimate
at $r = r^{\diamond}$ [i.e.\ $\widehat{\bs}_{\sML}(r^{\diamond})$, see
(\ref{eq:thetahatr})] are both {\em unique}\/ and {\em coincide}\/ with
$\bs^\diamond$.
\end{itemize}
\end{theorem}
\begin{IEEEproof}
See Appendix \ref{AppProofThm5}.
\end{IEEEproof}

Theorem \ref{theorem5} shows that the USS objective function {\em
transforms}\/ the constrained optimization problem $({\rm P}_0)$ in
(\ref{eq:P0}) into an equivalent unconstrained problem (\ref{eq:USS})
and that USS optimally selects the signal sparsity level $r$ that
allows accurate signal representation with as few nonzero signal
elements as possible.

In the practical scenarios where $r^{\diamond} \ge 3$, condition (2)
of Theorem \ref{theorem5} reduces to $N \ge 2 \, r^{\diamond}$,
which is the condition required to ensure the uniqueness of the (${\rm P}_0$)
problem, see \cite[Theorem 2]{BrucksteinDonohoElad}.

In the following, we use DORE to approximately evaluate the USS objective
function and apply this approximate USS criterion to automatically
select the signal sparsity level.

\subsection{The ADORE Algorithm for Unknown Sparsity Level $r$}
\label{ADORE}

\noindent We approximate the USS objective function (\ref{eq:USS}) by
replacing the computationally intractable ML estimate
$\widehat{\sigma}_{\sML}^2(r)$ with its DORE estimate.  Maximizing
this approximate USS objective function with respect to $r$ by an
exhaustive search may be computationally expensive because we need to
apply a full DORE iteration for each sparsity level $r$ in the set of
integers between $0$ and $N/2$.\footnote{Note that $N/2$ is the largest value
of the sparsity level $r$ for which reasonable reconstruction is
possible from $N$ measurements; otherwise, the (${\rm P}_0$) and ML
estimates of the sparse signal may not be unique, see e.g. \cite[Theorem 2]{BrucksteinDonohoElad}.}
Here, we propose the ADORE
algorithm that applies the {\em golden-section search}\/
\cite[Sec.\ 4.5.2.1]{Thisted}
to maximize  the approximate USS objective function with respect to $r$,
 with the initial search boundaries set to $0$ and $\lceil N/2
\rceil$.
Note that
${\rm USS}(0) = 0$ assuming that $\by \neq {\bf 0}_{\sN \times 1}$,
which is of practical interest.
For each candidate $0 < r \le \lceil N/2
\rceil$, we estimate $\widehat{\sigma}_{\sML}^2(r)$ using the DORE iteration.
After running one golden sectioning step, the length of the new search
interval is approximately $0.618$ of the previous interval (rounded to
the closest integer).  The search process ceases when the desired
resolution $L$ is reached, i.e.\ when the searching interval becomes
shorter than the prescribed resolution level $L$.  Therefore,
ADORE requires roughly $1.4 \, [\log_2 (N/L) - 1]$ full
DORE iterations.  For the golden-section search to find the
exact maximum of (\ref{eq:USS}), ${\rm USS}(r)$ must be {\em
unimodal}\/ in $r$, which is not true in general. Hence,
ADORE maximizes (\ref{eq:USS})  only approximately, yielding
$r_{\sADORE}$; then, our ADORE
sparse-signal estimate is equal to the corresponding
DORE estimate at $r = r_{\sADORE}$.

\newsection{Numerical Examples}
\label{NumEx}

\noindent We now compare our proposed methods in Sections \ref{DORE} and
\ref{USSADORE} with existing large-scale sparse reconstruction techniques
using two image recovery experiments, with purely and approximately
sparse signals, respectively.
In particular, we compare
\begin{itemize}
\item the DORE and ADORE schemes initialized by the zero sparse signal
      estimate:
\be
\label{eq:s0}
\bs^{(0)} = {\bf 0}_{m \times 1}
\ee
with ADORE search resolution set to
$L=500$ and \textsc{Matlab} implementations
      available at \url{http://home.eng.iastate.edu/~ald/DORE.htm};
\item the IHT and NIHT schemes in
\cite{BlumensathDavies} and \cite{BlumensathDavies2}, initialized by
the zero sparse signal estimate
$\bs^{(0)}$ in (\ref{eq:s0});
\item the automatic hard thresholding (AHT) method in
      \cite{QNDE09} using the moving-average
window length $100$, initialized with
$\bz_{\sinit} = {\bf 0}_{m \times 1}$ and $r_{\sinit} = 1$;
\item the debiased gradient-projection for
sparse reconstruction method in
\cite[Sec.\ III.B]{FigueiredoNowakWright} with the
convergence threshold $\texttt{tolP} = 10^{-5}$ and
regularization parameter set to
\begin{description}
\item[(i)]
$\tau = 0.1 \, \| H^T \, \by \|_{\ell_{\infty}}$, suggested in \cite[e.q. (22)]{FigueiredoNowakWright} (labeled GPSR$_0$)
and
\item[(ii)]
$\tau = 0.001 \, \| H^T \, \by \|_{\ell_{\infty}}$, obtained by
manual tuning for good performance in the following two numerical examples
(labeled GPSR);
\end{description}
\item the minimum-norm signal estimate (labeled MN):
\be
\label{eq:sMN}
\widehat{\bs}_{\sMN} = H^T \, (H \, H^T)^{-1} \, \by
\ee
which achieves zero squared residual error by ignoring sparsity.
\end{itemize}

For the DORE, ADORE, IHT and NIHT iterations, we use
the following convergence criterion\footnote{To implement the
IHT and NIHT schemes, we incorporated the
convergence criterion (\ref{eq:convcrit}) into the corresponding
\textsc{Matlab} codes from the \texttt{sparsify} toolbox at
\url{http://www.see.ed.ac.uk/~tblumens/sparsify/sparsify.html}.}:
\be
\label{eq:convcrit}
\| \bs^{(p+1)} - \bs^{(p)} \|_{\ell_2}^2 \,/\, m < 10^{-14}.
\ee

The sensing matrix $H$ has the following structure (see
e.g. \cite[eq.\ (2) and Fig. 1]{BaraniukSpMag}):
\be
H = \Phiit \, \Psiit
\ee
where $\Phiit$ is an $N \times m$ {\em sampling matrix}\/ and $\Psiit$ is an
appropriate $m \times m$ orthogonal {\em sparsifying transform matrix}\/.
In our examples presented here, $\Psiit$ are
 inverse discrete wavelet transform (DWT) matrices \cite{DaubechiesBook}.

For an underlying image $\Psiit \, \bs$, the signal vector $\bs$ is the wavelet coefficient vector of the
 image.
Our performance metric is the peak signal-to-noise ratio (PSNR) of a
reconstructed image $\Psiit \, \widehat{\bs}$,
where $\widehat{\bs}$ is the estimated wavelet coefficients vector:
\be
\label{eq:PSNRdef}
\PSNR~\mbox{(dB)}
=  10 \, \log_{10} \Big\{ \frac{ [(\Psiit \, \bs)_{\sMAX} - (\Psiit \, \bs)_{\sMIN}]^2}{\| \Psiit \, \widehat{\bs} - \Psiit \, \bs \|_{\ell_2}^2 / m}  \Big\}
=  10 \, \log_{10} \Big\{ \frac{ [(\Psiit \, \bs)_{\sMAX} - (\Psiit \, \bs)_{\sMIN}]^2}{\| \widehat{\bs} - \bs \|_{\ell_2}^2 / m}  \Big\}
\ee
where
%$\Psiit$ is an appropriate $m \times m$ orthogonal (sparsifying)
% inverse discrete wavelet transform
%(DWT) matrix \cite{DaubechiesBook},
% $\bs$ and $\widehat{\bs}$ are the true signal vector and its estimate
% whose performance .
%wavelet coefficient vector of the
%underlying image (signal vector) and an estimate of $\bs$, and
$(\Psiit \,
\bs)_{\sMIN}$ and $(\Psiit \, \bs)_{\sMAX}$ denote the smallest and
largest elements of $\Psiit \,
\bs$.
%Note that (\ref{eq:PSNRdef}) is also the PSNR of an estimate
%$\Psiit \, \widehat{\bs}$ of the underlying image.

\subsection{Tomographic Image Reconstruction}
\label{NumEx1}

\begin{figure}
\vspace{-0.1in}
\hspace{0.2in}
\parbox{2in}{
    \centering \includegraphics[width=2in]{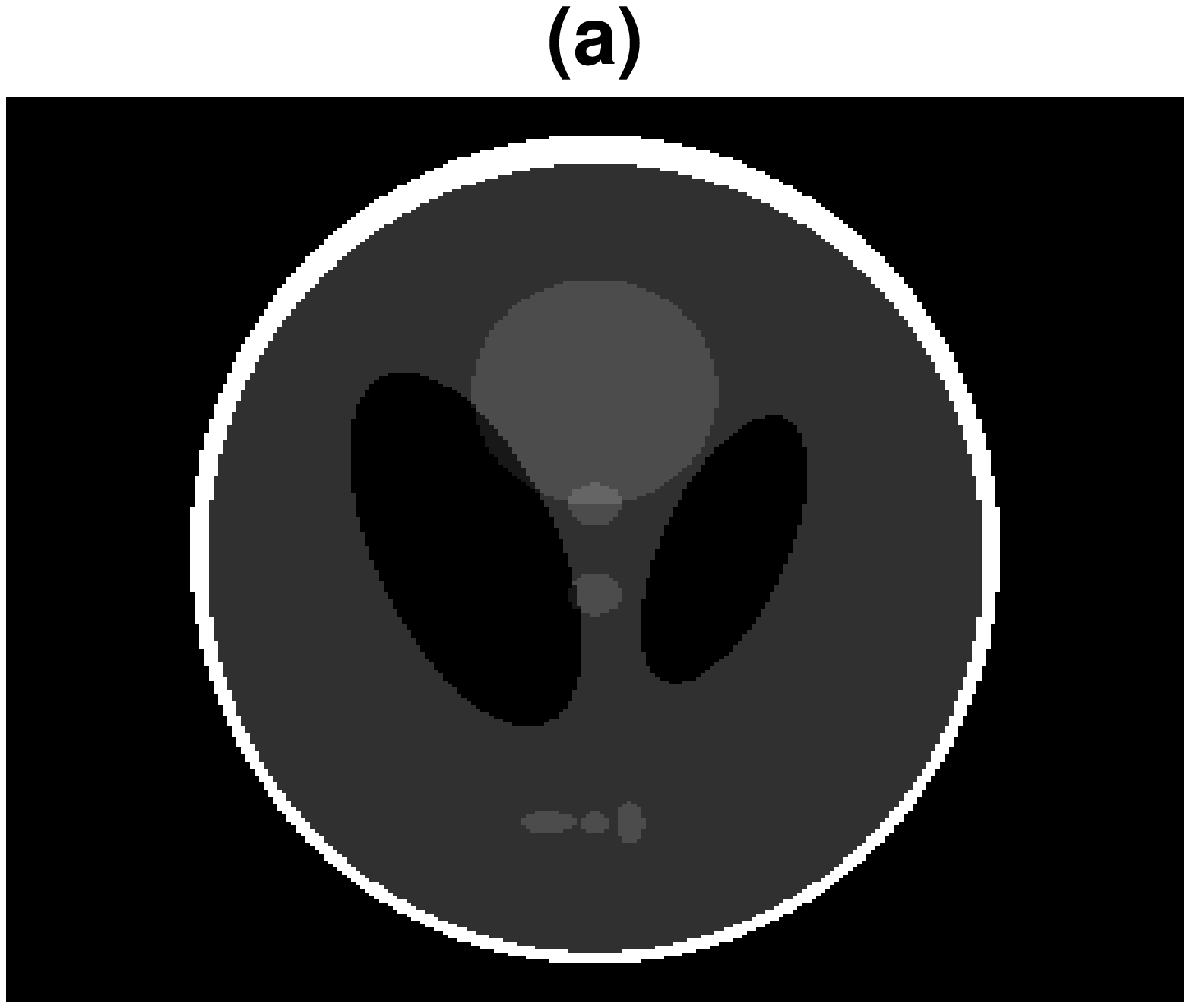}
}
 \hspace{0.1in}
\parbox{2in}{
    \centering \includegraphics[width=2in]{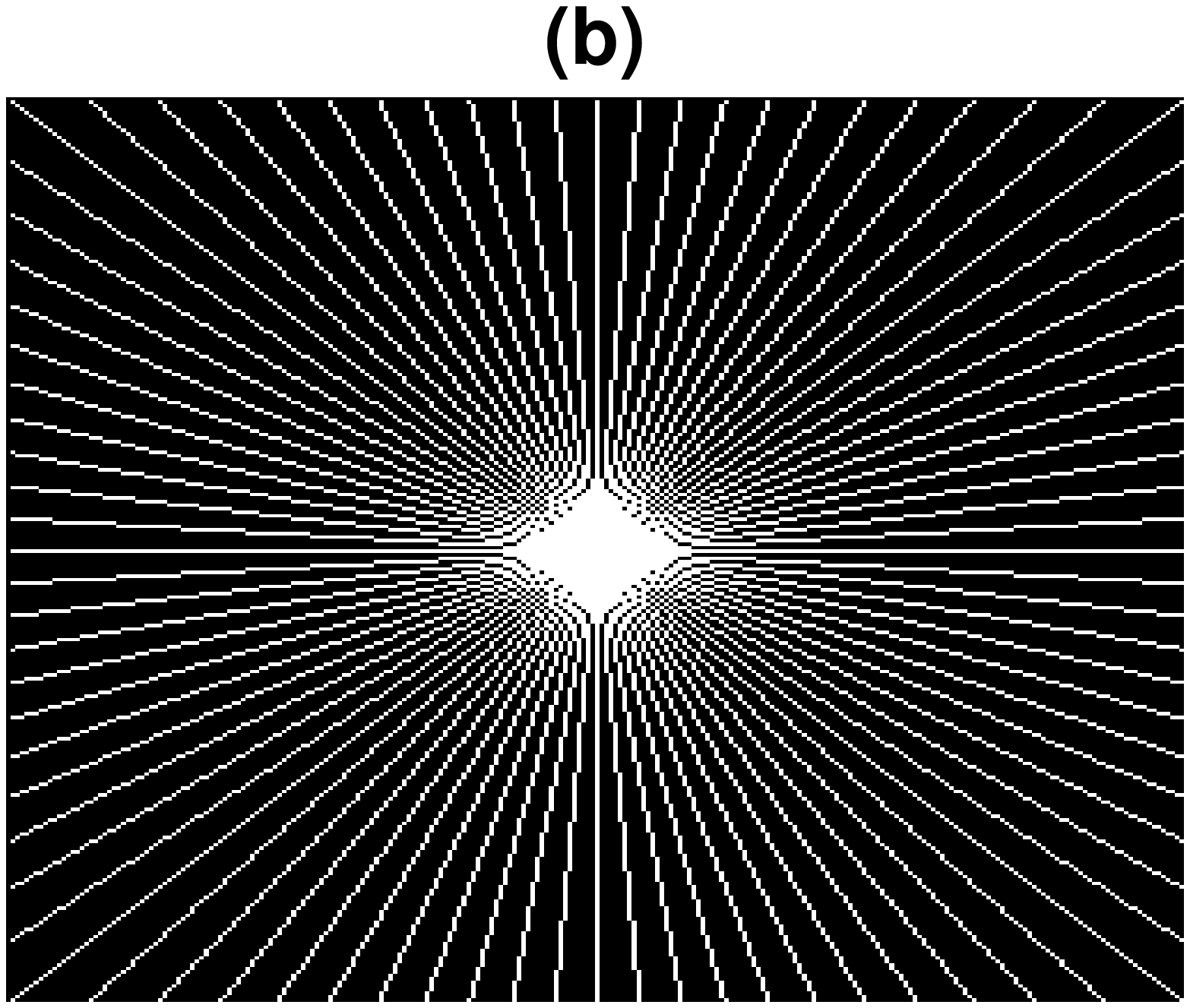}
}
 \hspace{0.1in}
\parbox{2in}{
    \centering \includegraphics[width=2in]{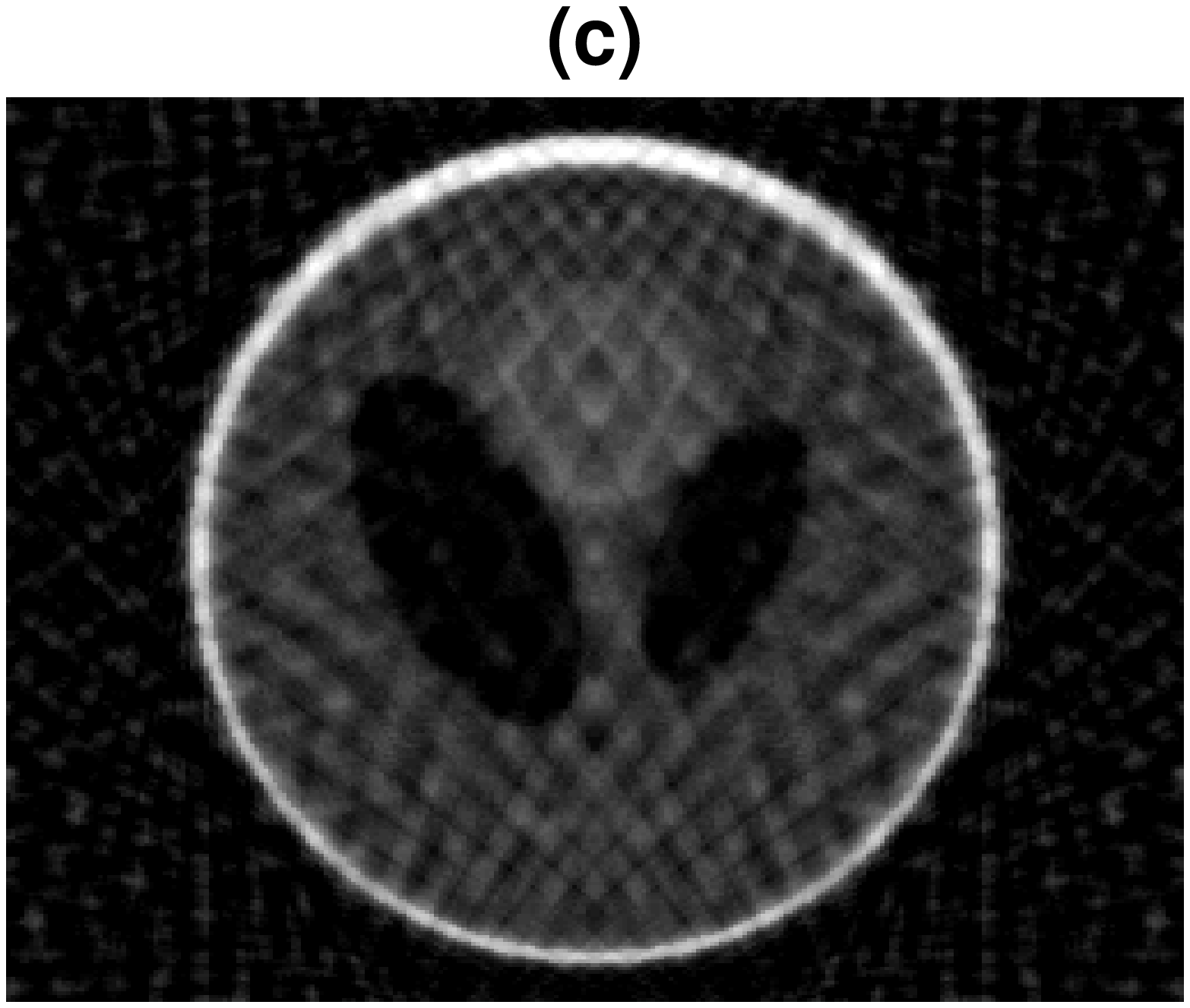}
}

\hspace{0.2in}
\parbox{2in}{
    \centering \includegraphics[width=2in]{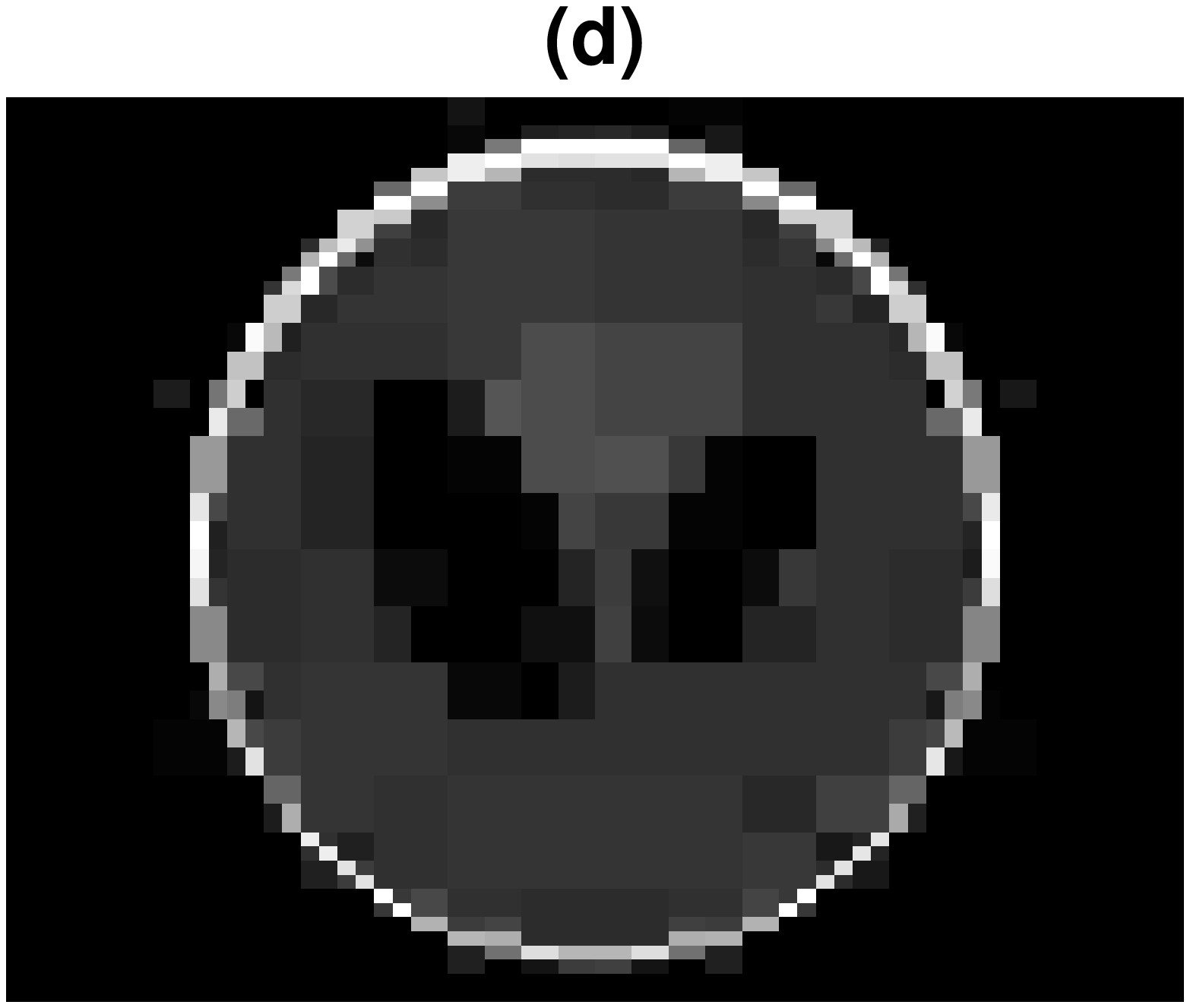}
}
\hspace{0.1in}
\parbox{2in}{
    \centering \includegraphics[width=2in]{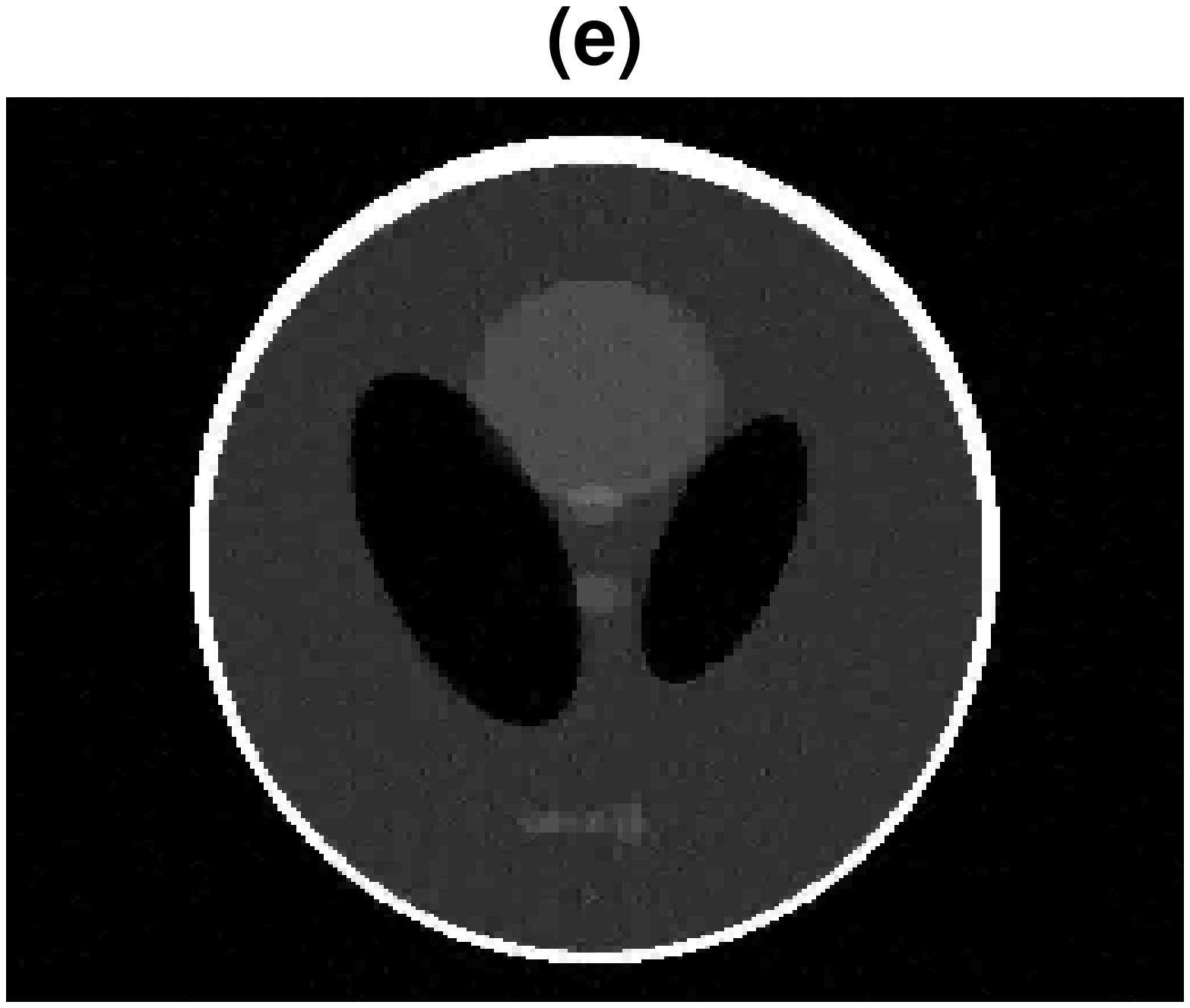}
}
 \hspace{0.1in}
\parbox{2in}{
    \centering \includegraphics[width=2in]{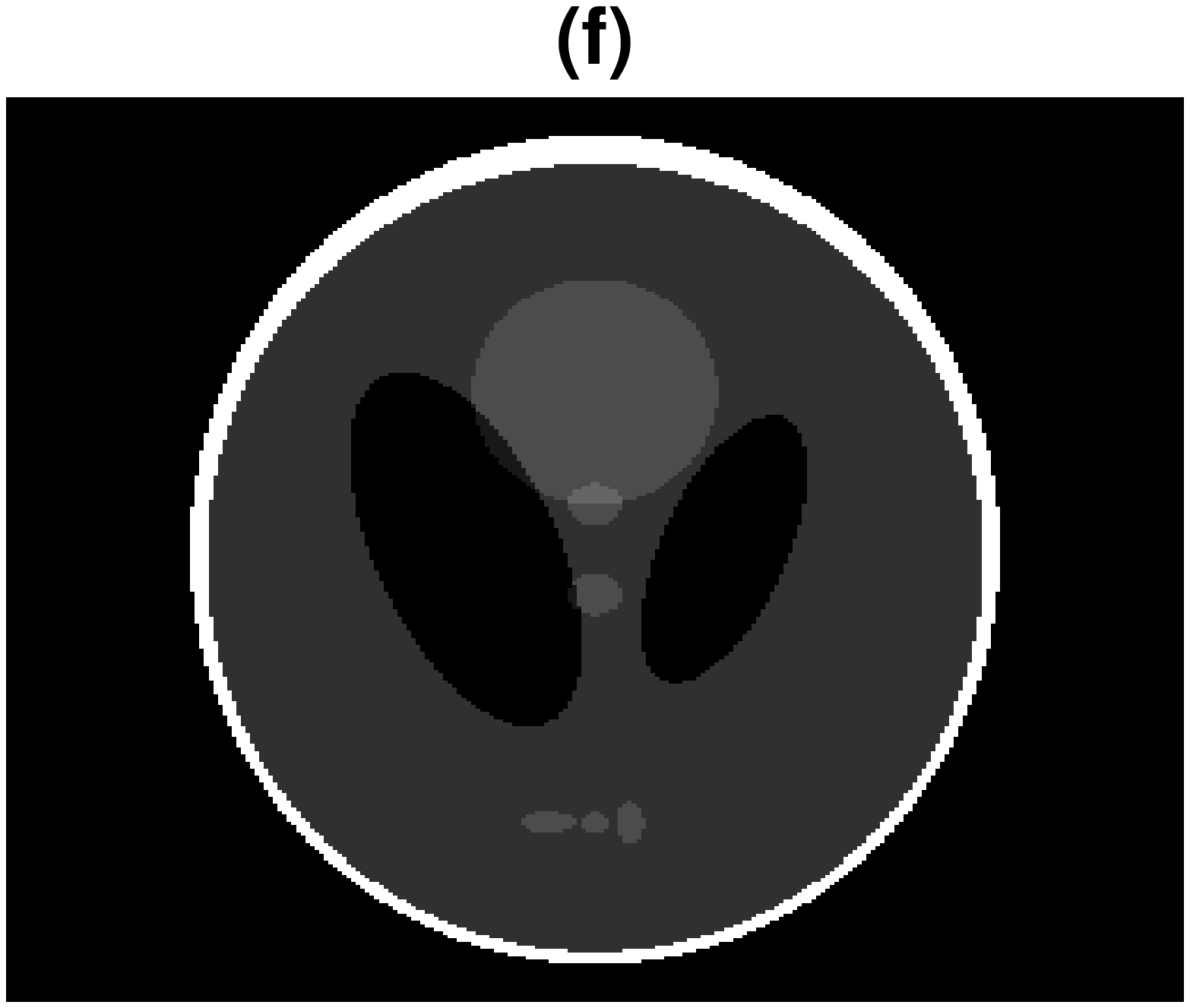}
}
\caption{(a) The size-$256^2$ Shepp-Logan phantom, (b) a
star-shaped sampling domain in the frequency plane
containing 44 radial lines, and
 (c) the filtered back-projection (minimum-norm) reconstruction, (d)
     GPSR$_0$ reconstruction, (e) GPSR reconstruction,
and (f) the almost perfect reconstruction achieved by all
hard-thresholding schemes, for the sampling pattern in (b).}
\label{fig:numex1}
\end{figure}

\noindent
Consider the reconstruction of the Shepp-Logan phantom of
size $m=256^2$ in Fig.\ \ref{fig:numex1}~(a) from tomographic
projections. The elements of $\by$ are 2-D discrete Fourier transform (DFT)
coefficients of the phantom sampled over a star-shaped domain, as
illustrated in Fig.\ \ref{fig:numex1}~(b); see also  \cite{CandesRombergTao_ITpaper},
 \cite{BlumensathDavies2}, and \cite{QNDE09}.
Therefore, the sampling matrix $\Phiit$ is constructed using selected
rows of the DFT matrix that yield the corresponding DFT coefficients
of the phantom image within the star-shaped domain. In this example, we select
 the inverse Haar (Daubechies-2) DWT matrix to be
the orthogonal sparsifying transform matrix $\Psiit$.
The Haar wavelet
transform coefficients of the phantom image in Fig.\
\ref{fig:numex1}~(a) are sparse, with $\| \bs \|_{\ell_0} = 3769
\approx 0.06 \, m$, where the true signal vector $\bs$ consists of the
Haar wavelet transform coefficients of the phantom in Fig.\
\ref{fig:numex1}~(a).

For our choices of $\Phiit$ and $\Psiit$, the rows of $H$ are
orthonormal, i.e.\ (\ref{eq:orthsensingmatrixcond}) holds, implying
that IHT is equivalent to the ECME iteration in Section \ref{ECME}.
  DORE, IHT, and NIHT require knowledge of the signal
sparsity level $r$; in this example, we set $r$ to the true signal
support size:
\be
r=3769.
\ee
In contrast, the ADORE and AHT methods are
{\em automatic}\/ and {\em estimate}\/ $r$ from the measurements using
the USS and GML selection criteria, respectively.

Figs.\ \ref{fig:numex1}~(c)--(f) show the  images reconstructed by
the above methods using the 44 radial-line sampling
pattern in Fig.\
\ref{fig:numex1}~(b), which corresponds to $N/m = 0.163$.  In this example, the
MN signal estimate (\ref{eq:sMN}) is also the {\em filtered
back-projection estimate}\/ obtained by setting the unobserved DFT
coefficients to zero and taking the inverse DFT, see
\cite{CandesRombergTao_ITpaper}.  Here, all hard-thresholding methods
(DORE, IHT, NIHT, ADORE, and AHT) achieve almost perfect
reconstructions of
 the original phantom image with PSNRs over 100~dB, in
contrast with the MN (filtered back-projection) and GPSR methods that
achieve inferior reconstructions with PSNRs 20.2~dB for the MN, 33.0~dB
for GPSR, and 17.9~dB for GPSR$_0$ estimates.
%Interestingly, the GPSR$_0$
%reconstruction is poorer than that of the filtered back-projection,
%compare Figs.\ \ref{fig:numex1}~(c) and (d).

\begin{figure}
\vspace{-2ex}
\begin{center}
\hspace{0in}
\parbox{3.75in}{
    \centering \includegraphics[width=3.75in]{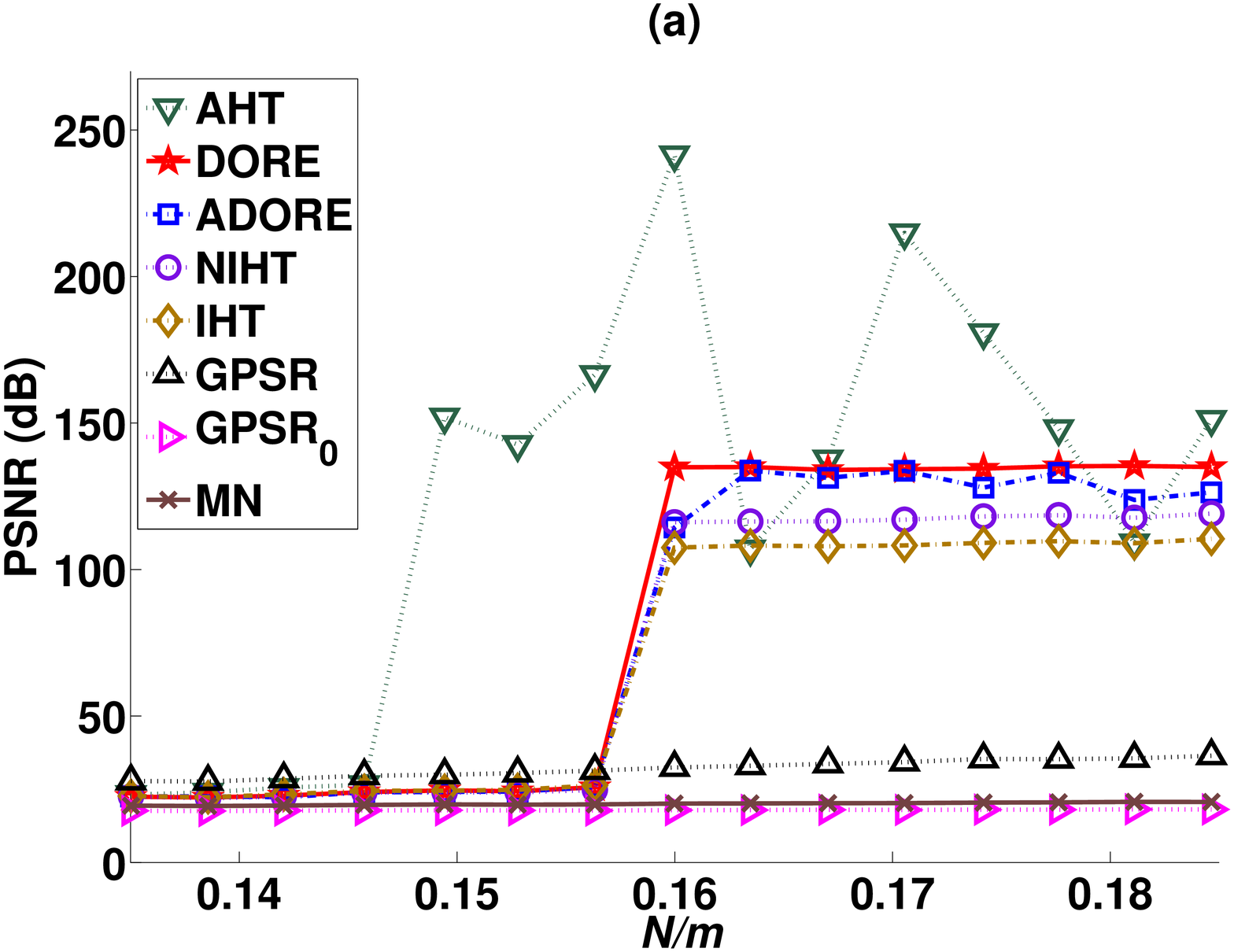}
}

\vspace{0ex}
\hspace{-0.25in}
\parbox{3.75in}{
    \centering \includegraphics[width=3.75in]{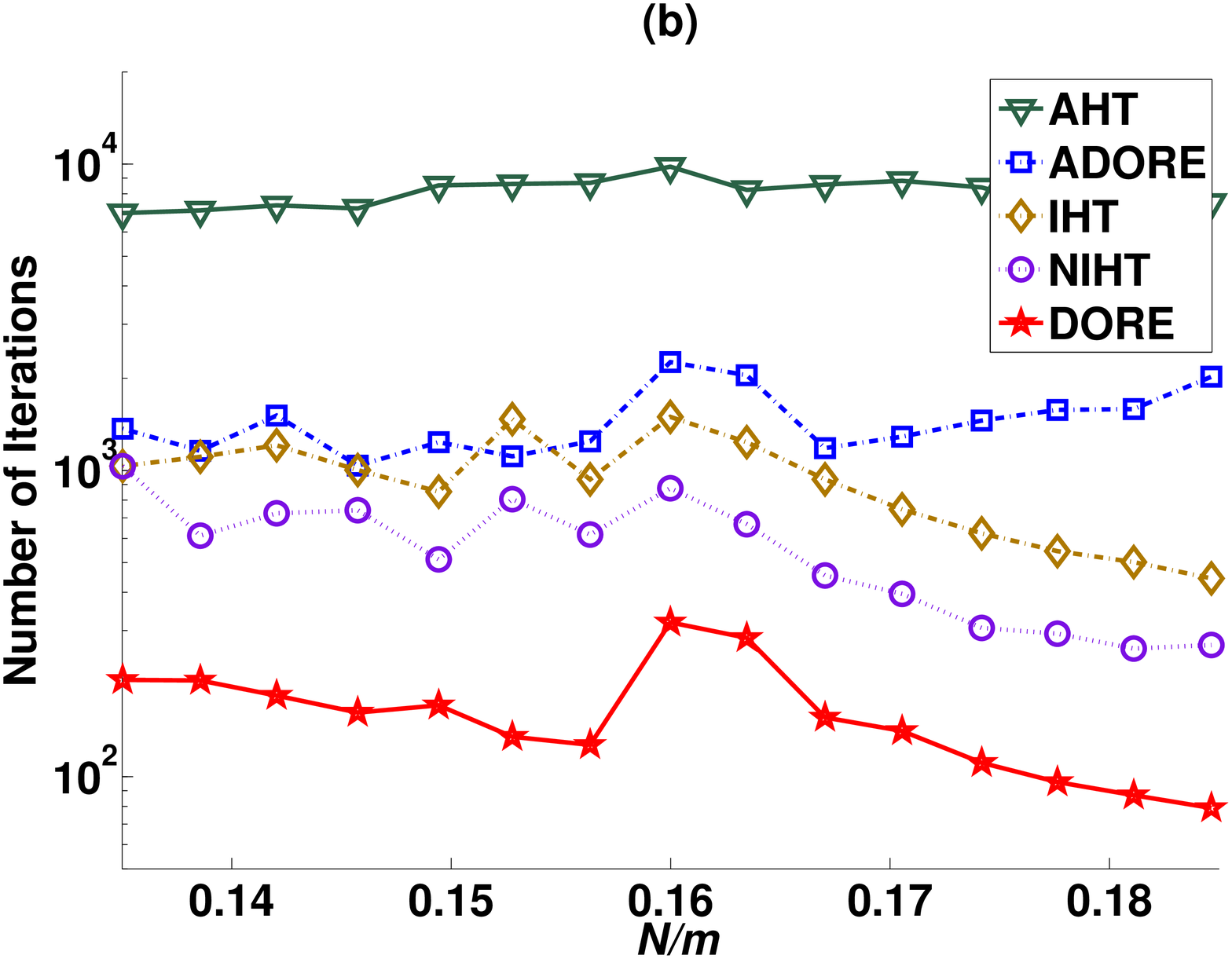}
}
\hspace{-0.2in}
\vspace{0ex}
\parbox{3.75in}{
    \centering \includegraphics[width=3.75in]{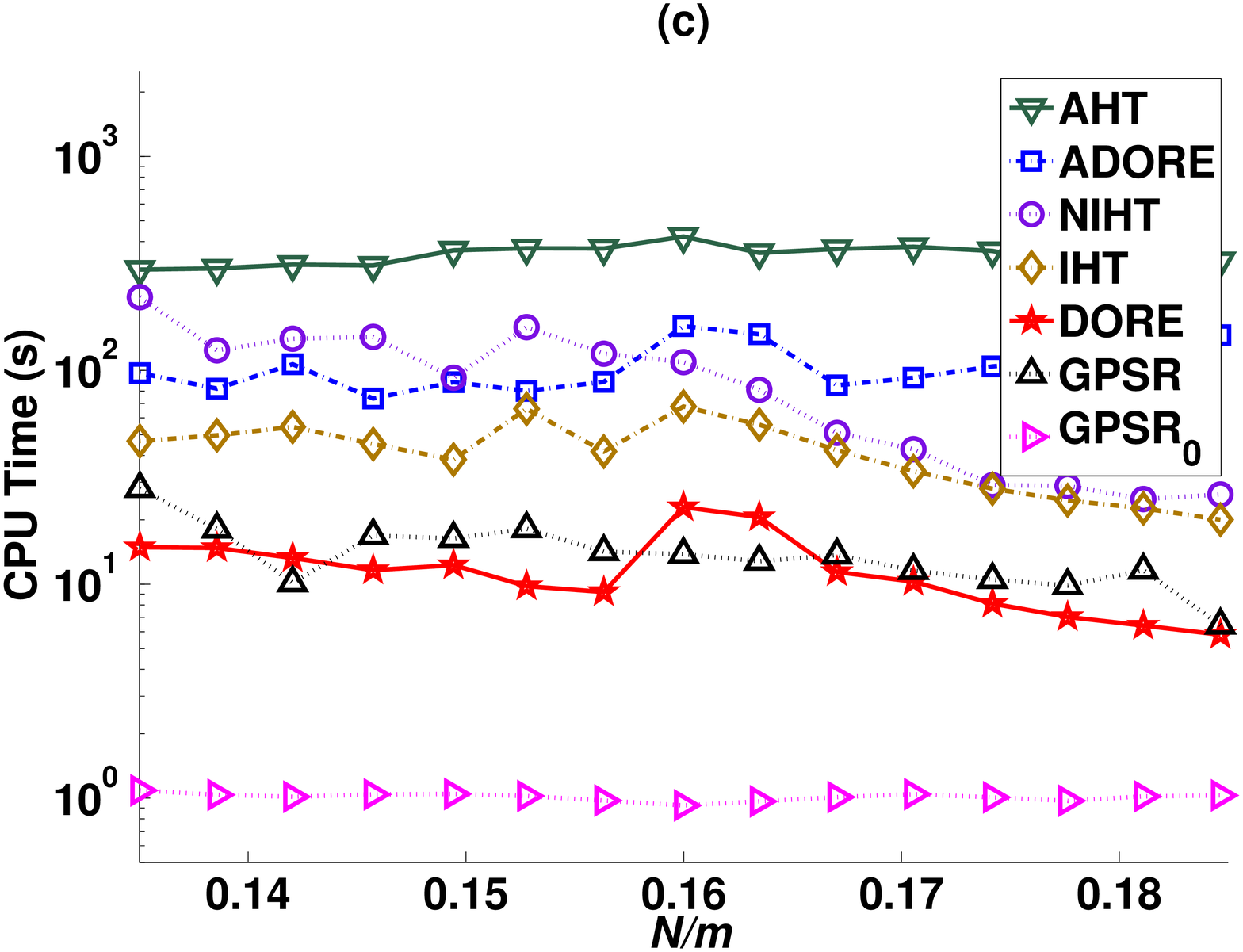}
}
\end{center}

\caption{(a) PSNR, (b) number of iterations, and (c) CPU time
as the functions of the normalized number of measurements $N/m$ for phantom image reconstruction.}
\label{fig:PSNRnumiterCPUvsNoverm}
\end{figure}

Fig.\ \ref{fig:PSNRnumiterCPUvsNoverm} shows (a) the PSNRs, (b) numbers of
iterations, and (c) CPU times of the above methods as we change $N/m$ by
varying the number of radial lines in our star-shaped partial Fourier sampling
pattern.
 In this example,
all hard-thresholding methods have significantly sharper phase
transitions than the manually tuned GPSR, and outperform
GPSR after the phase transitions. \textsc{AHT} exhibits the
phase transition at $N/m
\approx 0.15$; the phase transitions of the other hard thresholding
methods occur at $N/m
\approx 0.16$. ADORE performs as well
as the DORE, IHT, and NIHT methods that require prior knowledge of the
signal sparsity level.  Indeed, the USS criterion accurately selects
the signal sparsity level in this case, which is consistent with the
essence of Theorem \ref{theorem5}.  Among all hard-thresholding
methods, DORE needs the smallest number of iterations to converge and
is also the fastest in terms of the CPU time.  DORE needs
4.4 to 10.9 times less iterations
than IHT and 2.3 to 6 times less iterations
than
NIHT; in terms of the CPU time, DORE is 2.7 to 6.7
times
faster than IHT and 3.6 to 16.3 times faster than
NIHT.  The CPU
times of DORE, IHT, and ADORE {\em per iteration}\/ are
approximately constant as $N/m$ varies. One DORE step is about twice
slower than one IHT step, validating the computational
complexity analysis in Section \ref{Complexity}.

 We now compare the two automatic thresholding methods
(AHT and ADORE) in this example: ADORE requires 3.7 to 7.7 times less
iterations and is 2.2 to 4.6 times faster than AHT.  We note that
AHT's computational complexity does not scale well with the increase
of the signal size and its support, which is the case considered in
the following example, where we increase the signal size four times,
to $m=512^2$.

\subsection{Lena Reconstruction From Compressive Samples}
\label{NumEx2}

\begin{figure}
\begin{center}
\hspace{0.4in}
\parbox{2.5in}{
    \centering \includegraphics[width=2.5in]{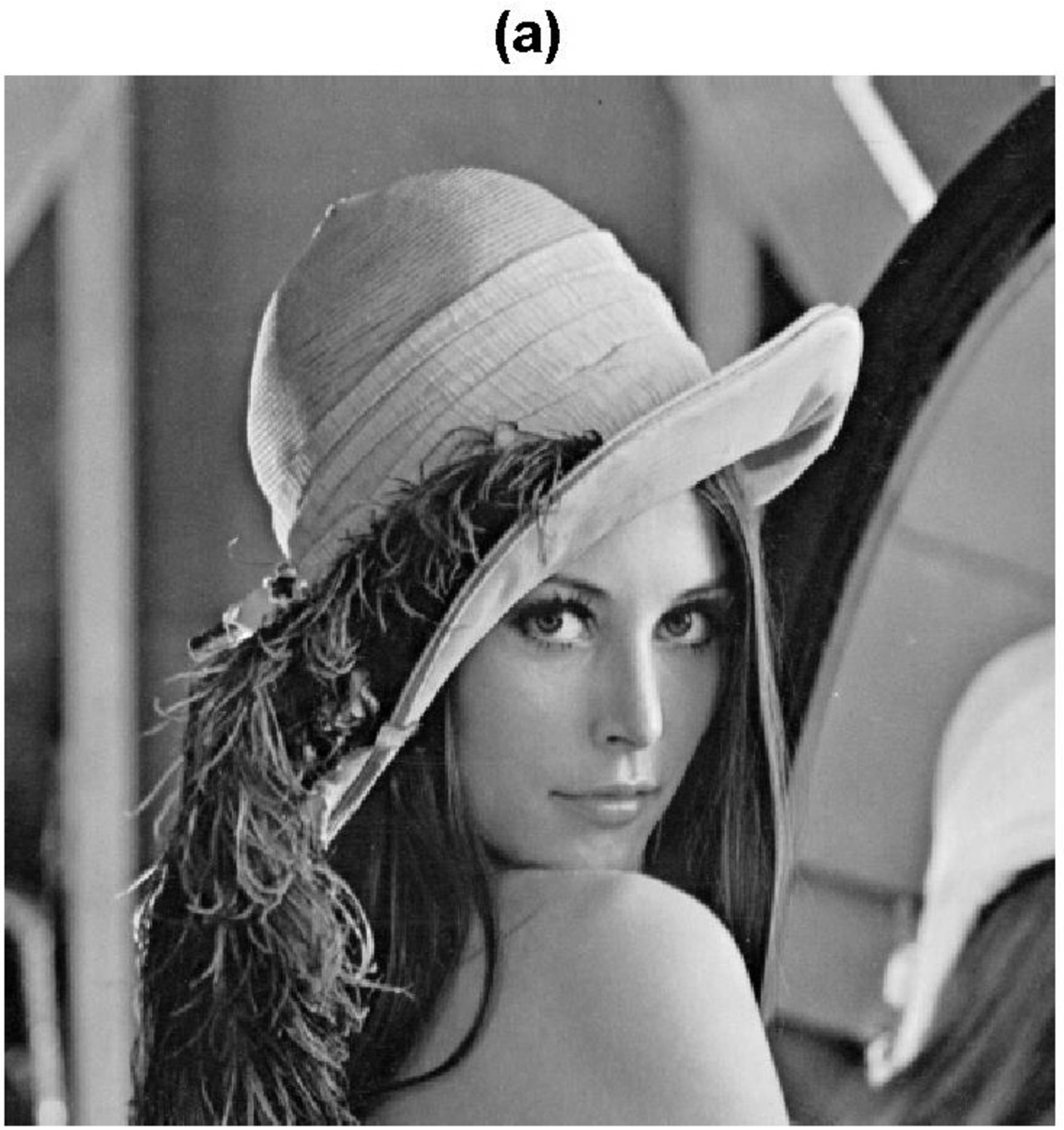}
}
\hspace{0.4in}
\parbox{3.75in}{
    \centering \includegraphics[width=3.75in]{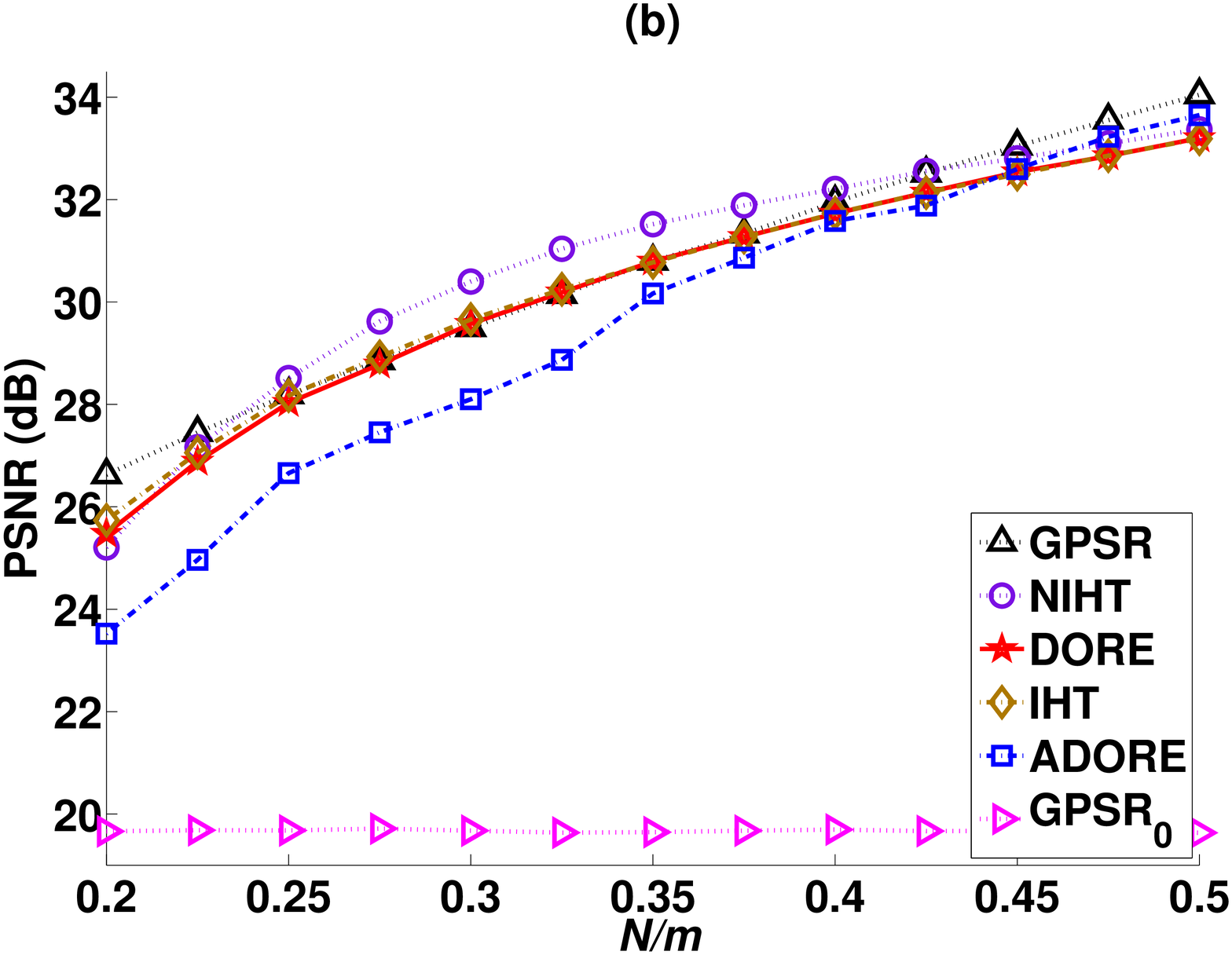}
}

\vspace{-0ex}
\hspace{-0.25in}
\parbox{3.75in}{
    \centering \includegraphics[width=3.75in]{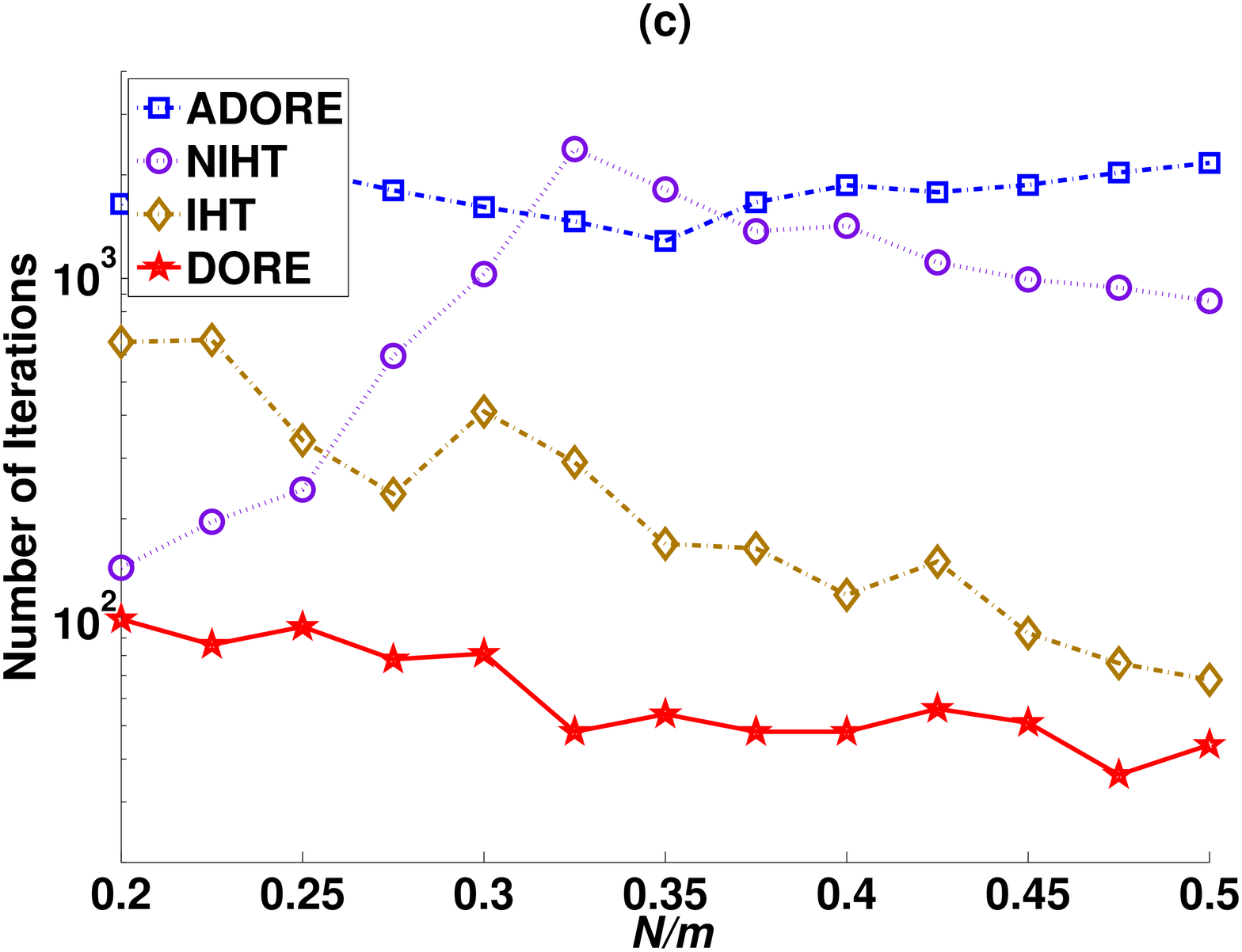}
}
\hspace{-0.2in}
\vspace{-0ex}
\parbox{3.75in}{
    \centering \includegraphics[width=3.75in]{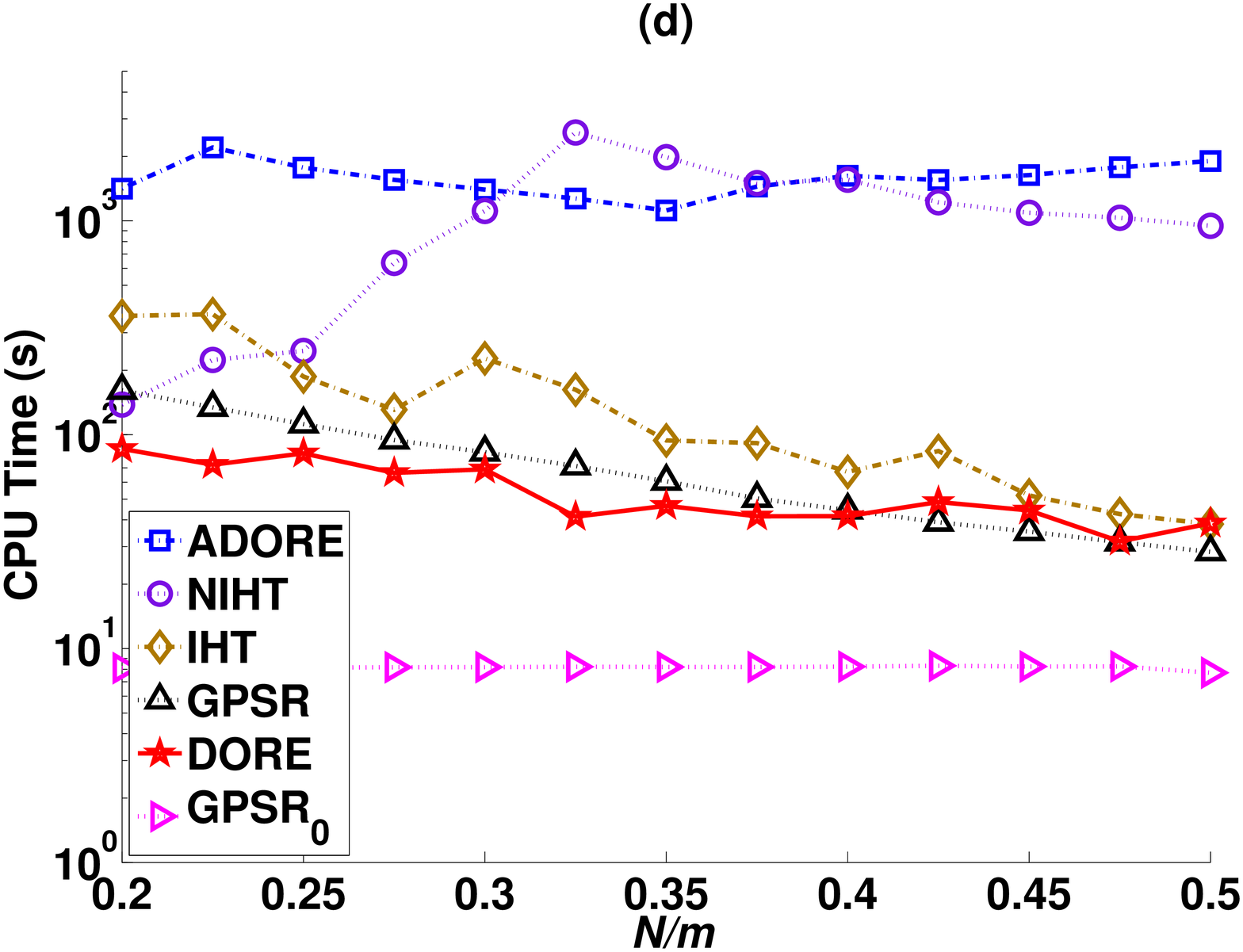}
}
\end{center}

\caption{(a) The $512 \times 512$ Lena image, and (b) PSNR, (c) number of iterations, and (d) CPU time
as the functions of the normalized number of measurements $N/m$ for Lena image reconstruction.}
\label{fig:numex2}
\end{figure}

\noindent
We now reconstruct the standard Lena image of size
$m=512^2$ in Fig.\ \ref{fig:numex2}~(a) from compressive samples.
In this example, we select
the  {\em structurally random
sampling matrices}\/
 $\Phiit$
 proposed in \cite{DoTranGan} and
 the  inverse  Daubechies-6 DWT matrix to be
the orthogonal sparsifying transform matrix $\Psiit$.  The wavelet
coefficients of the Lena image are only approximately sparse.
 If we have a parameter estimate  $\widehat{\btheta}(r)=\big(
\widehat{\bs}(r), \widehat{\sigma}^2( r ) \big)$, we
can construct the following empirical Bayesian estimate of the missing
data vector $\bz$:
\be
\label{eq:empiricalBayesian}
\Exp_{\sbz \,| \sby, \,
\sbtheta }[\bz \,|\, \by, \widehat{\btheta}(r) ] = \widehat{\bs}(r) +
H^T (H H^T)^{-1} [\by - H \, \widehat{\bs}(r) ].
\ee
Unlike $\widehat{\bs}(r)$,
the empirical Bayesian estimate (\ref{eq:empiricalBayesian})
is not $r$-sparse in general, and is therefore preferable for
reconstructing approximately sparse signals that have many small-magnitude
signal coefficients.

For our choices of $\Phiit$ and $\Psiit$, the rows of $H$ are
orthonormal, i.e.\ (\ref{eq:orthsensingmatrixcond}) holds, and
 IHT is equivalent to the ECME iteration in Section \ref{ECME}.
For all hard thresholding methods, we apply the empirical Bayesian
estimate (\ref{eq:empiricalBayesian}), with $\widehat{\btheta}(r)$
equal to the parameter estimates obtained upon their convergence. We
chose the sparsity level
\be
r=10000 \approx 0.038 \, m
\ee
to implement the
DORE, IHT, and NIHT iterations.

Figs.\ \ref{fig:numex2}~(b)--(d) show the PSNRs, numbers of
iterations, and CPU times of various methods as functions of the
normalized number of measurements (subsampling factor) $N/m$. Here, we do not include the
AHT and MN estimates in the simulation results because, in
this example, AHT is very slow
compared with the other approaches and does not outperform them in
terms of reconstruction performance,
and the MN estimates
(\ref{eq:sMN}) are poor; indeed, the PSNRs of the minimum-norm
estimates vary between 14.21~dB and 16.24~dB for the range of
$N/m$ in Figs.\ \ref{fig:numex2}~(b)--(d).

Unlike the phantom reconstruction example in Section \ref{NumEx1},
here the underlying signal (the vector of the wavelet coefficients of
the Lena image) is {\em not}\/ strictly sparse and, consequently,
\begin{itemize}
\item
the difference in
reconstruction accuracy between the {\em manually tuned}\/ convex GPSR
method and hard thresholding methods
 is significantly smaller: compare Fig.\
\ref{fig:numex2}~(b) with Fig.\
\ref{fig:PSNRnumiterCPUvsNoverm}~(a);
\item the achieved PSNRs of all methods are significantly smaller as
      well, even though the subsampling factor $N/m$ ranges over a fairly
      wide interval, between $0.2$ and $0.5$.
\end{itemize}
The importance of tuning the GPSR's regularization parameter $\tau$
is evident from  Fig.\
\ref{fig:numex2}~(b):  GPSR$_0$ reconstructs the signal
poorly compared with the tuned GPSR. We point out that it is {\em not
known}\/ how to manually tune $\tau$ in practical cases where the
ground-truth image in Fig.\ \ref{fig:numex2}~(a) is not available. In
contrast, our ADORE algorithm automatically selects the sparsity level
and performs similarly to the other methods that require careful
tuning; ADORE is particularly competitive when the number of
measurements is fairly large, see Fig.\
\ref{fig:numex2}~(b). Therefore, the USS model selection criterion is
quite effective in this practical example where the underlying signal
is not strictly sparse. Figs.\
\ref{fig:numex2}~(c) and (d) show that DORE requires the smallest
number of iterations and CPU time among the hard
thresholding methods, and that it is also faster than the manually
tuned GPSR method for $N/m < 0.4$.  When $N/m>0.3$, the CPU time of
the {\em automatic}\/ ADORE method (which employs multiple DORE
iterations) is comparable to that of the {\em tuned}\/ NIHT
method.

\newsection{Concluding Remarks}
\label{Conclusion}

\noindent We proposed a probabilistic framework for sparse signal
reconstruction in underdetermined linear models where the regression
coefficient vector consists of a sparse deterministic component and a random
Gaussian component. We developed three hard thresholding methods based
on this framework: ECME, DORE, and ADORE.  We showed that, under
certain mild conditions, ECME converges to a local maximum of the
concentrated marginal likelihood for the above probabilistic
model. Our ECME convergence conditions are
invariant to invertible linear transforms of either the rows
or the columns of
 the sensing matrix. To
develop our near-optimal recovery results for the ECME and DORE
methods, we introduced new measures of the sensing matrix's
reconstruction ability: sparse subspace quotient (SSQ) and minimum
SSQ. Minimum SSQ is more flexible than the well-established restricted
isometry property (RIP) and coherence measures: it is invariant to
invertible linear transforms of the rows of the sensing matrix. When
the minimum $2r$-SSQ is sufficiently large, our ECME for sparsity
level $r$ {\em perfectly}\/ recovers the true $r$-sparse signal
 from the noiseless measurements and estimates the best
$r$-term approximation of an arbitrary non-sparse signal from
noisy measurements within a bounded error.  The DORE algorithm
interleaves two overrelaxation steps with one ECME step and
significantly accelerates the convergence of the ECME iteration.  To
automatically estimate the sparsity level from the data, we proposed
the unconstrained sparsity selection (USS) criterion and utilized it to
develop the automatic ADORE scheme that does not require prior
knowledge of the signal sparsity level.

Since only a single choice (\ref{eq:s0}) is used to initialize DORE and ADORE, their PSNR curves in Section
\ref{NumEx} are only lower bounds on the PSNRs achievable
by these methods.  The reconstruction performances of these methods
can be improved by using multiple initial values, where the
improvement is particularly significant for purely sparse signals: our
preliminary results indicate that, in terms of reconstruction
accuracy, DORE with multiple initial values outperforms AHT in the
phantom example [see Fig.\
\ref{fig:PSNRnumiterCPUvsNoverm}~(a)] and can
slightly outperform the manually tuned GPSR in the Lena example [see Fig.\
\ref{fig:numex2}~(b)]. Full details of the multiple initialization
scheme and its reconstruction performance will be published
elsewhere. Further research will also include:
\begin{itemize}
\item
analyzing the
convergence speed of the DORE algorithm,
\item
looking for systematic means
of generating sensing matrices that have large minimum SSQ, and
\item
applying our probabilistic framework to develop sparse signal
reconstruction methods for quantized measurements.
\end{itemize}

\appendices

\section*{Appendix}

 \renewcommand
    {\theequation}
    {\Alph{section}.\arabic{equation}\alph{subequation}}

\appsection{ECME Algorithm Derivation}
\label{Appder}

\noindent Consider the following hierarchical two-stage model:
\bsea
\baselabel{eq:pyzpz}
\label{eq:pyz}
p_{\sby \,|\, \sbz}( \by \,|\, \bz) &=& {\cal N}(  \by \,;\, H \, \bz, C )
\Nextline
\label{eq:pz}
p_{\sbz \,|\, \sbtheta}( \bz \,|\, \btheta  ) &=&
{\cal N}(  \bz \,;\,  \bs, \sigma^2 \, I_m )
\esea
where $\bz$ is the vector of missing data and
$C$ is a known noise covariance
matrix.
%The marginal likelihood function of $\btheta$ is obtained by {\em
%integrating $\bz$ out}\/:
%\be
%\label{eq:likelihoodfunctionApp}
%    p_{ \sby \,|\, \sbtheta }( \by \,|\, \btheta  ) = {\cal N}(  \by
%    \,;\,  H \, \bs, \sigma^2 \, H \, H^T + C ).
%\ee
For $C = 0_{\sN \times \sN}$, this model reduces to that in
(\ref{eq:model_1})--(\ref{eq:model_2}) in Section \ref{MM}.

We will first derive an EM step for estimating $\bs$ under the above
general model and then set $C = 0_{\sN \times \sN}$ to reduce it to
 the EM step in Section \ref{MM}.
The {\em complete-data likelihood
function}\/ of the measurements $\by$ and the missing data $\bz$ given
$\btheta = ( \bs, \sigma^2 ) \in \Theta_{r}$
 follows
from (\ref{eq:pyz}) and (\ref{eq:pz}):
\begin{eqnarray}
\label{eq:completelf}
    p_{\sbz, \sby \,|\, \sbtheta}( \bz, \by \,|\,  \btheta  ) &=&
 \frac{ \exp[ - \half \, ( \by - H \, \bz )^T \, C^{-1} \, ( \by - H \, \bz )  ] }{ \sqrt{  \det(  2 \, \pi \, C ) }  }
\cdot \frac{ \exp( - \half \, \| \bz - \bs \|_{\ell_2}^2
/  \sigma^2   ) }{  \sqrt{  ( 2 \, \pi \, \sigma^2 )^m }  }.
\end{eqnarray}
From (\ref{eq:completelf}), the conditional pdf
of $\bz$ given $\by$ and $\btheta$ is
\be
p_{\sbz \,|\, \sby ,\sbtheta}( \bz \,|\, \by  , \btheta ) =  {\cal N} \Big( \bz \,;\,
 \bs + \sigma^2 \, H^T \, ( C + \sigma^2 \,
H \, H^T )^{-1} \, ( \by -  H \, \bs ),
\sigma^2 \, I_m -  (\sigma^2)^2 \, H^T \, ( C + \sigma^2 \,
H \, H^T )^{-1} \, H
\Big)
\ee
see \cite[Theorem 11.1]{Kay}.
Assume that the
parameter estimate $\btheta^{(p)} = \big( \bs^{(p)},
(\sigma^2)^{(p)} \big)$ is available; then, in {\em
Iteration $p+1$}, the E and M steps for estimating $\bs$
simplify to
\bsea
\baselabel{eq:ECMEonestepApp}
\label{eq:zpplusoneApp}
\bz^{(p+1)} = \Exp_{\sbz \,| \, \sby,  \, \sbtheta }[\bz \,|\, \by,
\btheta^{(p)}] = \bs^{(p)}
+ (\sigma^2)^{(p)} \, H^T \, \big[ C + (\sigma^2)^{(p)} \, H \, H^T
  ]^{-1} \, ( \by - H \, \bs^{(p)} )
\esea
and
\addtocounter{equation}{-1}
\bsea
\stepcounter{subequation}
\label{eq:ECMEzdeltapplusoneApp}
\bs^{(p+1)} = \arg \min_{  \sbs \in {\cal S}_r } \|
\bz^{(p+1)} - \bs  \|_{\ell_2}^2  = {\cal T}_r( \bz^{(p+1)}  \big).
\esea
Setting $C = 0_{\sN \times \sN}$ in (\ref{eq:zpplusoneApp}) and
(\ref{eq:ECMEzdeltapplusoneApp}) yields (\ref{eq:zpplusone}) and
(\ref{eq:ECMEzdeltapplusone}), which are not dependent on $(\sigma^2)^{(p)}$.

\appsection{Proof of Theorem \ref{theorem1}}
\label{AppProofThm1}

\noindent We first prove Lemma \ref{lemma1}, which will be used in the proof of Theorem \ref{theorem1}.
\begin{lemma}
\label{lemma1}
Assume that the sensing matrix $H$ satisfies the URP condition, see
also (\ref{eq:URP}).  For an index set $A \subset \{1,2,\ldots,m\}$,
\begin{description}
\item[(a)]
if
\bsea
\label{eq:lemma1_cond1}
0 < {\rm dim}(A) \leq N
\esea
then
\addtocounter{equation}{-1}
\bsea
\stepcounter{subequation}
\label{eq:lemma1_1}
\lambda_{\min}\big( \, H_{\sA}^T \, (H \, H^T)^{-1} \, H_{\sA} \,  \big) > 0,
\esea
\item[(b)] if
\bsea
\label{eq:lemma1_cond2}
0 < {\rm dim}(A) \leq  m-N
\esea
then
\addtocounter{equation}{-1}
\bsea
\stepcounter{subequation}
\label{eq:lemma1_2}
\lambda_{\max}\big( \, H_{\sA}^T \, (H \, H^T)^{-1} \, H_{\sA} \,  \big) < 1.
\esea
\end{description}
\end{lemma}
\begin{IEEEproof}
The conditions (\ref{eq:URP}) and (\ref{eq:lemma1_cond1})
imply that all columns of $H_{\sA}$ are linearly independent; therefore,
$H_{\sA}^T (H H^T)^{-1} H_{\sA}$ is a full-rank positive definite
matrix, and (\ref{eq:lemma1_1}) follows.

We now assume (\ref{eq:lemma1_cond2})
 and show
(\ref{eq:lemma1_2}). Observe that
\be
\label{eq:lemma1_proof1}
\lambda_{\max} \big( H_{\sA}^T \, (H \, H^T)^{-1} \, H_{\sA}  \big) =
\lambda_{\max} \big(  (H \, H^T)^{-1} \, H_{\sA} H_{\sA}^T   \big) =
\lambda_{\max} \big(  I_{\sN} - (H \, H^T)^{-1} \, H_{\sA^\sc}
H_{\sA^\sc}^T  \big)
\ee
where
\be
\label{eq:complementdef}
A^\sc \df \{1,2,\ldots,m\} \, \backslash \, A
\ee
defines the index index set complementary to $A$. Since
${\rm dim}(A^\sc) = m- {\rm dim}(A) \ge N$, $(H \, H^T)^{-1} \, H_{\sA^\sc} H_{\sA^\sc}^T$ is positive definite;
therefore,
\be
\lambda_{\max} \big( H_{\sA}^T \, (H \, H^T)^{-1} \, H_{\sA} \big) = 1
- \lambda_{\min} \big( (H \, H^T)^{-1} H_{\sA^\sc} H_{\sA^\sc}^T \big) < 1
\ee
and (\ref{eq:lemma1_2}) follows.
\end{IEEEproof}

Now, we prove Theorem \ref{theorem1}.
\begin{IEEEproof}[Proof of Theorem \ref{theorem1}]
We now prove that our ECME iteration converges to its fixed point.  If
$\bs^{(p+1)} = \bs^{(p)}$, the convergence to a fixed point
immediately follows. Therefore, without loss of generality, we assume
$\bs^{(p+1)} \neq \bs^{(p)}$.  Since $\cE(\bs^{(p)})$ in (\ref{eq:cE})
converges to a limit, $\cE(\bs^{(p)})-\cE(\bs^{(p+1)})$ converges to
zero.  Now,
\bsea
\baselabel{eq:thm1_proof1}
 \cE(\bs^{(p)})-\cE(\bs^{(p+1)})
\label{eq:thm1_proof1_1}
&=& \cQ(\bs^{(p)} \,|\, \bs^{(p)}) - \cH(\bs^{(p)} \,|\, \bs^{(p)}) - [ \, \cQ(\bs^{(p+1)} \,|\, \bs^{(p)}) - \cH(\bs^{(p+1)} \,|\, \bs^{(p)}) \, ]
\Nextline
\label{eq:thm1_proof1_2}
&\ge& (\bs^{(p+1)} - \bs^{(p)})^T \, [I_m - H^T \, ( H  H^T)^{-1} \, H] \, (\bs^{(p+1)} - \bs^{(p)})
\Nextline
\label{eq:thm1_proof1_3}
&=& (\bs^{(p+1)}_A - \bs^{(p)}_A)^T \, [I_{{\rm dim}(\sA)} - H_A^T \, ( H
H^T)^{-1} \, H_A] \, (\bs^{(p+1)}_A - \bs^{(p)}_A)
\Nextline
\label{eq:thm1_proof1_4}
&=& \Big[ 1 - \frac{ (\bs^{(p+1)}_A - \bs^{(p)}_A)^T \,
H_{\sA}^T (H H^T)^{-1} H_{\sA} \, (\bs^{(p+1)}_A - \bs^{(p)}_A)
}{\| \bs^{(p+1)} - \bs^{(p)} \|_{\ell_2}^2} \Big] \, \| \bs^{(p+1)} -
\bs^{(p)} \|_{\ell_2}^2
\hspace{0.5in}
\Nextline
\label{eq:thm1_proof1_5}
&\ge& \big[1-\lambda_{\max} \big( \, H_{\sA}^T (H H^T)^{-1} H_{\sA} \,  \big) \big] \, \| \bs^{(p+1)} - \bs^{(p)} \|_{\ell_2}^2
\esea
where $A={\rm supp}(\bs^{(p)}) \cup {\rm supp}(\bs^{(p+1)})$.
Here, (\ref{eq:thm1_proof1_1}) follows from (\ref{eq:cE_identity}),
(\ref{eq:thm1_proof1_2}) follows by
(\ref{eq:Qineq}) and the fact that $\cH(\bs^{(p)}
\,|\, \bs^{(p)})=0$,
(\ref{eq:thm1_proof1_3}) is obtained by using the identities
$\| \bs^{(p+1)} - \bs^{(p)} \|_{\ell_2}^2=\| \bs^{(p+1)}_A - \bs^{(p)}_A \|_{\ell_2}^2$ and
$H \, (\bs^{(p+1)} - \bs^{(p)}) = H_{\sA} \, (\bs_{\sA}^{(p+1)} -
\bs_{\sA}^{(p)})$, and
(\ref{eq:thm1_proof1_5}) follows by using the Rayleigh-quotient
property
\cite[Theorem 21.5.6]{Harville}.
Note that $0 < {\rm dim}(A)
\le 2 \, r \le m-N$, where the second inequality follows from
(\ref{eq:rcondfixedpoint}). Therefore, (\ref{eq:lemma1_cond2}) holds
 and (\ref{eq:lemma1_2}) in Lemma \ref{lemma1} implies that the term
$1-\lambda_{\max} \big( \, H_{\sA}^T (H H^T)^{-1} H_{\sA} \,
\big)$ in (\ref{eq:thm1_proof1_5}) is strictly positive. Since
$\cE(\bs^{(p)})-\cE(\bs^{(p+1)})$ converges to zero, then $\|
\bs^{(p+1)} - \bs^{(p)} \|_{\ell_2}^2$ converges to zero as well.
Finally, the claim of monotonicity of convergence follows from the discussion in
Section \ref{Analysis} prior to Theorem \ref{theorem1}.
This completes the proof.
\end{IEEEproof}

\appsection{Proofs of Lemma \ref{lemma2}, Lemma \ref{lemma4} and Theorem \ref{theorem2}}
\label{AppProofThm2}

\noindent
\begin{IEEEproof}[Proof of Lemma \ref{lemma2}]
The proof is by contradiction.  Suppose that there exists an index $i
\in \{1,2,\ldots,m\}$ satisfying (\ref{eq:lemma2cond}), but not
(\ref{eq:lemma2}); without loss of generality, assume that
the following partial derivative is positive:
\be
\frac{\partial f(\bs^\star)}{\partial s_i} \df \frac{\partial f(\bs)}{\partial s_i}\Big|_{\bs=\bs^{\star}} =  \lim_{\epsilon \rightarrow 0} \frac{f(\bs^{\star} + \epsilon \,
 \bee_i) - f(\bs^{\star})}{\epsilon}  > 0
\ee
where $\bee_i$ is the $i$th column of $I_m$. By the definition of the
limit, there exists a
$\delta > 0$ such that, for all $\epsilon \in (0, \delta)$,
\bsea
\Big |\frac{f(\bs^{\star} + \epsilon \, \bee_i) -
f(\bs^{\star})}{\epsilon}
-
\frac{\partial f(\bs^\star)}{\partial s_i} \Big| <  \half  \, \frac{\partial f(\bs^\star)}{\partial s_i}
\Nextline
\Big |\frac{f(\bs^{\star} - \epsilon \, \bee_i) -
f(\bs^{\star})}{-\epsilon} -
\frac{\partial f(\bs^\star)}{\partial s_i} \Big| <  \half  \, \frac{\partial f(\bs^\star)}{\partial s_i}
\esea
and, therefore,
\bsea
f(\bs^{\star} + \epsilon \, \bee_i) >  f(\bs^{\star}) + \half \, \epsilon \, \frac{\partial f(\bs^\star)}{\partial s_i} > f(\bs^{\star})
\Nextline
f(\bs^{\star} - \epsilon \, \bee_i)< f(\bs^{\star}) - \half \, \epsilon \, \frac{\partial f(\bs^\star)}{\partial s_i} < f(\bs^{\star}).
\esea
For all $\epsilon
\in (0, \delta)$,
the vectors $\bs^{\star} + \epsilon \, \bee_i $ and
$\bs^{\star} - \epsilon \, \bee_i $ are $r$-sparse,  $f(\bs^{\star} + \epsilon \, \bee_i)$ is larger
than $f(\bs^{\star})$, and $f(\bs^{\star} - \epsilon \, \bee_i)$
is smaller than $f(\bs^{\star})$, which contradicts the assumption that
$\bs^{\star}$ is an $r$-local maximum or minimum point.
\end{IEEEproof}

Before proving Lemma \ref{lemma4}, we prove the following useful result that will be used in the
proof of Lemma \ref{lemma4}.

\begin{lemma}
\label{lemma3}
For any $r$-sparse vector $\bs' \in {\cal S}_r$, there exists a
$\delta > 0$ such that, for all $\bs \in {\cal S}_r$ satisfying $\|
\bs - \bs' \|_{\ell_2} < \delta$, we have
\be
\label{eq:lemma3}
{\rm dim}\big({\rm supp}(\bs)
\cup {\rm supp}(\bs')\big) \leq r.
\ee
\end{lemma}
\begin{IEEEproof}
The proof is by contradiction.
First, define $A = {\rm supp}(\bs)$ and  $A' = {\rm supp}(\bs')$.
Suppose that, for all $\delta > 0$,
there exists a $\bs \in {\cal S}_r$ satisfying
\bsea
\label{eq:lemma3proof1}
\| \bs - \bs' \|_{\ell_2} < \delta
\esea
and
\addtocounter{equation}{-1}
\bsea
\stepcounter{subequation}
\label{eq:lemma3proof2}
{\rm dim}(A \cup A') > r.
\esea
Since ${\rm dim}(A) \le r$,
\be
{\rm dim}\big(A' \cap A^\sc \big) = {\rm dim}\big(A \cup A'\big) - {\rm dim}\big(A\big)>r-r=0
\ee
implying that the set $A' \cap A^\sc$ is not empty, see also the definition of the complementary
index set in (\ref{eq:complementdef}). Choose $\delta$ to be half the
magnitude of the smallest nonzero element in $\bs'$:
\be
\label{eq:deltachoicelemma3}
\delta=\half \min_{i\in A'}
|s'_i|
\ee
Now,
\be
\| \bs - \bs' \|_{\ell_2} \ge \big\| \bs'_{A' \cap \sA^\sc} \big\|_{\ell_2} \ge \min_{i\in A'} |\bs'_{i}| > \delta
\ee
which contradicts (\ref{eq:lemma3proof1}).  Therefore, for a positive
number $\delta$ in (\ref{eq:deltachoicelemma3}), no $r$-sparse vector
$\bs$ can satisfy the conditions (\ref{eq:lemma3proof1}) and
(\ref{eq:lemma3proof2})
 simultaneously.
\end{IEEEproof}

Lemma \ref{lemma3} shows that, in a sufficiently small neighborhood of
an $r$-sparse vector $\bs'$, the support sets of all other $r$-sparse
vectors $\bs$ significantly overlap with the support set of $\bs'$. In
particular, if $\bs'$ has exactly $r$ nonzero elements (i.e.\ $\| \bs'
\|_{\ell_0} =r$), then all other $r$-sparse vectors in its
sufficiently small neighborhood must have the same support set as
$\bs'$. If $\bs'$ has less than $r$ nonzero elements,
an $r$-sparse
vector $\bs$ in this neighborhood can contain
a few
inconsistent elements that do not belong to the support set of $\bs'$
{\em as long as}\/ (\ref{eq:lemma3}) is satisfied.
%This result is
%useful in finding the following sufficient condition of $r$-local
%maximum or minimum points.
We now prove Lemma \ref{lemma4}.

\begin{IEEEproof}[Proof of Lemma \ref{lemma4}]
We first consider the case of an $r$-local maximum of $f(\bs)$ and
assume that conditions (1) and (2) hold for a point $\bs^{\star} \in
{\cal S}_r$.  By condition (2), for the positive number $\delta_1$,
the Hessian matrix is negative semidefinite around $\bs^{\star}$ for
all $\bs \in {\cal S}_r$ satisfying $\| \bs -
\bs^{\star} \|_{\ell_2} < \delta_1$.  By Lemma \ref{lemma3}, for any
$r$-sparse vector $\bs^{\star}$, there exists a $\delta_2 > 0$ such
that, for all $\bs
\in {\cal S}_r$ satisfying $\| \bs - \bs^{\star} \|_{\ell_2} <
\delta_2$, we have
\be
\label{eq:lemmafproofcond}
{\rm dim}( {\rm supp}(\bs) \cup {\rm
supp}(\bs^{\star}) ) \leq r.
\ee

Now, for $\delta=\min
\{\delta_1, \delta_2 \}$, consider any $\bs \in {\cal S}_r$ satisfying
$\| \bs - \bs^{\star} \|_{\ell_2} < \delta$, and expand $f(\bs)$ around
$\bs^{\star}$ using the Taylor series with Lagrange's form of the
remainder \cite[p.\ 243]{Colley}:
\bsea
\label{eq:lemma4_proof1}
f(\bs) - f(\bs^{\star})&=& (\bs - \bs^{\star})^T \, \frac{\partial f(\bs)}{\partial \bs}\Big|_{\sbs=\sbs^{\star}} + \half \, (\bs - \bs^{\star})^T \,
\frac{\partial^2 f(\bs)}{\partial \bs \partial \bs^T}\Big|_{\sbs=\sbs^{\star} + c \, (\sbs - \sbs^{\star})} \, (\bs - \bs^{\star})
\Nextline
\label{eq:lemma4_proof2}
&\leq& (\bs - \bs^{\star})^T \, \frac{\partial f(\bs)}{\partial \bs}\Big|_{\sbs=\sbs^{\star}}
\Nextline
\label{eq:lemma4_proof3}
&=& \sum_{i \in {\rm supp}(\sbs) \cup {\rm supp}(\sbs^{\star})} (\bs_i - \bs^{\star}_i) \, \frac{\partial f(\bs)}{\partial s_i}\Big|_{\sbs=\sbs^{\star}}
\Nextline
\label{eq:lemma4_proof4}
&=& 0
\esea
where $c \in (0,1)$. Since the vector $\bs^{\star} + c \, (\bs -
\bs^{\star})$ is $r$-sparse and satisfies $\| \bs^{\star} + c \, (\bs -
\bs^{\star}) - \bs^{\star} \|_{\ell_2} < \delta$,  the
Hessian in (\ref{eq:lemma4_proof1}) is negative-semidefinite
and
 (\ref{eq:lemma4_proof2}) follows.
Condition (1) of Lemma \ref{lemma4} and (\ref{eq:lemmafproofcond}) imply that the partial
derivatives in (\ref{eq:cond1}) are zero for all coordinates with
indices $i \in {\rm supp}(\bs)
\cup {\rm supp}(\bs^{\star})$, and (\ref{eq:lemma4_proof4}) follows.  Now, we have a
$\delta = \min \{\delta_1, \delta_2 \} > 0$ such that, for all
$\bs \in {\cal S}_r$ satisfying $\| \bs - \bs^{\star} \|_{\ell_2} <
\delta $, $f(\bs) \leq f(\bs^{\star})$; therefore $\bs^{\star}$
is an $r$-local maximum.

If the Hessian matrix $\frac{\partial^2 f(\bs)}{\partial \bs \partial \bs^T}$ is positive semidefinite around $\bs^{\star}$, then
$\frac{\partial^2 [-f(\bs)]}{\partial \bs \partial \bs^T}$ is negative
semidefinite around $\bs^{\star}$. Therefore, $\bs^{\star}$ is an
$r$-local maximum of $-f(\bs)$, and, by Definition \ref{def1},
$\bs^{\star}$ is an $r$-local minimum of $f(\bs)$.
\end{IEEEproof}

We are now ready to show Theorem \ref{theorem2}.
\begin{IEEEproof}[Proof of Theorem \ref{theorem2}]
Since $\btheta^{\star}=(\bs^{\star},(\sigma^2)^{\star})$ is a fixed
point of the ECME iteration, we have
\be
\label{eq:thm2_proof0}
\bs^{\star} = \arg \min_{  \sbs \in \scalS_r } \cQ(\bs \,|\, \bs^{\star}) =
\arg \min_{  \sbs \in \scalS_r } \| \bs - [ \bs^{\star} + H^{T} (H H^T)^{-1} (\by - H \bs^{\star} ) ]  \|^2_{\ell_2}
\ee
see (\ref{eq:ECMEzdeltapplusone}) and (\ref{eq:Qfunction}).

We first show that the conditions of Lemma \ref{lemma4} hold for the
function $f(\bs) = \cE(\bs)$ in (\ref{eq:cE}) and the $r$-sparse
vector $\bs^{\star}$ in (\ref{eq:thm2_proof0}).
%, i.e.\ for all indices $i \in \{1,2,\ldots,m\}$ that
%satisfies ${\rm dim} \big( \{i\} \cup {\rm supp}(\bs^{\star}) \big)
%\le r$, the partial derivative in the direction of $i$-th coordinate
%$\frac{\partial \cE}{\partial s_i} (\bs^{\star}) = 0$.
The proof is by contradiction.
 Suppose that condition (1) of Lemma \ref{lemma4} is not satisfied,
 i.e. there exists an index $i
\in \{1,2,\ldots,m\}$ such that ${\rm dim} \big( \{i\} \cup {\rm
supp}(\bs^{\star}) \big) \le r$, but the corresponding partial derivative
\be
\frac{\partial \cE(\bs^{\star})}{\partial s_i} \df \frac{\partial \cE(\bs)}{\partial s_i}\Big|_{\bs=\bs^{\star}}
\ee
is not zero;
without loss of generality, assume that this partial derivative is positive:
\be
\frac{\partial \cE(\bs^{\star})}{\partial s_i}  > 0.
\ee
By the definitions of the partial derivative and limit, for the real
number $\half \, \frac{\partial \cE(\bs^{\star})}{\partial s_i}$,
there exists a positive number $\delta > 0$ such that, for all $\epsilon \in (0, \delta)$,
the vector $\bs_{\epsilon}=\bs^{\star} - \epsilon \, \bee_i$ satisfies
\bsea
\Big |\frac{\cE(\bs^{\star} - \epsilon \, \bee_i) - \cE(\bs^{\star})}{-\epsilon} - \frac{\partial \cE(\bs^{\star})}{\partial s_i} \Big| <  \half  \, \frac{\partial \cE(\bs^{\star})}{\partial s_i}
\esea
and, therefore,
\addtocounter{equation}{-1}
\bsea
\stepcounter{subequation}
\label{eq:thm2_proof1}
\cE(\bs^{\star} - \epsilon \, \bee_i) <  \cE(\bs^{\star}) - \half  \,  \epsilon \, \frac{\partial \cE(\bs^{\star})}{\partial s_i}.
\esea
Now, compute [see  (\ref{eq:Hfunction})]
\bsea
\baselabel{eq:thm2_proof2}
\cH(\bs^{\star} - \epsilon \, \bee_i \,|\, \bs^{\star}) - \cH(\bs^{\star} \,|\, \bs^{\star})
\label{eq:thm2_proof2_1}
&=&
(\bs^{\star} - \epsilon \, \bee_i - \bs^{\star})^T \, [I_m - H^T \, (
H  H^T)^{-1} \, H] \, (\bs^{\star} - \epsilon \, \bee_i - \bs^{\star})
\hspace{0.5in}
\Nextline
\label{eq:thm2_proof2_2}
&\le& \| \bs^{\star} - \epsilon \, \bee_i - \bs^{\star} \|_{\ell_2}^2 = \epsilon^2
\esea
where (\ref{eq:thm2_proof2_2}) follows by observing that $H^T \, ( H
H^T)^{-1} \, H$ is positive semidefinite.  Therefore, we have [see
(\ref{eq:cE_identity})]
\bsea
\label{eq:thm2_proof3_1}
\cQ( \bs^{\star} - \epsilon \, \bee_i \,|\, \bs^{\star}) &=& \cE(\bs^{\star} - \epsilon \, \bee_i) + \cH(\bs^{\star} - \epsilon \, \bee_i \,|\, \bs^{\star})
\Nextline
\label{eq:thm2_proof3_2}
&<& \cE(\bs^{\star}) -  \half  \, \epsilon \, \frac{\partial \cE(\bs^{\star})}{\partial s_i} + \cH(\bs^{\star}\,|\, \bs^{\star}) + \epsilon^2
\Nextline
&=& \cQ(\bs^{\star} \,|\, \bs^{\star}) - \Big( \half \frac{\partial \cE(\bs^{\star})}{\partial s_i}-\epsilon \Big) \, \epsilon
\esea
where (\ref{eq:thm2_proof3_2}) follows from (\ref{eq:thm2_proof1}) and
(\ref{eq:thm2_proof2}).  Note that the vector
$\bs_{\epsilon}=\bs^{\star} - \epsilon \, \bee_i $ is $r$-sparse.  For
any
\be
\epsilon \in \Big(0, \min \Big\{  \delta, \half\, \frac{\partial
\cE(\bs^{\star})}{\partial s_i} \Big\} \Big)
\ee
we have
$\cQ(\bs_{\epsilon}\,|\, \bs^{\star}) < \cQ(\bs^{\star} \,|\,
\bs^{\star})$, which contradicts (\ref{eq:thm2_proof0}).
Hence, the condition (1) of Lemma \ref{lemma4} holds.

The condition (2) of Lemma \ref{lemma4} holds because, for any $\bs
\in {\cal R}^m$, the Hessian of $\cE(\bs)$ is
\be
\label{eq:thm2_proof4}
\frac{\partial^2 \cE (\bs) }{\partial \bs \, \partial \bs^T} = 2 \, H^{T} (H H^T)^{-1} H
\ee
which is clearly a positive semidefinite matrix.

Since the conditions of Lemma \ref{lemma4} hold for
the function $f(\bs) = \cE(\bs)$ in (\ref{eq:cE}) and fixed point
$\bs^{\star}$, we apply Lemma \ref{lemma4} and conclude that
$\bs^{\star}$ is an $r$-local minimum point of $\cE(\bs)$.
Consequently, $\bs^{\star}$ is an $r$-local maximum point of the
concentrated marginal likelihood function
(\ref{eq:profilelikelihood}), which follows from the fact that
(\ref{eq:profilelikelihood}) is a monotonically decreasing function of
$\cE(\bs)= N \, \widehat{\sigma}^2(\bs)$, see also
(\ref{eq:sigma2_hat}).
\end{IEEEproof}

\appsection{Proofs of Lemma \ref{lemma5}, Lemma \ref{lemma6}, Theorem \ref{theorem3} and Theorem \ref{theorem4}}
\label{AppProofThm34}

We first show Lemma \ref{lemma5}.
\begin{IEEEproof}[Proof of Lemma \ref{lemma5}]
Equation (\ref{eq:OPQvar}) in part~(a) follows by noting that $H
\, \bs = H_{\sA} \bs_{\sA}$,
%\be
%(GH)^T \, [(GH)  (GH)^T]^{-1} \, GH = H^T G^T (G^T)^{-1} \, (H  H^T)^{-1} \, G^{-1} G H = H^T \, (H  H^T)^{-1} \, H.
%\ee
and (\ref{eq:lemma5_1_1}) in part (a) follows by using
 (\ref{eq:OPQvar}) and the Rayleigh quotient property \cite[Theorem 21.5.6]{Harville}, respectively:
\begin{eqnarray}
\rho_{r,\min}(H)&=&\min_{\bs \in {\cal S}_r \backslash {\bf 0}_{m \times 1}} \rho_r(\bs,H)
\nonumber
\\
&=& \min_{A \subseteq \{1,2,\ldots,m\} , \,  {\rm dim}(A)=r } \Big[
\min_{\bs_{\sA} \in {\cal R}^r \backslash {\bf 0}_{r \times 1} } \frac{ \bs_{\sA}^T H_{\sA}^T \, (H  H^T)^{-1} \, H_{\sA} \bs_{\sA}}{\|\bs_{\sA}\|_{\ell_2}^2} \Big]
\nonumber
\\
\label{eq:lemma5_proof2}
&=& \min_{A \subseteq \{1,2,\ldots,m\} , \,  {\rm dim}(A)=r} \lambda_{\min} \big( H_{\sA}^T \, (H  H^T)^{-1} \, H_{\sA} \big)
\end{eqnarray}

Part~(b) holds because the row spaces of $H$ and $G \, H$ coincide.

The inequalities (\ref{eq:lemma5_3}) in part~(c) follow by applying
the Rayleigh-quotient property and the fact that the projection matrix
$H^T \, (H H^T)^{-1} \, H$ only has eigenvalues $0$ and $1$.  When
$r>N$, $H_{\sA}^T \, (H H^T)^{-1} \, H_{\sA}$ is not a full-rank
matrix for any index set $A$ with dimension $r$; consequently,
$\rho_{r,\min}(H)=0$ follows by using (\ref{eq:lemma5_1_1}) in
part~(a) of this lemma.
 When $N=m$, $r$-SSQ in (\ref{eq:def_SSQ_1}) is equal to one
for any $0 < r
\leq m$, and, therefore, $\rho_{r,\min}(H)=1$.

In part (d), we first show that (\ref{eq:lemma5_4_1}) implies
(\ref{eq:lemma5_4}). When $\spark(H)>r$, the matrix $H_{\sA}^T (H
H^T)^{-1} H_{\sA}$ is positive definite for any $A \subset
\{1,2,\ldots,m\}$ with ${\rm dim}(A) = r$, and (\ref{eq:lemma5_4})
follows by using  (\ref{eq:lemma5_1_1}) in part~(a) of this
lemma. We now show the `only if'
direction of part (d) by contradiction. Suppose that (\ref{eq:lemma5_4})
holds but $\spark(H) \leq r$. By the definition of $\spark$, there
exists an index set $A$ with ${\rm dim}(A) = r$ such that the columns
of $H_{\sA}$ are linearly dependent.  Therefore, the minimum
eigenvalue of the matrix $H_{\sA}^T (H H^T)^{-1} H_{\sA}$ is zero and
(\ref{eq:lemma5_1_1}) implies that $\rho_{r,\min}(H)=0$, which leads to
contradiction.
%applying (\ref{eq:lemma5_1_1})
%Part (d) follows by
%applying (\ref{eq:lemma5_1_1}) and Lemma
%\ref{lemma1}~(a).

Finally, part~(e) follows because the $r_1$-sparse
vector that minimizes $\rho_{r_1}(\bs,H)$ is also $r_2$-sparse and
does not minimize $\rho_{r_2}(\bs,H)$ in general.
%We now apply (c) to prove (d). First assume $0 < r \leq N$.
%By (\ref{eq:lemma1_1}) of lemma \ref{lemma1}, for any $A \in \{1,2,\ldots,m\}$ such that ${\rm dim}(A)=r$,
%$\lambda_{\min} \big( H_{\sA}^T \, (H  H^T)^{-1} \, H_{\sA} \big)>0$.
%From (\ref{eq:lemma5_1_1}), we conclude (\ref{eq:lemma5_4}).
%Now assume $r \le m-N$. By (\ref{eq:lemma1_2}) of lemma \ref{lemma1}, for any $A \in \{1,2,\ldots,m\}$ such that ${\rm dim}(A)=r$,
%$\lambda_{\max} \big( H_{\sA}^T \, (H  H^T)^{-1} \, H_{\sA} \big) < 1$.
%(\ref{eq:lemma5_4_2}) follows from (\ref{eq:lemma5_3_2}).
\end{IEEEproof}

Now, we prove Lemma \ref{lemma6}.
\begin{IEEEproof}[Proof of Lemma \ref{lemma6}]
By part~(d) of Lemma \ref{lemma5}, the condition (\ref{eq:lemma6cond})
holds if and only if $\spark(H)>2 r^\diamond$, which is exactly the
condition required in \cite[Theorem 2]{BrucksteinDonohoElad} to
develop the same claim about the uniqueness of the $({\rm P}_0)$
problem. This concludes the proof.
%The proof is by contradiction. Suppose that there exists another
%$r^\diamond$-sparse signal $\bs' \neq \bs^{\diamond}$
%%(i.e.\
%%$\bs' \in {\cal S}_r$ and $\bs' \neq \bs^{\diamond}$)
% that satisfies
%$\by = H \, \bs'$. Then,
%\be
%H \, (\bs'-\bs^{\diamond}) = \by - \by = {\bf 0}_{\sN \times 1}
%\ee
%where $\bs'-\bs^{\diamond}$ is a nonzero vector that is $2 r^{\diamond}$-sparse,
%leading to
%\be
%\rho_{2 r^{\diamond}}(\bs'-\bs^{\diamond},H)=\frac{ \| H^T \, (H  H^T)^{-1} \, H (\bs'-\bs^{\diamond}) \|_{\ell_2}^{2}}{\|\bs'-\bs^{\diamond}\|_{\ell_2}^2} = 0.
%\ee
%By (\ref{eq:lemma5_3}) in Lemma \ref{lemma5}~(c),
%\be
%0 \leq \rho_{2r^{\diamond},\min}(H) \leq \rho_{2r^{\diamond}}(\bs'-\bs^{\diamond},H) = 0
%\ee
%which contradicts the assumption (\ref{eq:lemma6cond}).  Therefore,
%if (\ref{eq:lemma6cond}) holds, $\bs^{\diamond}$ is the unique $r^{\diamond}$-sparse
%vector that satisfies the linear system $\by = H \bs^{\diamond}$ and,
%by the definition (\ref{eq:P0}), must be the unique solution to the
%$({\rm P}_0)$ problem.
\end{IEEEproof}

The proof of Theorem \ref{theorem3} is shown as follows.
\begin{IEEEproof}[Proof of Theorem \ref{theorem3}]
Let $\bs^{(p)}$ be the estimate of $\bs$ obtained in {\em
Iteration $p$} of our ECME iteration.
We assume  $\bs^{(p)} \neq \bs^{\diamond}$ without loss of generality;
otherwise,  the claim follows immediately. Now,
\bsea
\label{eq:thm3_proof1}
\|\bs^{\diamond}-\bs^{(p+1)}\|_{\ell_2}^2 &=& \frac{1}{\rho_{2 r}(\bs^{\diamond}-\bs^{(p+1)},H)} \| H^T \, (H  H^T)^{-1} \, H  (\bs^{\diamond}-\bs^{(p+1)}) \|_{\ell_2}^{2}
\Nextline
\label{eq:thm3_proof2}
&\le& \frac{1}{\rho_{2 r,\min}(H)} \, \cE(\bs^{(p+1)})
\Nextline
\label{eq:thm3_proof3}
&=& \frac{1}{\rho_{2 r,\min}(H)} \, [ \, \cQ(\bs^{(p+1)} \,|\, \bs^{(p)}) - \cH(\bs^{(p+1)} \,|\, \bs^{(p)}) \, ]
\Nextline
\label{eq:thm3_proof4}
&\le& \frac{1}{\rho_{2 r,\min}(H)} \,  \cQ(\bs^{(p+1)} \,|\, \bs^{(p)})
\Nextline
\label{eq:thm3_proof5}
&\leq& \frac{1}{\rho_{2 r,\min}(H)} \,  \cQ(\bs^{\diamond} \,|\, \bs^{(p)})
\Nextline
\label{eq:thm3_proof6}
&=& \frac{1}{\rho_{2 r,\min}(H)} \,\| \bs^{\diamond} - \bs^{(p)} -
H^{T} (H H^T)^{-1} H \, (\bs^{\diamond} -  \bs^{(p)} )  \|^2_{\ell_2}
\Nextline
&=& \frac{1}{\rho_{2 r,\min}(H)} \, ( \bs^{\diamond} - \bs^{(p)}
)^T [I_m - H^{T} (H H^T)^{-1} H] \, (\bs^{\diamond} -  \bs^{(p)} )
\label{eq:thm3_proof7}
\Nextline
&=& \frac{1}{\rho_{2 r,\min}(H)} \big[ \| \bs^{\diamond}  -
\bs^{(p)} \|^2_{\ell_2} - \rho_{2 r}(\bs^{\diamond}-\bs^{(p)},H)
\, \| \bs^{\diamond}  - \bs^{(p)} \|^2_{\ell_2}  \big]
\label{eq:thm3_proof7'}
\Nextline
\label{eq:thm3_proof8}
&\leq& \zeta(H) \, \| \bs^{\diamond}  - \bs^{(p)} \|^2_{\ell_2}
\esea
where
\be
\label{eq:zeta}
\zeta(H) \df \frac{1-\rho_{2 r,\min}(H)}{\rho_{2 r,\min}(H)}
\ee
and (\ref{eq:thm3_proof1}) follows from the definition
(\ref{eq:def_SSQ_1}) and the fact that $\bs^{\diamond}-\bs^{(p+1)}$ is
at most $2r$-sparse since $\bs^{\diamond}$ and $\bs^{(p+1)}$ are
$r$-sparse; (\ref{eq:thm3_proof2}) is due to the definitions
(\ref{eq:cE}) and (\ref{eq:def_SSQ_2}), (\ref{eq:thm3_proof3}) results
from the identity (\ref{eq:cE_identity}); (\ref{eq:thm3_proof4}) holds
because $\cH(\bs^{(p+1)} \,|\, \bs^{(p)})$ is nonnegative [see (\ref{eq:Hfunction})];
(\ref{eq:thm3_proof5}) follows due to the M step of the
ECME algorithm (\ref{eq:ECMEzdeltapplusone}); (\ref{eq:thm3_proof6})
uses the definition (\ref{eq:Qfunction}) and the condition
(\ref{eq:thm3cond1}); (\ref{eq:thm3_proof7})--(\ref{eq:thm3_proof8})
follow by expanding (\ref{eq:thm3_proof6}) and using the definitions
(\ref{eq:def_SSQ_1}), (\ref{eq:def_SSQ_2}), and (\ref{eq:zeta}),
respectively.

We now apply the condition (\ref{eq:thm3cond3}) and conclude that
$\zeta(H)$ in (\ref{eq:zeta}) is nonnegative and smaller than one:
\be
\label{eq:zeta2}
0 \leq \zeta(H) = \frac{1-\rho_{2 r,\min}(H)}{\rho_{2 r,\min}(H)} < 1.
\ee
Therefore, the sequence $\|\bs^{\diamond}-\bs^{(p)}\|_{\ell_2}^2$
monotonically shrinks to zero and the claim follows.
\end{IEEEproof}

Finally, we prove Theorem \ref{theorem4}.
\begin{IEEEproof}[Proof of Theorem \ref{theorem4}]
Denote by $\bs^{(p)}$ the sparse signal estimate in {\em Iteration
$p$} of our ECME iteration. Now, for $p \geq 0$
\bsea
\label{eq:thm4_proof1}
& & \| \bs^{(p+1)} - \bs_r^{\diamond} \|_{\ell_2}
\nosenumber
\Nextline
&\le& \frac{ \| H^T \, (H  H^T)^{-1} \, H \, (\bs^{(p+1)}-\bs_r^{\diamond}) \|_{\ell_2} }{\sqrt{\rho_{2r,\min}(H)}}
\Nextline
\label{eq:thm4_proof2}
&=& \frac{  \| H^T \, (H  H^T)^{-1} \, (\by  - H \bs^{(p+1)}) + H^T \, (H  H^T)^{-1} \,
H \, ( \bs_r^{\diamond}  - \bs^{\diamond} ) \, - \, H^T \, (H  H^T)^{-1} \bn
\|_{\ell_2}  }{\sqrt{\rho_{2r,\min}(H)}}
\Nextline
\label{eq:thm4_proof3}
&\leq& \frac{  \sqrt{\cE(\bs^{(p+1)})} + \| \bs_r^{\diamond} -  \bs^{\diamond}  \|_{\ell_2} + \|H^T \, (H  H^T)^{-1} \bn \|_{\ell_2}  }{\sqrt{\rho_{2r,\min}(H)}}
\Nextline
\label{eq:thm4_proof4}
&\leq& \frac{  \sqrt{\cQ(\bs_r^{\diamond} \,|\, \bs^{(p)})} + \| \bs_r^{\diamond} -  \bs^{\diamond}  \|_{\ell_2} + \|H^T \, (H  H^T)^{-1} \bn \|_{\ell_2}  }{\sqrt{\rho_{2r,\min}(H)}}
\Nextline
\label{eq:thm4_proof5}
&\leq& \frac{  \| \bs_r^{\diamond} - \bs^{(p)} - H^{T} (H H^T)^{-1} [  H \bs_r^{\diamond} + H (\bs^{\diamond}-\bs_r^{\diamond}) + \bn - H \, \bs^{(p)} ]  \|_{\ell_2} + \| \bs_r^{\diamond} -  \bs^{\diamond}  \|_{\ell_2} + \|H^T \, (H  H^T)^{-1} \bn \|_{\ell_2}  }{\sqrt{\rho_{2r,\min}(H)}}
\nosenumber
\Nextline
& &
\Nextline
\label{eq:thm4_proof6}
&\leq& \frac{  \| \bs_r^{\diamond} - \bs^{(p)} - H^{T} (H H^T)^{-1} H \, (\bs_r^{\diamond} - \bs^{(p)} )  \|_{\ell_2} + 2 \| \bs_r^{\diamond} -  \bs^{\diamond}  \|_{\ell_2} +  2 \|H^T \, (H  H^T)^{-1} \bn \|_{\ell_2}  }{\sqrt{\rho_{2r,\min}(H)}} \quad
\Nextline
\label{eq:thm4_proof7}
&\le& [\zeta(H)]^{1/2}  \, \|
 \bs^{(p)} - \bs_r^{\diamond}  \|_{\ell_2}
+ \frac{ 2 \| \bs_r^{\diamond} - \bs^{\diamond} \|_{\ell_2} + 2 \|H^T \, (H  H^T)^{-1} \bn \|_{\ell_2}  }{\sqrt{\rho_{2r,\min}(H)}}
\esea
where (\ref{eq:thm4_proof1}) follows from the definition
(\ref{eq:def_SSQ_2}) and the fact that $\bs^{(p+1)}-\bs_r^{\diamond}$
is $2r$-sparse; (\ref{eq:thm4_proof2}) follows by using
(\ref{eq:thm4cond1}); in (\ref{eq:thm4_proof3}), we use the triangle
inequality
($\| \ba + \bb\|_{\ell_2} \leq \| \ba \|_{\ell_2} + \| \bb\|_{\ell_2}$),
definition (\ref{eq:cE}), and the fact that the
eigenvalues of $H^T \, (H H^T)^{-1} \, H$ are $0$ and $1$;
(\ref{eq:thm4_proof4}) follows along the same lines as
(\ref{eq:thm3_proof2})--(\ref{eq:thm3_proof5}) with
$\bs^{\diamond}$ replaced by $\bs_r^{\diamond}$;
(\ref{eq:thm4_proof5}) follows from (\ref{eq:Qfunction}) and (\ref{eq:thm4cond1});
(\ref{eq:thm4_proof6}) holds due to the the triangle inequality
and the fact that the eigenvalues of $H^T \, (H H^T)^{-1} \, H$ are $0$ and $1$;
finally, (\ref{eq:thm4_proof7}) follows from the same lines as
(\ref{eq:thm3_proof6})--(\ref{eq:thm3_proof8}) with
$\bs^{\diamond}$ replaced by $\bs_r^{\diamond}$.

From (\ref{eq:thm4_proof7}), we can see by induction that,  for $p \geq 1$
\begin{eqnarray}
\label{eq:thm4_proof8}
& & \!\!\!\!\!\!\!\!\!\!\!\!\!\!
\|\bs^{(p)}-\bs_r^{\diamond}\|_{\ell_2}
\leq [ \zeta(H) ]^{p/2} \, \| \bs^{(0)}  - \bs_r^{\diamond} \|_{\ell_2}
+ 2 \frac{ \sum_{i=0}^{p-1} [ \zeta(H) ]^{i/2}
  }{\sqrt{\rho_{2r,\min}(H)}} \,  \big[ \| \bs^{\diamond} - \bs_r^{\diamond} \|_{\ell_2} + \|H^T \, (H  H^T)^{-1} \bn \|_{\ell_2} \big]
\nonumber
\\
& & =  [ \zeta(H) ]^{p/2} \,
\| \bs^{(0)}  - \bs_r^{\diamond} \|_{\ell_2}
+ 2 \frac{1}{\sqrt{\rho_{2r,\min}(H)}} \, \frac{1 -  [ \zeta(H) ]^{p/2}
}{1 -  [\zeta(H)]^{1/2} } \big[ \| \bs^{\diamond} - \bs_r^{\diamond}
\|_{\ell_2} + \|H^T \, (H  H^T)^{-1} \bn \|_{\ell_2} \big] \hspace{0.35in}
\end{eqnarray}
where $\bs^{(0)}$ is the initial signal estimate. Since the condition
 (\ref{eq:thm4cond}) implies
that
$\zeta(H)$ in (\ref{eq:zeta}) is nonnegative and smaller than one
[see (\ref{eq:zeta2})], we have
\be
\lim_{p \nearrow +\infty }  [ \zeta(H) ]^{p/2} = 0
\ee
the first term in (\ref{eq:thm4_proof8}) disappears (i.e.\ the effect
of the initial signal estimate washes out), and the claim follows.
\end{IEEEproof}

\appsection{Proof of Theorem \ref{theorem5}}
\label{AppProofThm5}

\begin{IEEEproof}[Proof of Theorem \ref{theorem5}]
When conditions (1) and (2) of Theorem \ref{theorem5} hold, we have
$\spark(H)=N+1 > 2 \, r^{\diamond}$ and the condition of
\cite[Theorem 2]{BrucksteinDonohoElad} is satisfied. Therefore, $\bs^\diamond$ is the unique solution of the $({\rm P}_0)$ problem, according to
\cite[Theorem 2]{BrucksteinDonohoElad}.
We now consider the USS function under different sparsity level $r$.

For $r=r^{\diamond}$, the ML estimate of $\btheta$ is
 $\widehat{\btheta}_{\sML}(r^{\diamond})=  \big( \widehat{\bs}_{\sML}(r^{\diamond}),
\widehat{\sigma}^2_{\sML}(r^{\diamond}) \big) = ( \bs^\diamond, 0 )$ and unique, since it
leads to infinite likelihood function (\ref{eq:likelihoodfunction}) and
no other $\btheta$ yields infinite likelihood, due to the fact that $\bs^\diamond$ is the unique solution of
the $({\rm P}_0)$ problem. Furthermore, since
(\ref{eq:theorem_cond}) holds, we have $N - r^{\diamond} - 2 >0$ and therefore ${\rm USS}(r)$ is infinite as well.
 Note that
\be
{\rm USS}(r) = \widetilde{{\rm USS}} \big( r,\widehat{\sigma}^2_{\sML}(r) \big)
\ee
where
\be
\label{eq:USS_tilde}
\widetilde{{\rm USS}}(r,\sigma^2) = - \half \, r \, \ln\big( \frac{N}{m} \big)
- \half \, ( N - r - 2 ) \, \ln\Big( \frac{ \sigma^2 }{ \by^T \, (H \, H^T)^{-1} \, \by / N  } \Big).
\ee
Now,
\begin{eqnarray}
\label{eq:theorem_proof}
\lim_{\sigma^2\ \searrow 0}  \frac{\widetilde{{\rm USS}}(r^\diamond,\sigma^2) }{\ln (1/\sigma^2)}
= \half \, (N - r^\diamond - 2) > 0
\end{eqnarray}
specifying the rate of growth to infinity of $\widetilde{{\rm
USS}}(r^{\diamond},\sigma^2)$ as $\sigma^2$ approaches the ML estimate
$\widehat{\sigma}^2_{\sML}(r^{\diamond})=0$.
%i.e.\ $\widetilde{{\rm USS}}(r^\diamond,\sigma^2)$ goes to infinity at
%the speed of $\half \, (N - r^\diamond - 2) \ln (1/\sigma^2)$ as
%$\sigma^2$ approaches the ML estimate
%$\widehat{\sigma}^2_{\sML}(r)=0$.

For $r < r^{\diamond}$, $\by \neq H \, \bs$ for any $r$-sparse vector
$\bs$; consequently, $\sigma^2_{\sML}(r) > 0$ and ${\rm USS}(r)$ is finite.

For $r > r^{\diamond}$, the ML estimate of $\sigma^2$ must be
$\widehat{\sigma}^2_{\sML}(r)=0$, which leads to infinite likelihood.
However, in this case,
\begin{eqnarray}
\label{eq:theorem_proof2}
\lim_{\sigma^2\ \searrow 0}  \frac{\widetilde{{\rm USS}}(r,\sigma^2) }{\ln (1/\sigma^2)}
= \half \, (N - r - 2) < \half \, (N - r^\diamond - 2).
\end{eqnarray}
Therefore, if $r \ge N-2$, ${\rm USS}(r)$ is either finite or goes to negative infinity.
For $r^{\diamond}<r<N-2$, ${\rm USS}(r)$ is infinitely large,
but the rate at which $\widetilde{{\rm
USS}}(r,\sigma^2)$ grows to infinity as $\sigma^2$ approaches the ML estimate
$\widehat{\sigma}^2_{\sML}(r)=0$ is smaller than that specified by
 (\ref{eq:theorem_proof}).
%
%In summary, we have shown that, for $r \neq r^{\diamond}$, ${\rm USS}(r)$ is either finite or,
%if infinitely large, the rate at which $\widetilde{{\rm
%USS}}(r,\sigma^2)$ grows to infinity is smaller than that specified by
% (\ref{eq:theorem_proof}).

The claim follows by combining the above conclusions.
\end{IEEEproof}

%********************************************************************
%***************************** Bibliography *************************
%********************************************************************


\begin{thebibliography}{99}



\bibitem{GorodnitskyRao} I.F.\ Gorodnitsky and B.D.\ Rao, ``Sparse
signal reconstruction from limited data using FOCUSS: A re-weighted
minimum norm algorithm,'' {\em IEEE Trans.\ Signal Processing},\/
vol.\ 45, pp.\ 600--616, Mar.\ 1997.




\bibitem{CandesRombergTao_ITpaper} E.J.\ Cand\`es, J.\ Romberg, and
T.\ Tao, ``Robust uncertainty principles: exact signal reconstruction
from highly incomplete frequency information,'' {\em IEEE Trans.\
Inform.\ Theory}\/, vol.\ 52, pp.\ 489--509, Feb.\ 2006.

\bibitem{BajwaHauptSayeedNowak} W.U.\ Bajwa, J.D.\ Haupt, A.M.\
Sayeed, and R.D.\ Nowak, ``Joint source-channel communication for
distributed estimation in sensor networks,'' {\em IEEE Trans.\
Inform.\ Theory}\/, vol.\ 53, pp.\ 3629--3653, Oct.\ 2007.

\bibitem{LustigDonohoPauly} M.\ Lustig, D.\ Donoho, and J.M.\ Pauly,
``Sparse MRI: The application of compressed sensing for rapid MR
imaging,'' {\em Magnetic Resonance in Medicine},\/ vol.\ 58, pp.\
1182--1195, Dec.\ 2007.


\bibitem{BaraniukSpMag} R.G.\ Baraniuk,
``Compressive sensing,'' {\em IEEE Signal Processing Mag.},\/ vol.\ 24, pp.\ 118--121, Jul.\ 2007.


\bibitem{IEEESpMag} {\em IEEE Signal Processing Mag.\ Special
Issue on Sensing, Sampling, and Compression},\/ Mar.\ 2008.


\bibitem{CandesTao}   E.J.\ Cand\`es and T.\ Tao,
``Decoding by linear programming,'' {\em IEEE Trans.\ Inform.\
Theory}\/, vol.\ 51, pp.\ 4203--4215, Dec.\ 2005.




\bibitem{BrucksteinDonohoElad}
A.M.\ Bruckstein, D.L.\ Donoho, and M.\ Elad, ``From sparse solutions
of systems of equations to sparse modeling of signals and images,''
{\em SIAM Review},\/ vol.\ 51, pp.\ 34--81,
Mar.\ 2009.





\bibitem{Natarajan} B.K.\ Natarajan, ``Sparse approximate solutions
to linear systems,'' {\em SIAM J.\ Comput.},\/ vol.\ 24, pp. 227--234,
1995.



\bibitem{ChenDonohoSaunders} S.\ Chen, D.\ Donoho, and M. Saunders,
``Atomic decomposition by basis pursuit,'' {\em SIAM J.\ Sci.\ Comp.}\/,
vol.\ 20, no.\ 1, pp. 33--61, 1998.



\bibitem{CandesRombergTao}
E.J.\ Cand\`es, J. Romberg, and T. Tao, ``Stable signal recovery
from incomplete and inaccurate information,'' {\em  Commun.\
Pure and Applied Mathematics}\/, vol.\ 59, pp.\ 1207--1233, Aug.\ 2006.





\bibitem{CandesTaoDantzig} E.\ Cand\`es and T.\ Tao, ``The Dantzig
selector: statistical estimation when $p$ is much larger than $n$,''
{\em Ann.\ Stat.}\/, vol.\ 35, pp.\ 2313--2351, Dec.\ 2007.





\bibitem{FigueiredoNowakWright} M.A.T.\ Figueiredo, R.D.\ Nowak, and
S.J.\ Wright, ``Gradient projection for sparse reconstruction:
application to compressed sensing and other inverse problems,'' {\em
 IEEE J.\ Select.\ Areas Signal Processing}\/, vol.\ 1,
pp.\ 586--597, Dec.\
 2007.



\bibitem{MallatZhang} S. Mallat, Z. Zhang, ``Matching pursuits with
time-frequency dictionaries,'' {\em IEEE Trans.\ Signal Processing}, \/
vol.\ 41, pp.\ 3397--3415, 1993.



\bibitem{Tropp} J.A.\ Tropp, ``Greed is good: Algorithmic results for sparse approximation'',
{\em IEEE Trans.\ Inform.\ Theory}\/, vol.\ 50, pp.\ 2231--2242, Oct.\ 2004.



\bibitem{TroppGilbert} J.A.\ Tropp and A.C.\ Gilbert, ``Signal
recovery from random measurements via orthogonal matching pursuit,''
{\em IEEE Trans.\ Inform.\ Theory}\/, vol.\ 53, pp.\ 4655--4666,
Dec. 2007.




\bibitem{NeedelTroppCoSaMP} D.\ Needell and J.A.\ Tropp,
``\textsc{CoSaMP}: Iterative signal recovery from
incomplete and inaccurate samples,''  {\em
 Appl.\ Comp.\ Harmonic Anal.},\/  vol.\ 26, pp.\ 301--321, May 2009.




\bibitem{HerrityGilbertTropp} K.K.\ Herrity, A.C.\ Gilbert, and J.A.\
Tropp, ``Sparse approximation via iterative thresholding,'' in {\em
Proc.\ Int.\ Conf.\ Acoust., Speech, Signal Processing,}\/ Toulouse,
France, May 2006, pp.\ 624--627.




\bibitem{BlumensathDavies0} T.\ Blumensath and M.E.\ Davies,
``Iterative thresholding for sparse approximations,'' {\em J.\ Fourier
Anal.\ Appl.}, vol.\ 14, pp.\ 629--654, Dec. 2008.



\bibitem{BlumensathDavies} T.\ Blumensath and M.E.\ Davies,
``Iterative hard thresholding for compressed sensing,''
{\em Appl.\ Comp.\ Harmonic Anal.},\/ vol.\ 27, pp.\ 265--274, Nov.\ 2009.


\bibitem{BlumensathDavies2} T.\ Blumensath and M.E.\ Davies,
``Normalized iterative hard thresholding; guaranteed stability and
performance,'' {\em IEEE J.\ Select.\ Areas Signal Processing}\/,
vol.\ 4, pp.\ 298--309, Apr.\ 2010.





\bibitem{WipfRao} D.P.\ Wipf and B.D.\ Rao, ``Sparse Bayesian learning
for basis selection,'' {\em IEEE Trans.\ Signal
Processing},\/ vol.\ 52, pp.\ 2153--2164, Aug.\ 2004.



\bibitem{JiXueCarin} S.\ Ji, Y.\ Xue, and L.\ Carin, ``Bayesian
compressive sensing,'' {\em  IEEE
Trans.\ Signal Processing},\/ vol.\ 56, pp.\ 2346--2356, Jun.\ 2008.



\bibitem{QiuDogandzic10} K.\ Qiu and A.\ Dogand\v{z}i\'c,
``Variance-component based sparse signal reconstruction and model
selection,'' {\em IEEE Trans.\ Signal Processing}\/, vol.\ 58, pp.\
2935--2952, Jun.\ 2010.



\bibitem{MalekiDonoho} A.\ Maleki and D.L.\ Donoho, ``Optimally tuned
iterative thresholding algorithms for compressed sensing,''
{\em IEEE J.\ Select.\ Areas Signal Processing}\/, vol.\ 4, pp.\
330--341, Apr.\ 2010.


%
%\bibitem{DaubechiesDefriesDeMol} I.\ Daubechies, M.\ Defries,C. De Mol,
%``An iterative thresholding algorithm for linear inverse problems
%with a sparsity constraint,'' {\em Commun. Pure Appl. Math.},\/ vol.\ 57, pp.\ 1413¨C1457, 2004.




\bibitem{QNDE09} A.\ Dogand\v{z}i\'c and K.\ Qiu,
 ``Automatic hard
thresholding for sparse signal reconstruction from NDE measurements,"
  in {\em Rev.\ Progress Quantitative Nondestructive Evaluation},\/
  D.O.\ Thompson and D.E.\
Chimenti (Eds.), Melville NY: Amer.\ Inst.\ Phys., vol. 29, 2010, pp.\
806--813.



\bibitem{CISSDORE}
 K.\ Qiu and A.\ Dogand\v{z}i\'c, ``Double
overrelaxation thresholding methods for sparse signal
reconstruction,'' in {\em Proc.\ 44th Annu.\ Conf.\ Inform.\ Sci.\
Syst.},\/ Princeton, NJ, Mar.\ 2010.






\bibitem{HeLiu} Y.X.\ He and C.H.\ Liu, ``The dynamic ECME algorithm,''
Dept.\ Statistics, Purdue Univ., Tech.\ Report, 2009.







\bibitem{DLR} A.P.\ Dempster, N.M.\ Laird, and D.B.\ Rubin,
``Maximum likelihood from incomplete data via the EM algorithm,''
{\em J.\ R.\ Stat.\ Soc., Ser.\ B}\/, vol.\ 39, pp.\ 1--38, July 1977.





\bibitem{JeffWu} C.F.J.\ Wu,
``On the convergence properties of the EM algorithm,''
{\em Ann.\ Stat.}\/, vol.\ 11, pp.\ 95--103, Mar. 1983.






\bibitem{McLachlanKrishnan} G.J.\ McLachlan and T.\ Krishnan, {\em
The EM Algorithm and Extensions},\/  2nd.\ ed.,
New York: Wiley, 2008.



\bibitem{LiuRubin} C.H.\ Liu and D.B.\ Rubin, ``The ECME algorithm: A simple extension of
EM and ECM with fast monotone convergence,'' {\em  Biometrika},\/ vol.\
81, pp.\ 633--648, Dec.\ 1994.




\bibitem{Kay} S.M.\ Kay, {\it Fundamentals of
Statistical Signal Processing: Estimation Theory},\/ Englewood
Cliffs, NJ: Prentice-Hall, 1993.








\bibitem{Thisted} R.A.\ Thisted,
{\it Elements of Statistical Computing: Numerical Computation},\/
%New York:
Chapman \& Hall, 1988.




\bibitem{Harville} D.A.\ Harville,
{\em Matrix Algebra From a Statistician's Perspective},\/
New York: Springer-Verlag, 1997.






\bibitem{Colley} S.J.\ Colley, {\em Vector Calculus},\/ 3rd.\ ed.,
Upper Saddle River, NJ: Prentice Hall, 2006.


\bibitem{DonohoHuo2001} D.L.\ Donoho and X.\ Huo, ``Uncertainty principles and ideal atomic decomposition'',
{\em IEEE Trans.\ Inform.\ Theory}\/, vol.\ 47, pp.\ 2845--2862, Nov.\ 2001.


\bibitem{DonohoElad2003} D.L.\ Donoho and M.\ Elad, ``Optimally sparse
representation in general (nonorthogonal) dictionaries via $\ell_1$
minimization'', {\em Proc. Nat.\ Acad.\ Sci. USA}\/, vol.\ 100, pp.\
2197--2202, Mar.\ 2003.



\bibitem{CandesTao_NearOptimal}   E.J.\ Cand\`es and T.\ Tao,
``Near-optimal signal recovery from random
projections: universal encoding strategies?'' {\em IEEE Trans.\ Inform.\
Theory}\/, vol.\ 52, pp.\ 5406--5425, Dec.\ 2006.

\bibitem{OppenheimSchafer} A.V.\ Oppenheim and R.W.\ Schafer {\em
Discrete-time Signal Processing},\/ 3rd ed., Upper Saddle River, NJ:
Prentice Hall, 2010.

\bibitem{DaubechiesBook} I.\ Daubechies, {\em Ten Lectures on Wavelets},\/ Philadelphia: SIAM, 1992.

\bibitem{DoTranGan} T.T.\ Do, T.D.\ Tran, and L.\ Gan, ``Compressive
sampling with structurally random matrices,'' in {\em Proc.\ Int.\
Conf.\ Acoust., Speech, Signal Processing,}\/ Las Vegas, NV, pp.\ 3369--3372, Apr.\
2008.


\end{thebibliography}
\end{document}